\documentclass[twocolumn]{aastex6}

\pdfoutput=1
\usepackage{epstopdf}
\usepackage{amsmath}
\usepackage{comment} 
\usepackage{appendix}
\usepackage{enumerate}

\usepackage{color}

\usepackage{lineno}
\slugcomment{\today}

\newcommand{\sigMINUS}{\sigma_{-}}
\newcommand{\sigPLUS}{\sigma_{+}}

\newcommand{\plus}{\texttt{+}}

\newcommand{\LCDM}{$\Lambda \text{CDM}$}
\newcommand{\OCDM}{$o \text{CDM}$} 
\newcommand{\WCDM}{$w$\text{CDM}}
\newcommand{\WACDM}{$w_0w_a$\text{CDM}}

\newcommand{\mumodel}{\mu_{\rm model}}

\begin{document}

\newcommand{\massallstepa}{\ensuremath{0.048}}
\newcommand{\massallstepe}{\ensuremath{0.009}}
\newcommand{\massalllow}{\ensuremath{411}}
\newcommand{\massallhigh}{\ensuremath{611}}
\newcommand{\masscallstepa}{\ensuremath{0.053}}
\newcommand{\masscallstepe}{\ensuremath{0.009}}
\newcommand{\masscalllow}{\ensuremath{411}}
\newcommand{\masscallhigh}{\ensuremath{611}}
\newcommand{\masspsstepa}{\ensuremath{0.034}}
\newcommand{\masspsstepe}{\ensuremath{0.016}}
\newcommand{\masspslow}{\ensuremath{116}}
\newcommand{\masspshigh}{\ensuremath{163}}
\newcommand{\masscpsstepa}{\ensuremath{0.037}}
\newcommand{\masscpsstepe}{\ensuremath{0.016}}
\newcommand{\masscpslow}{\ensuremath{116}}
\newcommand{\masscpshigh}{\ensuremath{163}}
\newcommand{\masssdssstepa}{\ensuremath{0.048}}
\newcommand{\masssdssstepe}{\ensuremath{0.016}}
\newcommand{\masssdsslow}{\ensuremath{126}}
\newcommand{\masssdsshigh}{\ensuremath{209}}
\newcommand{\masscsdssstepa}{\ensuremath{0.051}}
\newcommand{\masscsdssstepe}{\ensuremath{0.016}}
\newcommand{\masscsdsslow}{\ensuremath{126}}
\newcommand{\masscsdsshigh}{\ensuremath{209}}
\newcommand{\masssnlsstepa}{\ensuremath{0.058}}
\newcommand{\masssnlsstepe}{\ensuremath{0.018}}
\newcommand{\masssnlslow}{\ensuremath{140}}
\newcommand{\masssnlshigh}{\ensuremath{96}}
\newcommand{\masscsnlsstepa}{\ensuremath{0.052}}
\newcommand{\masscsnlsstepe}{\ensuremath{0.017}}
\newcommand{\masscsnlslow}{\ensuremath{140}}
\newcommand{\masscsnlshigh}{\ensuremath{96}}
\newcommand{\masslowzstepa}{\ensuremath{0.042}}
\newcommand{\masslowzstepe}{\ensuremath{0.031}}
\newcommand{\masslowzlow}{\ensuremath{29}}
\newcommand{\masslowzhigh}{\ensuremath{143}}
\newcommand{\massclowzstepa}{\ensuremath{0.034}}
\newcommand{\massclowzstepe}{\ensuremath{0.031}}
\newcommand{\massclowzlow}{\ensuremath{29}}
\newcommand{\massclowzhigh}{\ensuremath{143}}
\newcommand{\masshststepa}{\ensuremath{nan}}
\newcommand{\masshststepe}{\ensuremath{nan}}
\newcommand{\masshstlow}{\ensuremath{0}}
\newcommand{\masshsthigh}{\ensuremath{0}}
\newcommand{\masschststepa}{\ensuremath{nan}}
\newcommand{\masschststepe}{\ensuremath{nan}}
\newcommand{\masschstlow}{\ensuremath{0}}
\newcommand{\masschsthigh}{\ensuremath{0}}
\newcommand{\ruta}{\ensuremath{365}}
\newcommand{\rutxa}{\ensuremath{332}}
\newcommand{\rutxb}{\ensuremath{303}}
\newcommand{\rutxc}{\ensuremath{303}}
\newcommand{\rutxd}{\ensuremath{293}}
\newcommand{\rutxe}{\ensuremath{288}}
\newcommand{\rutxf}{\ensuremath{288}}
\newcommand{\rutxg}{\ensuremath{282}}
\newcommand{\rutya}{\ensuremath{33}}
\newcommand{\rutyb}{\ensuremath{29}}
\newcommand{\rutyc}{\ensuremath{0}}
\newcommand{\rutyd}{\ensuremath{10}}
\newcommand{\rutye}{\ensuremath{5}}
\newcommand{\rutyf}{\ensuremath{0}}
\newcommand{\rutyg}{\ensuremath{6}}
\newcommand{\allPS}{\ensuremath{365}}

\newcommand{\alphanumn}{\ensuremath{(0.154\pm0.005)}}
\newcommand{\betanumn}{\ensuremath{(3.030\pm0.063)}}
\newcommand{\gammanumn}{\ensuremath{(0.053\pm0.009)}}
\newcommand{\alphanuma}{\ensuremath{(0.156\pm0.007)}}
\newcommand{\betanuma}{\ensuremath{(3.030\pm0.064)}}
\newcommand{\gammanuma}{\ensuremath{(0.053\pm0.009)}}
\newcommand{\alphanumza}{\ensuremath{(-0.007\pm0.024)}}
\newcommand{\alphanumb}{\ensuremath{(0.154\pm0.006)}}
\newcommand{\betanumb}{\ensuremath{(3.139\pm0.099)}}
\newcommand{\gammanumb}{\ensuremath{(0.052\pm0.009)}}
\newcommand{\betanumzb}{\ensuremath{(-0.348\pm0.289)}}
\newcommand{\alphanumc}{\ensuremath{(0.155\pm0.005)}}
\newcommand{\betanumc}{\ensuremath{(3.028\pm0.063)}}
\newcommand{\gammanumc}{\ensuremath{(0.075\pm0.015)}}
\newcommand{\gammanumzc}{\ensuremath{(-0.079\pm0.041)}}
\newcommand{\alphanumd}{\ensuremath{(0.158\pm0.008)}}
\newcommand{\betanumd}{\ensuremath{(3.138\pm0.098)}}
\newcommand{\gammanumd}{\ensuremath{(0.076\pm0.015)}}
\newcommand{\alphanumzd}{\ensuremath{(-0.015\pm0.024)}}
\newcommand{\betanumzd}{\ensuremath{(-0.348\pm0.285)}}
\newcommand{\gammanumzd}{\ensuremath{(-0.082\pm0.041)}}
\newcommand{\nalphanumd}{\ensuremath{(0.153\pm0.007)}}
\newcommand{\nbetanumd}{\ensuremath{(3.129\pm0.091)}}
\newcommand{\ngammanumd}{\ensuremath{(0.088\pm0.016)}}
\newcommand{\nalphanumzd}{\ensuremath{(-0.025\pm0.024)}}
\newcommand{\nbetanumzd}{\ensuremath{(-0.492\pm0.297)}}
\newcommand{\ngammanumzd}{\ensuremath{(-0.067\pm0.049)}}

\newcommand{\numCSP}{\ensuremath{26}}
\newcommand{\rumCSP}{\ensuremath{19}}
\newcommand{\numCFAt}{\ensuremath{78}}
\newcommand{\rumCFAt}{\ensuremath{7}}
\newcommand{\numCFAf}{\ensuremath{41}}
\newcommand{\rumCFAf}{\ensuremath{2}}
\newcommand{\numCFAo}{\ensuremath{9}}
\newcommand{\rumCFAo}{\ensuremath{-4}}
\newcommand{\numCFAw}{\ensuremath{18}}
\newcommand{\rumCFAw}{\ensuremath{1}}
\newcommand{\numSDSS}{\ensuremath{335}}
\newcommand{\numPS}{\ensuremath{279}}
\newcommand{\numSNLS}{\ensuremath{236}}
\newcommand{\numSCP}{\ensuremath{3}}
\newcommand{\numGOOD}{\ensuremath{15}}
\newcommand{\numCAN}{\ensuremath{8}}
\newcommand{\numTOT}{\ensuremath{1048 }}

\newcommand{\salphanumg}{\ensuremath{0.154\pm0.006}}
\newcommand{\sbetanumg}{\ensuremath{3.02\pm0.06}}
\newcommand{\sgammanumg}{\ensuremath{0.053\pm0.009}}
\newcommand{\sintnumg}{\ensuremath{0.09}}
\newcommand{\sdispnumg}{\ensuremath{0.14}}
\newcommand{\salphanums}{\ensuremath{0.139\pm0.013}}
\newcommand{\sbetanums}{\ensuremath{3.01\pm0.14}}
\newcommand{\sgammanums}{\ensuremath{0.045\pm0.020}}
\newcommand{\sintnums}{\ensuremath{0.09}}
\newcommand{\sdispnums}{\ensuremath{0.14}}
\newcommand{\salphanumd}{\ensuremath{0.159\pm0.010}}
\newcommand{\sbetanumd}{\ensuremath{3.08\pm0.13}}
\newcommand{\sgammanumd}{\ensuremath{0.057\pm0.015}}
\newcommand{\sintnumd}{\ensuremath{0.09}}
\newcommand{\sdispnumd}{\ensuremath{0.14}}
\newcommand{\salphanuml}{\ensuremath{0.154\pm0.011}}
\newcommand{\sbetanuml}{\ensuremath{2.99\pm0.15}}
\newcommand{\sgammanuml}{\ensuremath{0.076\pm0.030}}
\newcommand{\sintnuml}{\ensuremath{0.10}}
\newcommand{\sdispnuml}{\ensuremath{0.15}}
\newcommand{\salphanump}{\ensuremath{0.167\pm0.012}}
\newcommand{\sbetanump}{\ensuremath{3.02\pm0.12}}
\newcommand{\sgammanump}{\ensuremath{0.039\pm0.016}}
\newcommand{\sintnump}{\ensuremath{0.08}}
\newcommand{\sdispnump}{\ensuremath{0.14}}
\newcommand{\calphanumg}{\ensuremath{0.156\pm0.005}}
\newcommand{\cbetanumg}{\ensuremath{3.69\pm0.09}}
\newcommand{\cgammanumg}{\ensuremath{0.054\pm0.009}}
\newcommand{\cintnumg}{\ensuremath{0.11}}
\newcommand{\cdispnumg}{\ensuremath{0.14}}
\newcommand{\calphanums}{\ensuremath{0.139\pm0.013}}
\newcommand{\cbetanums}{\ensuremath{3.59\pm0.17}}
\newcommand{\cgammanums}{\ensuremath{0.037\pm0.020}}
\newcommand{\cintnums}{\ensuremath{0.10}}
\newcommand{\cdispnums}{\ensuremath{0.14}}
\newcommand{\calphanumd}{\ensuremath{0.156\pm0.009}}
\newcommand{\cbetanumd}{\ensuremath{4.04\pm0.18}}
\newcommand{\cgammanumd}{\ensuremath{0.059\pm0.015}}
\newcommand{\cintnumd}{\ensuremath{0.11}}
\newcommand{\cdispnumd}{\ensuremath{0.14}}
\newcommand{\calphanuml}{\ensuremath{0.156\pm0.011}}
\newcommand{\cbetanuml}{\ensuremath{3.53\pm0.20}}
\newcommand{\cgammanuml}{\ensuremath{0.067\pm0.030}}
\newcommand{\cintnuml}{\ensuremath{0.12}}
\newcommand{\cdispnuml}{\ensuremath{0.15}}
\newcommand{\calphanump}{\ensuremath{0.167\pm0.012}}
\newcommand{\cbetanump}{\ensuremath{3.51\pm0.16}}
\newcommand{\cgammanump}{\ensuremath{0.041\pm0.016}}
\newcommand{\cintnump}{\ensuremath{0.10}}
\newcommand{\cdispnump}{\ensuremath{0.14}}
\newcommand{\oalphanump}{\ensuremath{0.161\pm0.011}}
\newcommand{\obetanump}{\ensuremath{2.93\pm0.11}}
\newcommand{\ogammanump}{\ensuremath{0.064\pm0.018}}
\newcommand{\ointnump}{\ensuremath{0.09}}
\newcommand{\odispnump}{\ensuremath{0.16}}
\newcommand{\oalphanums}{\ensuremath{0.128\pm0.013}}
\newcommand{\obetanums}{\ensuremath{3.08\pm0.14}}
\newcommand{\ogammanums}{\ensuremath{0.054\pm0.023}}
\newcommand{\ointnums}{\ensuremath{0.09}}
\newcommand{\odispnums}{\ensuremath{0.18}}
\newcommand{\oalphanumd}{\ensuremath{0.149\pm0.009}}
\newcommand{\obetanumd}{\ensuremath{3.11\pm0.12}}
\newcommand{\ogammanumd}{\ensuremath{0.078\pm0.016}}
\newcommand{\ointnumd}{\ensuremath{0.09}}
\newcommand{\odispnumd}{\ensuremath{0.15}}
\newcommand{\oalphanuml}{\ensuremath{0.147\pm0.011}}
\newcommand{\obetanuml}{\ensuremath{3.00\pm0.13}}
\newcommand{\ogammanuml}{\ensuremath{0.077\pm0.032}}
\newcommand{\ointnuml}{\ensuremath{0.11}}
\newcommand{\odispnuml}{\ensuremath{0.16}}
\newcommand{\oalphanumg}{\ensuremath{0.148\pm0.005}}
\newcommand{\obetanumg}{\ensuremath{3.02\pm0.06}}
\newcommand{\ogammanumg}{\ensuremath{0.072\pm0.010}}
\newcommand{\ointnumg}{\ensuremath{0.10}}
\newcommand{\odispnumg}{\ensuremath{0.17}}
\newcommand{\ogammanumeg}{\ensuremath{0.086\pm0.016}}
\newcommand{\egammanumeg}{\ensuremath{-0.056\pm0.050}}
\newcommand{\pdispnump}{\ensuremath{0.144}}
\newcommand{\pdispnumn}{\ensuremath{0.149}}

\newcommand{\sysstat}{\ensuremath{0.031}}
\newcommand{\syssys}{\ensuremath{0.025}}
\newcommand{\sysbeta}{\ensuremath{0.007}}
\newcommand{\sysint}{\ensuremath{0.005}}
\newcommand{\syssalt}{\ensuremath{0.014}}
\newcommand{\syssurv}{\ensuremath{0.009}}

\newcommand{\CMBBAOxOM}{\ensuremath{0.312 \pm 0.013}}
\newcommand{\CMBBAOxW}{\ensuremath{-0.991 \pm 0.074}}
\newcommand{\CMBBAOxH}{\ensuremath{67.508 \pm 1.633}}
\newcommand{\CMBHSTxOM}{\ensuremath{0.265 \pm 0.013}}
\newcommand{\CMBHSTxW}{\ensuremath{-1.188 \pm 0.062}}
\newcommand{\CMBHSTxH}{\ensuremath{73.332 \pm 1.729}}
\newcommand{\CMBBAOHSTxOM}{\ensuremath{0.289 \pm 0.011}}
\newcommand{\CMBBAOHSTxW}{\ensuremath{-1.119 \pm 0.068}}
\newcommand{\CMBBAOHSTxH}{\ensuremath{70.539 \pm 1.425}}
\newcommand{\SNCMBxOM}{\ensuremath{0.307 \pm 0.012}}
\newcommand{\SNCMBxW}{\ensuremath{-1.026 \pm 0.041}}
\newcommand{\SNCMBxH}{\ensuremath{68.183 \pm 1.114}}
\newcommand{\SNCMBBAOxOM}{\ensuremath{0.307 \pm 0.008}}
\newcommand{\SNCMBBAOxW}{\ensuremath{-1.014 \pm 0.040}}
\newcommand{\SNCMBBAOxH}{\ensuremath{68.027 \pm 0.859}}
\newcommand{\SNCMBHSTxOM}{\ensuremath{0.293 \pm 0.010}}
\newcommand{\SNCMBHSTxW}{\ensuremath{-1.056 \pm 0.038}}
\newcommand{\SNCMBHSTxH}{\ensuremath{69.618 \pm 0.969}}
\newcommand{\SNCMBBAOHSTxOM}{\ensuremath{0.299 \pm 0.007}}
\newcommand{\SNCMBBAOHSTxW}{\ensuremath{-1.047 \pm 0.038}}
\newcommand{\SNCMBBAOHSTxH}{\ensuremath{69.013 \pm 0.791}}
\newcommand{\CMBBAOwOM}{\ensuremath{0.343 \pm 0.025}}
\newcommand{\CMBBAOwWO}{\ensuremath{-0.616 \pm 0.262}}
\newcommand{\CMBBAOwWA}{\ensuremath{-1.108 \pm 0.771}}
\newcommand{\CMBBAOwH}{\ensuremath{64.614 \pm 2.447}}
\newcommand{\CMBHSTwOM}{\ensuremath{0.265 \pm 0.015}}
\newcommand{\CMBHSTwWO}{\ensuremath{-1.024 \pm 0.347}}
\newcommand{\CMBHSTwWA}{\ensuremath{-0.789 \pm 1.338}}
\newcommand{\CMBHSTwH}{\ensuremath{73.397 \pm 1.961}}
\newcommand{\CMBBAOHSTwOM}{\ensuremath{0.343 \pm 0.026}}
\newcommand{\CMBBAOHSTwWO}{\ensuremath{-0.619 \pm 0.270}}
\newcommand{\CMBBAOHSTwWA}{\ensuremath{-1.098 \pm 0.781}}
\newcommand{\CMBBAOHSTwH}{\ensuremath{64.666 \pm 2.526}}
\newcommand{\SNCMBwOM}{\ensuremath{0.308 \pm 0.018}}
\newcommand{\SNCMBwWO}{\ensuremath{-1.009 \pm 0.159}}
\newcommand{\SNCMBwWA}{\ensuremath{-0.129 \pm 0.755}}
\newcommand{\SNCMBwH}{\ensuremath{68.188 \pm 1.768}}
\newcommand{\SNCMBBAOwOM}{\ensuremath{0.308 \pm 0.008}}
\newcommand{\SNCMBBAOwWO}{\ensuremath{-0.993 \pm 0.087}}
\newcommand{\SNCMBBAOwWA}{\ensuremath{-0.126 \pm 0.384}}
\newcommand{\SNCMBBAOwH}{\ensuremath{68.076 \pm 0.858}}
\newcommand{\SNCMBHSTwOM}{\ensuremath{0.287 \pm 0.011}}
\newcommand{\SNCMBHSTwWO}{\ensuremath{-0.905 \pm 0.101}}
\newcommand{\SNCMBHSTwWA}{\ensuremath{-0.742 \pm 0.465}}
\newcommand{\SNCMBHSTwH}{\ensuremath{70.393 \pm 1.079}}
\newcommand{\SNCMBBAOHSTwOM}{\ensuremath{0.300 \pm 0.008}}
\newcommand{\SNCMBBAOHSTwWO}{\ensuremath{-1.007 \pm 0.089}}
\newcommand{\SNCMBBAOHSTwWA}{\ensuremath{-0.222 \pm 0.407}}
\newcommand{\SNCMBBAOHSTwH}{\ensuremath{69.057 \pm 0.796}}
\newcommand{\CMBBAOoOM}{\ensuremath{0.310 \pm 0.008}}
\newcommand{\CMBBAOoOL}{\ensuremath{0.689 \pm 0.008}}
\newcommand{\CMBBAOoH}{\ensuremath{67.900 \pm 0.747}}
\newcommand{\CMBHSToOM}{\ensuremath{0.266 \pm 0.014}}
\newcommand{\CMBHSToOL}{\ensuremath{0.723 \pm 0.012}}
\newcommand{\CMBHSToH}{\ensuremath{73.205 \pm 1.788}}
\newcommand{\CMBBAOHSToOM}{\ensuremath{0.303 \pm 0.007}}
\newcommand{\CMBBAOHSToOL}{\ensuremath{0.694 \pm 0.007}}
\newcommand{\CMBBAOHSToH}{\ensuremath{68.723 \pm 0.675}}
\newcommand{\SNCMBoOM}{\ensuremath{0.299 \pm 0.024}}
\newcommand{\SNCMBoOL}{\ensuremath{0.698 \pm 0.019}}
\newcommand{\SNCMBoH}{\ensuremath{69.192 \pm 2.815}}
\newcommand{\SNCMBBAOoOM}{\ensuremath{0.309 \pm 0.007}}
\newcommand{\SNCMBBAOoOL}{\ensuremath{0.690 \pm 0.007}}
\newcommand{\SNCMBBAOoH}{\ensuremath{67.985 \pm 0.699}}
\newcommand{\SNCMBHSToOM}{\ensuremath{0.274 \pm 0.012}}
\newcommand{\SNCMBHSToOL}{\ensuremath{0.717 \pm 0.011}}
\newcommand{\SNCMBHSToH}{\ensuremath{72.236 \pm 1.572}}
\newcommand{\SNCMBBAOHSToOM}{\ensuremath{0.303 \pm 0.007}}
\newcommand{\SNCMBBAOHSToOL}{\ensuremath{0.695 \pm 0.007}}
\newcommand{\SNCMBBAOHSToH}{\ensuremath{68.745 \pm 0.684}}
\newcommand{\SNoOM}{\ensuremath{0.319 \pm 0.070}}
\newcommand{\SNoOL}{\ensuremath{0.733 \pm 0.113}}
\newcommand{\SNoH}{\ensuremath{60.256 \pm 23.168}}
\newcommand{\SNlOM}{\ensuremath{0.298 \pm 0.022}}
\newcommand{\SNlOL}{\ensuremath{0.702 \pm 0.022}}
\newcommand{\SNlH}{\ensuremath{60.086 \pm 22.921}}
\newcommand{\SNtsOM}{\ensuremath{0.284 \pm 0.012}}
\newcommand{\SNtsOL}{\ensuremath{0.716 \pm 0.012}}
\newcommand{\SNtkOM}{\ensuremath{0.348 \pm 0.040}}
\newcommand{\SNtkOL}{\ensuremath{0.827 \pm 0.068}}
\newcommand{\SNtwOM}{\ensuremath{0.350 \pm 0.035}}
\newcommand{\SNtwW}{\ensuremath{-1.251 \pm 0.144}}
\newcommand{\SNsOM}{\ensuremath{0.298 \pm 0.022}}
\newcommand{\SNsOL}{\ensuremath{0.702 \pm 0.022}}
\newcommand{\SNkOM}{\ensuremath{0.319 \pm 0.070}}
\newcommand{\SNkOL}{\ensuremath{0.733 \pm 0.113}}
\newcommand{\SNwOM}{\ensuremath{0.316 \pm 0.072}}
\newcommand{\SNwW}{\ensuremath{-1.090 \pm 0.220}}

\newcommand{\gps}{\ensuremath{g_{\rm p1}}}
\newcommand{\rps}{\ensuremath{r_{\rm p1}}}
\newcommand{\ips}{\ensuremath{i_{\rm p1}}}
\newcommand{\zps}{\ensuremath{z_{\rm p1}}}
\newcommand{\yps}{\ensuremath{y_{\rm p1}}}
\newcommand{\wps}{\ensuremath{w_{\rm p1}}}
\newcommand{\griz}{gri\zps}

\newcommand{\PS}{\protect \hbox {Pan-STARRS1}}

\newlength{\blackoutwidth}
\newcommand{\blackout}[1]
{
  \settowidth{\blackoutwidth}{#1}
  \rule[-0.3em]{\blackoutwidth}{1.125em}
}

\title{The Complete Light-curve Sample of Spectroscopically Confirmed Type Ia Supernovae from Pan-STARRS1 and Cosmological Constraints from The Combined Pantheon Sample}


\shorttitle{PS1 and the Pantheon Sample}
\shortauthors{Scolnic et al.}

\author{
D.~M.~Scolnic\altaffilmark{\uch,\hub},
D.~O.~Jones\altaffilmark{\jhu},
A.~Rest\altaffilmark{\jhu,\stsci},
Y.~C.~Pan\altaffilmark{\ucsc},
R.~Chornock\altaffilmark{\ohio},
R.~J.~Foley\altaffilmark{\ucsc},
M.~E.~Huber\altaffilmark{\hawaii},
R.~Kessler\altaffilmark{\uch},
G.~Narayan\altaffilmark{\stsci},
A.~G.~Riess\altaffilmark{\stsci,\jhu},
S.~Rodney\altaffilmark{\usc},
E.~Berger\altaffilmark{\CfA},
D.~J.~Brout\altaffilmark{\penn},
P.~J.~Challis\altaffilmark{\CfA},
M.~Drout\altaffilmark{\carneg},
D.~Finkbeiner\altaffilmark{\harvard},
R.~Lunnan\altaffilmark{\stock},
R.~P.~Kirshner\altaffilmark{\CfA},
N.~E.~Sanders\altaffilmark{\non},
E.~Schlafly\altaffilmark{\berkeley},
S.~Smartt\altaffilmark{\queens},
C.~W.~Stubbs\altaffilmark{\harvard,\CfA},
J.~Tonry\altaffilmark{\hawaii},
W.~M.~Wood-Vasey\altaffilmark{\pitt},
M.~Foley\altaffilmark{\notredame},
J.~Hand\altaffilmark{\pitt},
E.~Johnson\altaffilmark{\villanova},
W.~S.~Burgett\altaffilmark{\gmto},
K.~C.~Chambers\altaffilmark{\hawaii},
P.~W.~Draper\altaffilmark{\durham},
K.~W.~Hodapp\altaffilmark{\hawaii},
N.~Kaiser\altaffilmark{\hawaii},
R.~P.~Kudritzki\altaffilmark{\hawaii},
E.~A.~Magnier\altaffilmark{\hawaii},
N.~Metcalfe\altaffilmark{\durham},
F.~Bresolin\altaffilmark{\hawaii},
E.~Gall\altaffilmark{\queens},
R.~Kotak\altaffilmark{\queens},
M.~McCrum\altaffilmark{\queens},
K.~W.~Smith\altaffilmark{\queens}
}

\def\uch{1}
\def\hub{2}
\def\jhu{3}
\def\stsci{4}
\def\ucsc{5}
\def\ohio{6}
\def\hawaii{7}
\def\usc{8}
\def\CfA{9}
\def\penn{10}
\def\carneg{11}
\def\harvard{12}
\def\stock{13}
\def\non{14}
\def\berkeley{15}
\def\queens{16}
\def\pitt{17}
\def\notredame{18}
\def\villanova{19}
\def\gmto{20}
\def\durham{21}
\def\princeton{22}
\def\naval{23}
\def\max{24}

\altaffiltext{\uch}{Kavli Institute for Cosmological Physics, The University of Chicago, Chicago,IL 60637, USA.  Email: dscolnic@kicp.uchicago.edu}
\altaffiltext{\hub}{Hubble, KICP Fellow}
\altaffiltext{\jhu}{Department of Physics and Astronomy, Johns Hopkins University, 3400 North Charles Street, Baltimore, MD 21218, USA}
\altaffiltext{\stsci}{Space Telescope Science Institute, 3700 San Martin Drive, Baltimore, MD 21218}
\altaffiltext{\ucsc}{Department of Astronomy and Astrophysics, University of California, Santa Cruz, CA 92064, USA}
\altaffiltext{\ohio}{Astrophysical Institute, Department of Physics and Astronomy, 
251B Clippinger Lab, Ohio University, Athens, OH 45701, USA}
\altaffiltext{\hawaii}{Institute for Astronomy, University of Hawaii, 2680 Woodlawn Drive, Honolulu, HI 96822, USA}
\altaffiltext{\usc}{Department of Physics and Astronomy, University of South Carolina, 712 Main St., Columbia,SC 29208, USA}
\altaffiltext{\CfA}{Harvard-Smithsonian Center for Astrophysics, 60 Garden Street, Cambridge, MA 02138, USA}
\altaffiltext{\penn}{Department of Physics and Astronomy, University of Pennsylvania, Philadelphia, PA 19104, USA}
\altaffiltext{\carneg}{The Observatories of the Carnegie Institution for Science, 813 Santa Barbara St., Pasadena, CA 91101,USA}
\altaffiltext{\harvard}{Department of Physics, Harvard University, 17 Oxford Street, Cambridge MA 02138}
\altaffiltext{\stock}{Department of Astronomy, Stockholm University, SE}
\altaffiltext{\non}{No affiliation}
\altaffiltext{\berkeley}{ Lawrence Berkeley National Laboratory, One Cyclotron
Road, Berkeley, CA 94720, USA}
\altaffiltext{\queens}{Astrophysics Research Centre, School of Mathematics and Physics, Queens University Belfast, Belfast, BT71NN, UK}
\altaffiltext{\pitt}{PITT PACC, Department of Physics and Astronomy, University
of Pittsburgh, Pittsburgh, PA 15260, USA.}
\altaffiltext{\notredame}{Department of Physics, University of Notre Dame, 225
Nieuwland Science Hall, Notre Dame, IN, 46556-5670, USA.}
\altaffiltext{\villanova}{Department of Astrophysics and Planetary Science, Villanova University, Villanova, PA, 19085 USA} 
\altaffiltext{\gmto}{GMTO Corporation, 251 S. Lake Ave., Suite 300, Pasadena, CA 91101 USA}
\altaffiltext{\durham}{Department of Physics, University of Durham Science Laboratories, South Road Durham DH1 3LE, UK}
\altaffiltext{\princeton}{Department of Astrophysical Sciences, Princeton University, Princeton, NJ 08544, USA}
\altaffiltext{\naval}{US Naval Observatory, Flagstaff Station, Flagstaff, AZ 86001, USA}
\altaffiltext{\max}{Max-Planck-Institut für Astrophysik,
  Karl-Schwarzschild-Str. 1, DE-85748 Garching-bei-München, Germany}


\begin{abstract}

We present optical light curves, redshifts, and classifications for \allPS~spectroscopically confirmed Type Ia supernovae (SNe\,Ia) discovered by the Pan-STARRS1 (PS1) Medium Deep Survey. We detail improvements to the PS1 SN photometry, astrometry and calibration that reduce the systematic uncertainties in the PS1 SN\,Ia distances.  We combine the subset of \numPS~PS1 SN\,Ia ($0.03 < z < 0.68$) with useful distance estimates of SN\,Ia from SDSS, SNLS, various low-z and HST samples to form the largest combined sample of SN\,Ia consisting of a total of \numTOT~SN\,Ia ranging from $0.01<z<2.3$, which we call the `Pantheon Sample'.   When combining \textit{Planck 2015} CMB measurements with the Pantheon SN sample, we find $\Omega_m=\SNCMBxOM$
and $w=\SNCMBxW$ for the \WCDM~model.  When the SN and CMB constraints are combined with constraints from BAO and local $H_0$ measurements, the analysis yields  the most precise measurement of dark energy to date: $w_0=\SNCMBBAOHSTwWO$ and
$w_a=\SNCMBBAOHSTwWA$~for the \WACDM~model. Tension with a cosmological constant previously seen in an analysis of PS1 and low-z SNe has diminished after an increase of $2\times$ in the statistics of the PS1 sample, improved calibration and photometry, and stricter light-curve quality cuts.  We find the systematic uncertainties in our measurements of dark energy are almost as large as the statistical uncertainties, primarily due to limitations of modeling the low-redshift sample.  This must be addressed for future progress in using SN\,Ia to measure dark energy.
\end{abstract}
\section{Introduction}
\label{sec:intro}

Combining measurements of SN\,Ia distances \citep{Riess98,Perlmutter99} with
measurements of the baryon acoustic peak in the large-scale
correlation function of galaxies \citep[e.g.,][]{Eisenstein05,Anderson13}
and the power spectrum of fluctuations in the cosmic microwave background (CMB)
\citep[e.g.,][]{Bennett03,Planck15} indicates that our universe is flat,
accelerating and primarily composed of baryons, dark matter, and dark
energy.  Together, this evidence points to a `standard model of cosmology', yet an understanding of the nature of dark energy remains elusive.  Due to improved determinations of cosmological distances, it is now possible to precisely constrain, to better than $10\%$, the equation-of-state of dark energy, characterized by the parameter $w = p/\rho$, where $p$ is its pressure and $\rho$ is its energy density.  Furthermore, new measurements (e.g., \citealp{Betoule14}, hereafter B14) have begun to place constraints on the evolution of the equation of state with redshift (e.g. with $w(z)=w_0+w_a\times z/(1+z)$).
However, some recent combinations of cosmological probes (e.g., \citealp{Planck15}, \citealp{Riess16}) do not appear to be consistent with the \LCDM~model.   To help understand this tension and make a direct measurement of $w$, SN analyses must both build up the statistics of their samples and examine in greater detail the nature of their systematics.

The leverage on cosmological constraints from SN samples stems from the combination of low-redshift SNe with high-redshift SNe. Over the last twenty years, there have been a number of SN surveys that together probe a large range in redshift.  Many groups have worked on assembling
large sets of low-redshift ($0.01<z<0.1$) SNe (e.g., CfA1-CfA4,
\cite{Riess99,Jha06,Hicken09a,Hicken09b,Hicken12}; CSP,
\cite{Contreras10,Folatelli10,Stritzinger11}; LOSS,
\cite{Ganeshalingam13}). There have been four main
surveys probing the $z>0.1$ redshift range: ESSENCE
\citep{Miknaitis07,Wood-Vasey07,Narayan16b}, SNLS \citep{Conley11,Sullivan11}, SDSS \citep{Frieman08, Kessler09, Sako16} and PS1 \citep{Rest14,S14b}.  These surveys have
overlapping redshift ranges from $0.1 \la z \la 0.4$ for SDSS, $0.2
\la z \la 0.7$ for ESSENCE, $0.03 \la z \la 0.68$ for PS1 and $0.3 \la z \la 1.1$ for SNLS. Furthermore, there is now high-z data ($z>1.0$) from the SCP survey \citep{Suzuki12} and both the GOODS \citep{Riess04,Riess07} and CANDELS/CLASH surveys (\citealp{Rodney14,Graur14,Riess17}).  These surveys extend the Hubble diagram out to $z=2.26$, from a dark-energy dominated universe to a dark-matter dominated universe.  

In this paper, we present the \textit{full} set of spectroscopically confirmed PS1 SN\,Ia and combine this sample with spectroscopically-confirmed SN\,Ia from CfA1-4, CSP, PS1, SDSS, SNLS and HST SN surveys.  The samples included in this analysis are ones that have been cross-calibrated with PS1 in (\citealp{Supercal}, hereafter S15) or have data from HST. While there have been many analyses that combine multiple SN\,Ia samples, this analysis reduces calibration systematics substantially by cross-calibrating all of the SN samples used (S15).  In \cite{Betoule14}, a cross calibration of SDSS and SNLS \citep{Betoule12} was used, but none of the other samples were cross-calibrated.  This is particularly important because calibration has been the dominant systematic uncertainty in all recent SN\,Ia cosmology analyses (B14).

The statistical and systematic uncertainties in recent SN\,Ia cosmology analyses have been roughly equal.  The growing size of the sample has motivated more focus on the systematic uncertainties and also allowed for an examination of various subsamples of the SN\,Ia population.  These tests include probing relations between luminosity and properties of the host galaxies of the SNe (\citealp[e.g.,][]{Kelly10,Sullivan10,Lampeitl10}) and analyses of the light-curve fit parameters of SNe and how these parameters relate to luminosity (\citealp[e.g.,][]{Scolnic16,Mandel16}).  Many of the associated systematic uncertainties of these effects are on the $1\%$ level, and considering a typical SN distance modulus is measured with roughly $15\%$ precision, it is difficult to properly analyze these effects without SN samples in the hundreds.

This analysis relies heavily on the work by \cite{Rest14} and \cite{S14b}, hereafter R14 and S14 respectively.  R14 and S14 analyzed the first 1.5 years of PS1 SN\,Ia data and combined it with a compilation of low-z surveys.  R14 and S14 chose not to analyze any of the higher-z surveys (SDSS, SNLS, HST) so as to focus on the PS1 data sample.  Almost every facet of those papers is improved in this analysis.  For one important example, the PS1 collaboration recently released photometry of all detected stellar sources \citep{Chambers16,Magniera,Magnierb,Magnierc,Flewelling16,Waters16} with sub-percent level relative calibration across $3\pi$ steradian of the sky; the photometry and calibration of our present analysis is ensured to be consistent with that of the public release.

The SN\,Ia presented in this paper include all SNe discovered during the PS1 survey (September 2009- January 2014) that have been spectroscopically confirmed as SN\,Ia.  The SN\,Ia presented in R14 make up roughly $40\%$ of the SN\,Ia presented in this paper.  Our sample does not include likely SN\,Ia in the PS1 sample without spectroscopic classifications.  The first effort to analyze these photometric-only SNe was presented in \cite{Jones16}, which is used to improve the PS1 survey simulations in this work.  Furthermore, a follow-up analysis of \cite{Jones16} that determines the cosmological parameters from the full PS1 photometric-only SN sample of $\sim1200$ SNe \citep{Jones17} is a companion analysis to ours and uses multiple pieces of our analysis.

\begin{figure}

\epsscale{1.15}  
\plotone{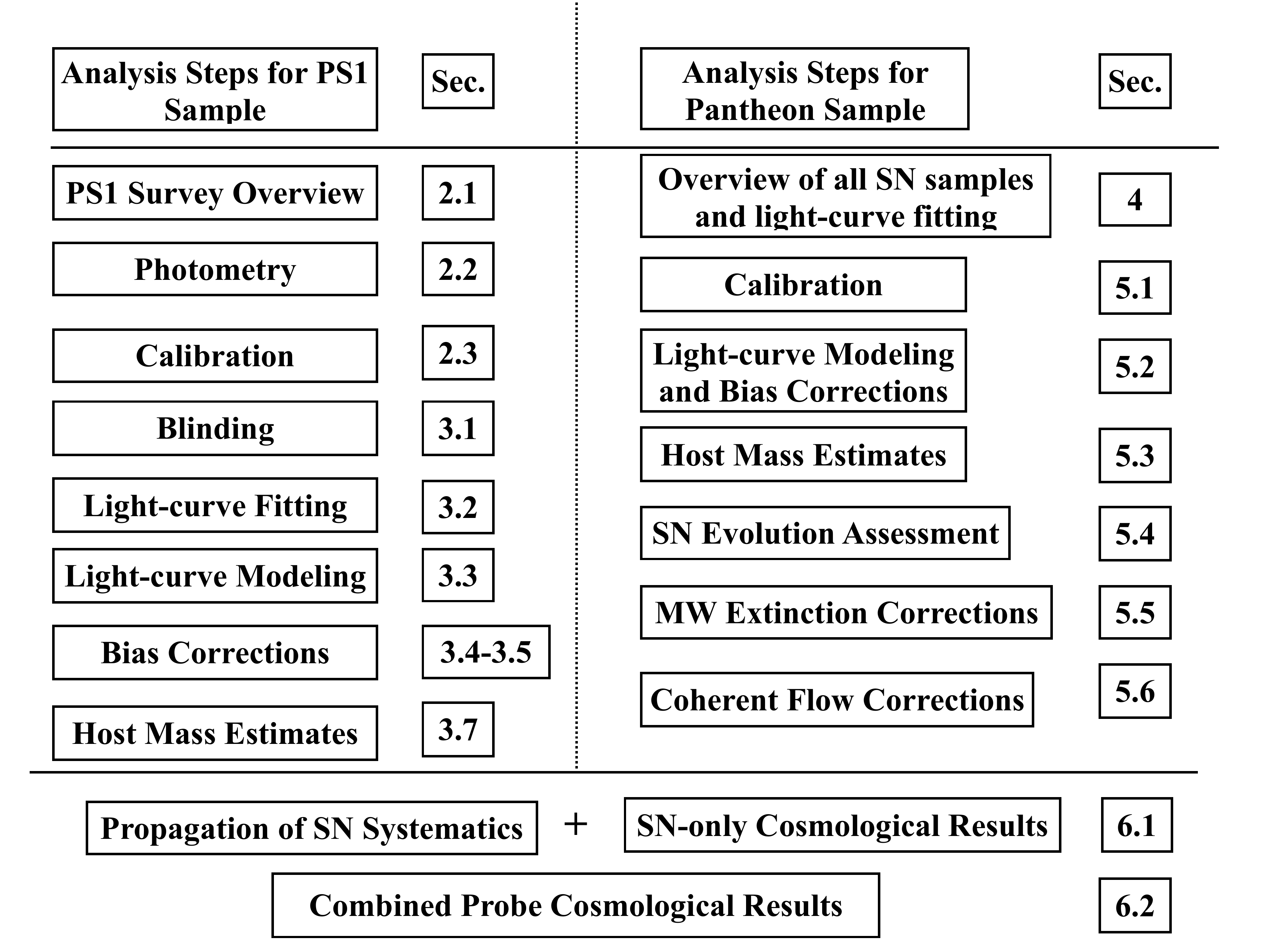}
\caption{An overview of the various analysis steps in this paper.  A common set of steps is done for both the PS1 sample and the combined Pantheon sample.}
\label{fig:ps1flow}
\end{figure}

With the set of spectroscopically confirmed SN\,Ia discovered by PS1 and multiple other subsamples, we analyze the combined sample to determine cosmological parameters.  Due to the number of steps and samples in the analysis, we show Fig.~\ref{fig:ps1flow} to demonstrate the analysis steps.  The paper is organized as follows.  In Section \ref{sec:pipeline}, we present improvements to the PS1 search, photometry and calibration pipelines.  In Section \ref{sec:lcfit}, we estimate distances from the PS1 SN sample and discuss simulations of the light-curves.  In Section \ref{sec:Data}, we combine the PS1 sample with other samples.  In Section \ref{sec:Analysis} and Section \ref{sec:Results}, the full assessment of systematic uncertainties and constraints on cosmology are given.  In Sections \ref{sec:Discussion} and \ref{sec:Conclusion}, we present our discussions and conclusions. 

\section{The PS1 Search, Photometry and Calibration Pipeline}
\label{sec:pipeline}

\subsection{Overview of the PS1 Survey}
\label{subsec:	Overview}

The PS1 data presented here is from the PS1 Medium Deep Survey which observes SNe in \griz with an average cadence of 7 days per filter\footnote{$y$ band observations are taken during bright time but are not used in this analysis}.  This cadence provides well-sampled, multi-band light-curves.  The description of the PS1 survey is given in \cite{Kaiser10}.   The PS1 Image Processing Pipeline (IPP) system \citep{Magnier13} performs
flat-fielding on each individual image and determines an initial astrometric solution.  The full description of these algorithms is given in \cite{Chambers16,Magniera,Magnierb,Magnierc,Flewelling16,Waters16}.  Once done, images are processed in \textit{Photpipe} \citep{Rest05} with updated methodology given in R14.

The discovery pipeline is explained in R14.  The main difference between the pipeline in the first and second half of the survey is that, as the survey went along, the average nightly seeing improved by $0.12''$ (due to camera/operation improvements) and better templates ($>0.5$ mag deeper) were used for the transient search.  The improved templates also had better artifact removal which significantly reduced the number of false positives in the transient candidate lists.

The spectroscopic selection over the full survey is similar to that outlined in R14.  Spectroscopic observations of PS1 targets were obtained with a variety of instruments: the Blue Channel Spectrograph
\citep{Schmidt89} and Hectospec \citep{Fabricant05} on the 6.5-m MMT,
the Gemini Multi-Object Spectrographs (GMOS; \citealt{Hook04}) on both
Gemini North and South, the Low Dispersion Survey Spectrograph-3
(LDSS3\footnote{http://www.lco.cl/telescopes-information/magellan/instruments-1/ldss-3-1})
and the Magellan Echellette (MagE; \citealt{Marshall08}) on the 6.5-m
Magellan Clay telescope, and the Inamori-Magellan Areal Camera and
Spectrograph (IMACS; \citealt{Dressler11}) on the 6.5-m Magellan Baade
telescope, the ISIS spectrograph on the WHT \footnote{http://www.ing.iac.es/}, and DEIMOS \citep{Keck} on the 10-m Keck telescope. 

Since there were a multitude of spectroscopic programs without a well-defined algorithm to determine which candidates to observe, an empirical algorithm is retroactively determined that best describes our selection of spectroscopic targets.  This is discussed further in section 3, but we note here that the spectroscopic selection for the full survey is very similar to that of the first 1.5 years of the PS1 survey described in R14.  The one exception was a program by PI Kirshner (GO-13046) to observe HST candidates for infrared follow-up specifically at $z\sim0.3$.  A table of the spectroscopically confirmed SN\,Ia that includes the dates of the observations and the telescopes used is given in Appendix A.

\begin{figure}

\epsscale{1.15}  
\plotone{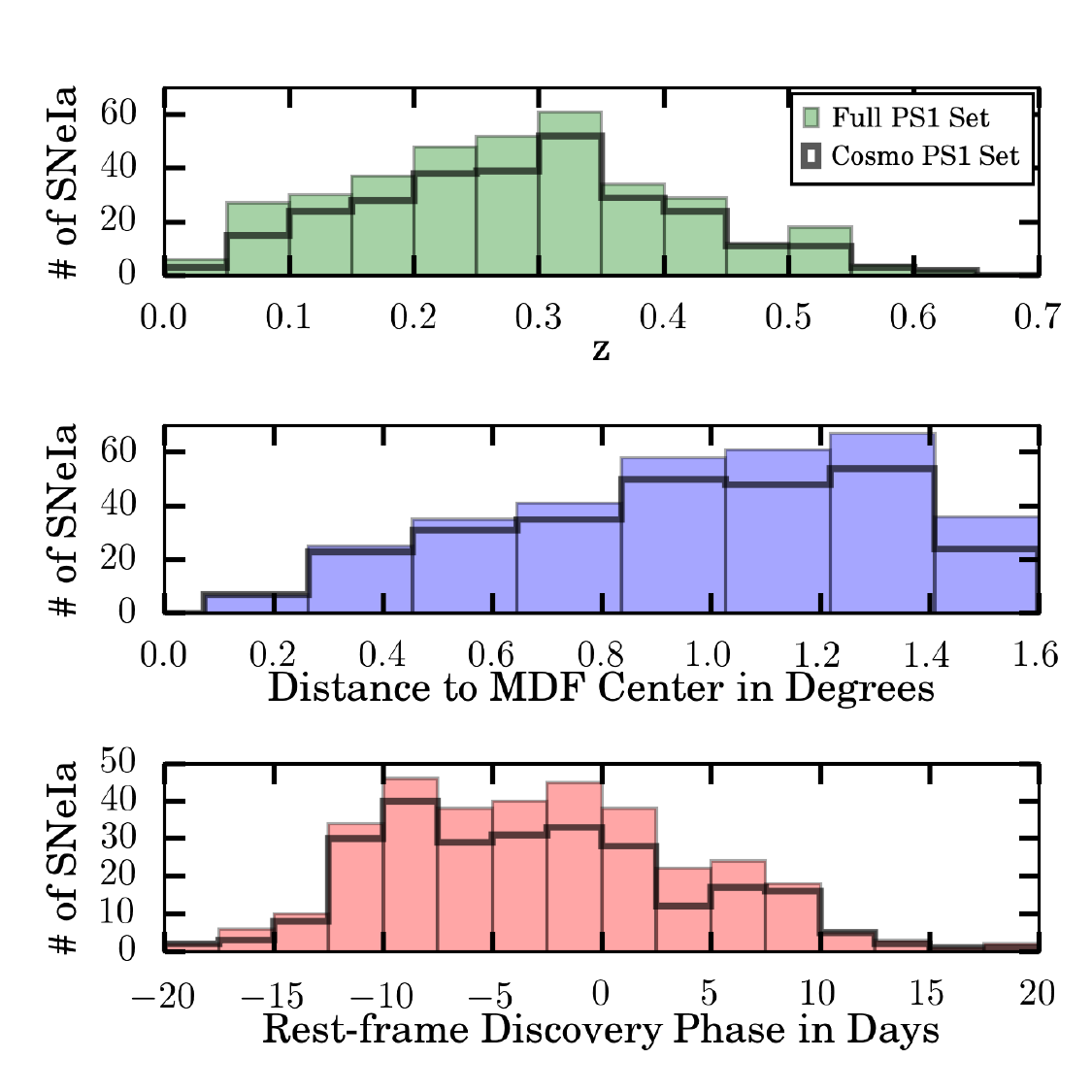}
\caption{Histograms comparing the set of all spectroscopically confirmed SN\,Ia against the subset that is deemed cosmologically useful.  Filled bars indicate the full spectroscopic sample of \allPS~SN\,Ia, while outlined bars indicate the \numPS~used for our cosmology analysis.   (Top) Distribution of redshift.  (Middle) Distribution of radial angular distance from center of focal plane. (Bottom) Distribution of the age at discovery as determined from the date of peak brightness subtracted from the discovery date. }
\label{fig:makeplot}
\end{figure}

The distribution of redshifts of the confirmed SN\,Ia is shown in Fig.~\ref{fig:makeplot}.  The median redshift is $0.3$, which is $\Delta z \sim 0.05$ smaller than the median redshift of likely photometric SN\,Ia discovered during the survey \citep{Jones16}.  As shown in Fig.~\ref{fig:makeplot}, the observed candidates are well dispersed over the focal plane with no systematic grouping at one focal position.  It is also shown in Fig.~ \ref{fig:makeplot} that the majority of candidates are discovered before peak.  A discovery (defined as 3 detections with $SNR>4$) after peak does not exclude the possibility that there were pre-explosion images acquired, only that the object was not detected at that time.

\subsection{Improvements to PS1 Photometry}
\label{subsec:	Photometry}

The photometry pipeline used in this analysis is a modified version of that described in R14.   The
overall process used is summarized as such: 
\begin{itemize}
\item \textit{Template Construction}.  For each PS1 chip, templates are constructed from stacking multiple, nightly, variance-weighted images from all but a single survey year around the SN explosion date.  The seasonal templates are made of $\sim60+$ images and reach $5\sigma$ depths of $25.00,25.1,25.15,24.80$ mag in \griz.  Excluding a particular year removes the possibility that $>1$ mmag of SN flux is included in the template.  We develop a scene-modeling pipeline (e.g., \citealp{Holtzman08}) as an independent cross-check on the template construction.  This is presented in Appendix B.
\item \textit{Astrometric Alignment}.  All nightly images and templates are astrometrically aligned with an initial catalog provided by the PS1 survey.  For all bright stars and galaxies observed on each CCD with a SN observation, an astrometry catalog is recreated with the average locations of each of the stars and galaxies over the full survey.  Then an astrometric solution for each nightly image and template is determined to match the improved catalog.
\item \textit{Stellar Zeropoints}. PSF photometry is performed on the image at the positions of the stars from the final catalog; there is no re-centroiding of the star's position per image, such that `forced' photometry is done.  The PSF module is based on a Python implementation \citep{Jones15} of the DAOPhot package \citep{Stetson87}.  A comparison of the photometry of these stars to updated PS1 stellar catalogs is used to find the zeropoint of each image.  Forced photometry on the stars is necessary so that a consistent procedure is done for both the stars and the SNe.  The PSF is determined for each epoch from neighboring stars of the SN.  Due to the fast-varying PSF on CCDs near the center of the focal plane ($<0.4$ deg), the region cutout to find neighboring stars is roughly 1/4 the area of the chip.  For CCDs away from the center of the focal plane ($>0.4$ deg), the full area of the 12.5' chip is used.

\item \textit{Template Matching}.  Templates are convolved with a PSF to match the nightly images \citep{Becker15}.  The convolved templates are then subtracted from the nightly images.
\item \textit{Forced SN Photometry}.  A flux-weighted centroid is found for each SN position.  Forced photometry is performed at the position of the SN.  The nightly zeropoint is applied to the photometry to determine the brightness of the SN for that epoch.  Small adjustments are made to the SN photometry based on the expectation value from the astrometric uncertainty of the SN centroid.  Forced photometry is also applied to random positions in the difference image to empirically determine the amount of correlated noise in the image.  The SN photometry uncertainties are then increased to account for this correlated noise.
\item \textit{Flux Adjustment}. The errors and the baseline flux of the SN measurements are adjusted so that the mean pre-explosion baseline flux level is 0 and the reduced $\chi^2$ is near unity.  The prescription for this step is described in R14.
\end{itemize}
 
The most significant changes relative to R14 are the additions of iterative astrometric alignment, forced photometry of stars with an updated PSF fitting routine, an updated Ubercal catalog, and a reduction in the area from which neighboring stars are drawn for building PSF models.  These steps improve the accuracy of the astrometric solution, alleviate systematic uncertainties in the photometry due to uncertainties in the astrometry, and account for the fast varying PSF near the center of the focal plane respectively.  

Improvements to understand the systematic uncertainties in this process are discussed below.  The systematic uncertainties in the photometry analysis are given in Table \ref{tab:systematics_cal}.

\begin{table}[h]
  \caption{}
  \begin{center}
  \begin{tabular}{p{3.5cm} l p{1cm}}
\hline
Source & Uncertainty \\
~ & [Millimag] \\
\hline
\textbf{SN Photometry} & ~  \\
Astrometric uncertainty            & 1         \\
Template construction              & 1         \\
Photometric Non-linearity      & 2      \\

\textbf{Internal Calibration} & ~  \\
Ubercal zeropoints          & 1           \\
Spatial variation              & 1    \\
Temporal variation             & 1           \\
Focal-plane variation      & 2      \\
\hline
    \end{tabular}
  \end{center}
\begin{flushleft} {Notes: The dominant systematic uncertainties in defining the
\PS\ photometric system. Each of the numbers given is the average over the four filters \griz.  The
bandpass uncertainties are $7$\AA~and are discussed in Section 4.1.}\end{flushleft}
  \label{tab:systematics_cal}

  \end{table}

\subsubsection{Astrometry}
\label{subsubsec:	Astrometry}

The recovered position of a SN detection can be different from the true SN centroid for the following reasons: accuracy of the WCS for a given image, the limited number of observations of the SN, poisson noise from sky, host galaxy, and SN and difference image artifacts.  Unlike in R14, forced photometry is performed on both stars and SNe, so the errors on SN positions and stellar position are similar and do not propagate to additional biases.   However, uncertainties that affect only the SN position are treated separately.  

\begin{figure}

\epsscale{1.15}  
\plotone{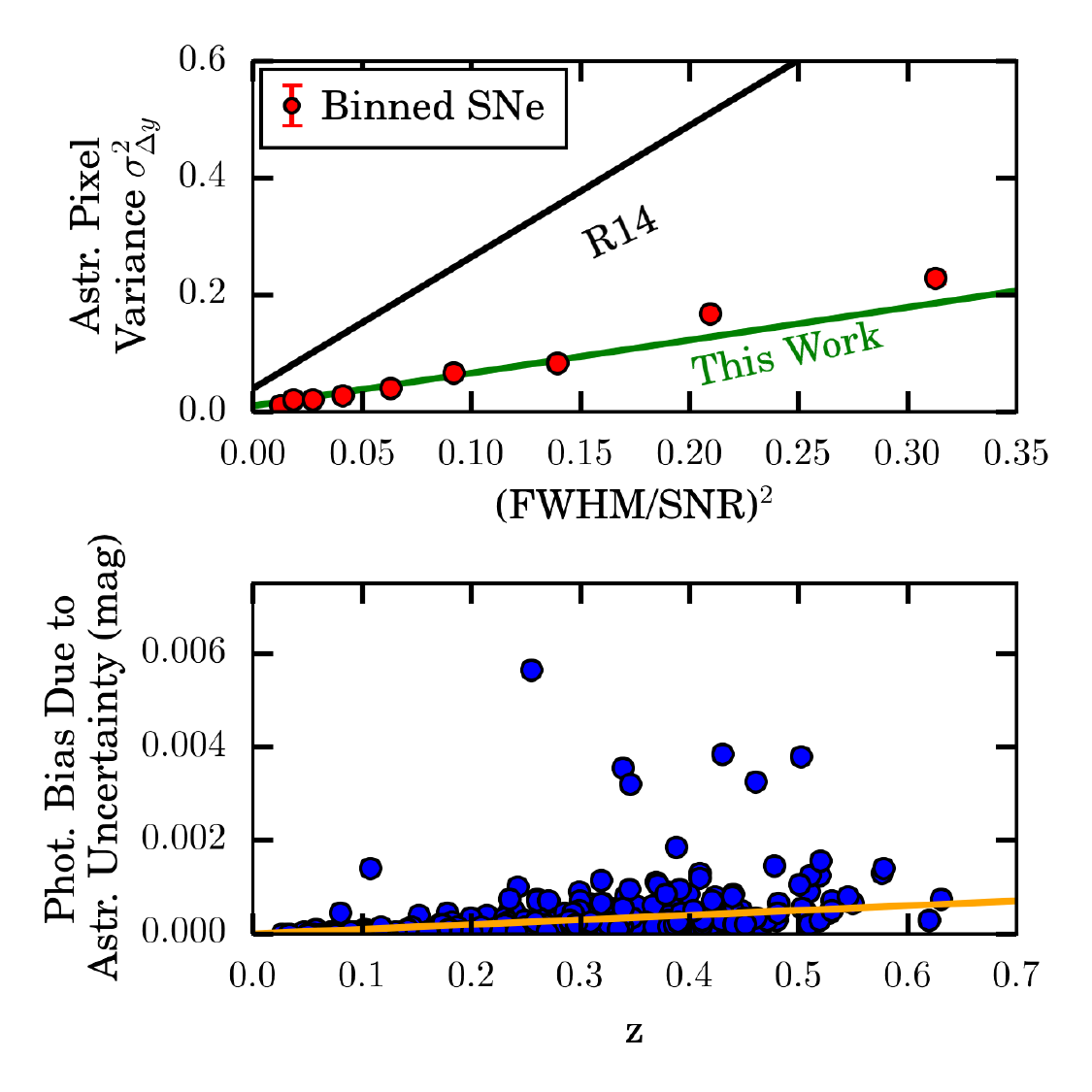}
\caption{(Top) A plot of the variance of recovered pixel offsets in one dimension ($y$) versus (FWHM/SNR)$^2$.  A similar overestimation of the astrometric error by R14 is seen in the $x$ direction as well.  (Bottom) The necessary photometric bias correction versus redshift of the SN due to the expected astrometric uncertainty of the central position of a SN from the combined series of images of that SN in one filter.  A best-fit line is overlaid in yellow.}
\label{fig:astrom}
\end{figure}

R14 shows that the astrometric uncertainty of objects depends on both the FWHM and SNR of the object.  Because of the SNR dependence, the astrometric uncertainty of the higher-redshift SNe is larger than the astrometric uncertainty of the lower-redshift SNe.  This astrometric uncertainty will propagate to a photometric bias because the expected average offset from the true centroid value causes biased photometric  measurements.  To understand this trend, the astrometric uncertainty of the individual detections is quantified.  This is done in R14 by first finding the linear relation between astrometric uncertainty (e.g. $\sigma^2_{\Delta x},\sigma^2_{\Delta y}$, here denoted as $\sigma^2_{a}$) and $(FWHM/SNR)^2$:

\begin{equation}
  \sigma_a^2 = \sigma_{a1}^2 + \sigma_{a2}^2\left( \frac{\rm FWHM}{\rm
      SNR}\right)^2,
\end{equation}
where the astrometric uncertainty $\sigma_a$ in pixels of a given
detection has a floor mostly due to pixelization ($\sigma_{a1}$), and
in addition a random error $\sigma_{a2}$.  R14 conservatively uses
$\sigma_{a1} = 0.20$ pixels and $\sigma_{a2} = 1.5$ to calculate the
astrometric uncertainty of a single detection.  In Fig.~\ref{fig:astrom}, it is clear that the quantified relation from R14 is too high by a factor of $2$ for our sample due to our improved astrometry such that we find the uncertainty of astrometry as we find a $\sigma_{a1}=0.1$ and $\sigma_{a2}=0.75$.  Much of this improvement is from the iterative astrometric alignment discussed above. 

The relation in Fig.~\ref{fig:astrom} is used to determine the astrometric uncertainty of each SN observation to properly determine the centroid accuracy of the SN.  With a more appropriate estimate of the astrometric uncertainty, the centroids and centroid errors are recalculated for each SN detection.  To remove the expected bias in the photometry from the centroid error, a conversion from R14 is applied between the astrometric uncertainty $\sigma_{\mathrm{SN,cent},a}$ to the bias in photometry $\Delta m_{Corr}$ from Eq. 1 such that:

\begin{eqnarray}
\Delta m_{\mathrm{Corr}} & = & \int h t^2 \mathrm{PDF}(\sigma_{\mathrm{SN,cent},a},t)\,\mathrm{d}t \label{eq:dmSNcen_expectation}
\end{eqnarray}
where PDF($\sigma_a$,t) is the probability density function with sigma
$\sigma_a$ and pixel variable $t$, assuming that $h$ is constant and independent of SNR.  The value of $h$ (0.043) is taken from R14 and is found to be a reliable first order approximation.  The corrections to the photometry for each SN are shown in Fig.~\ref{fig:astrom} (bottom).  The maximum correction is 6 mmag.

\subsection{Improvements to Photometric Calibration}
\label{subsubsec:	Error}

\begin{figure*}

\epsscale{1.15}  
\plotone{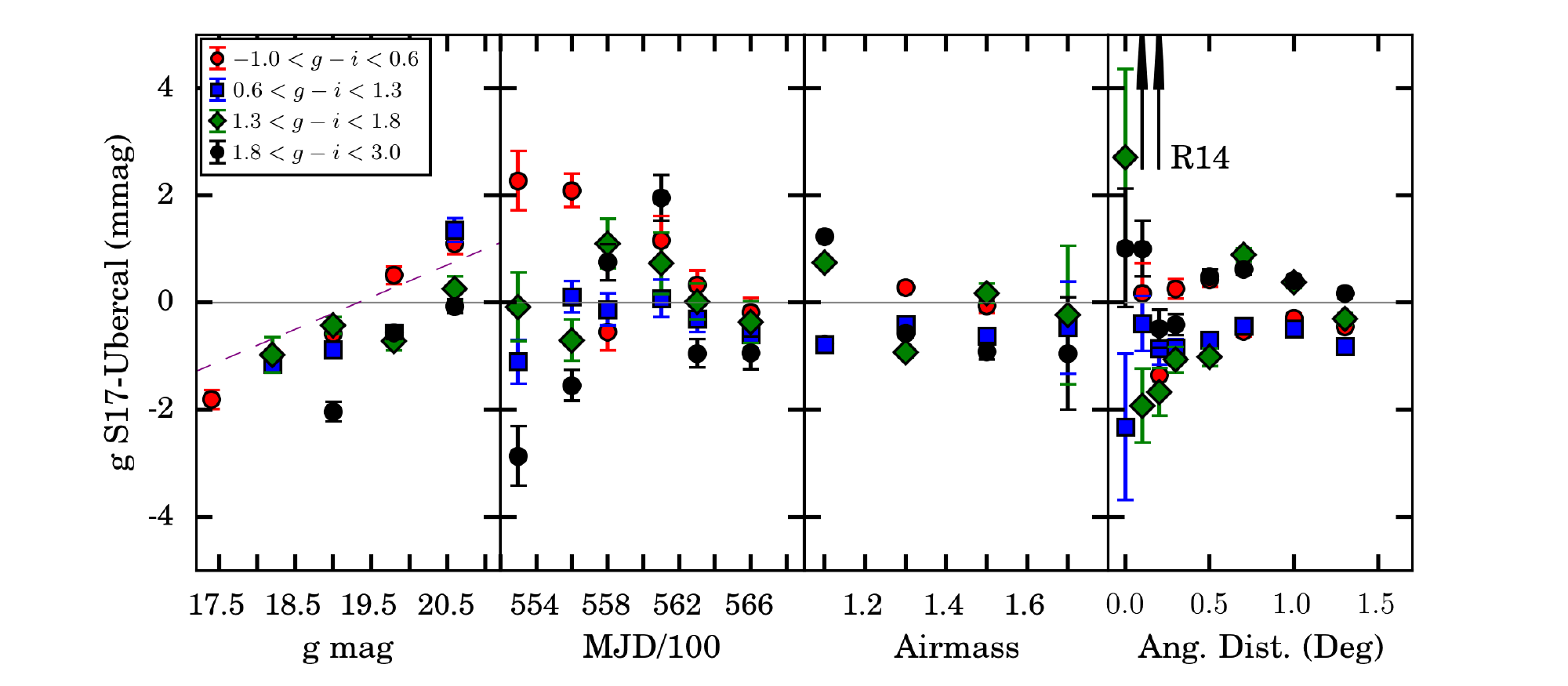}
\caption{Agreement between $g$ band nightly photometry and Ubercal photometry of $>1$ million stars and the dependence on magnitude, MJD, airmass and Focal plane position.  Different colors of the points represent bins of stellar colors.  In the rightmost panel, arrows indicate that the R14 discrepancy with the catalog photometry near the center of the focal plane was $>0.1$ mag.}
\label{fig:ps1star_plot}
\end{figure*}

The absolute calibration of the PS1 photometric system has been improved in a series of PS1 analyses.  The basis for the PS1 absolute calibration is first presented in \cite{Tonry12} and a full review of subsequent improvements is given in \cite{Supercal}.  The relative calibration across the sky of the PS1 survey is determined by the Ubercal process \citep{Schlafly12, Finkbeiner16}.  For the Medium Deep (MD) fields, we made a custom Ubercal star catalog of all of the data from
the MD fields in the same way as those produced in \cite{Schlafly12} but with a higher resolution nightly flat field, a lower threshold for masking of problematic areas of the focal plane, and a per-image zero point.  These catalogs are released with results from this paper at \dataset[10.17909/T95Q4X]{http://dx.DOI.org/10.17909/T95Q4X} and are consistent with \cite{Schlafly12} in relative calibration on degree scales to the 1 mmag level. \cite{Supercal}, which uses the same catalogs as in this current analysis, shows in comparisons with SDSS and SNLS that the likely systematic uncertainty of the zeropoints of each MD field for each filter is $\sim 3$ mmag.

The nightly photometry is transferred onto the PS1 system using a zeropoint measured by comparing the photometry with stellar magnitudes from the Ubercal catalog.  The result of this process is shown in Fig.~\ref{fig:ps1star_plot}, which presents differences between our PS1 catalogs and the final nightly photometry.   As the Ubercal catalogs are used to determine both the calibration of the HST Calspec standards and the MD fields, Fig.~\ref{fig:ps1star_plot} demonstrates the consistency of the nightly photometry with that used to create the PS1 calibration across $3\pi$ of the sky.  
We find that the nightly photometry and Ubercal catalogs are consistent across 4 mags to levels of $\sim2$ mmags, though with a trend in the discrepancy of $\sim1$ mmag per mag as shown with a linear fit overlaid on Fig.~\ref{fig:ps1star_plot} (left).  It is unclear what is causing this trend, and it is possible that this small trend is partly due to selection effects in the cuts made to make the catalogs. This trend is therefore included as part of our systematic error budget.  Image zeropoints for each observations are determined using stars brighter than $21.5, 21.0,21.0,21.0$ in \griz respectively.  Future analyses may try to use a SNR cut instead of a magnitude cut to reduce the Mamlquist bias.  Possible non-linearity has been tested in \cite{Supercal} in comparisons between PS1 with SDSS, SNLS and multiple low-z surveys: using our PS1 stellar catalogs, linearity behaves to better than $3$ mmag in \griz between mag of 15 and 21.  Further discussions of PS1 detector non-linearity are in \cite{Waters16}.

Systematic uncertainties in our photometry, due to spatial variation of the throughput across the focal plane as well as temporal variation of the filters over the entire survey, are examined here as well.  There is no evidence ($<1$ mmag) of differences in the system photometry over the full course of the survey.  There is also excellent agreement ($<1$ mmag) between the stellar photometry from our pipeline and the Ubercal catalogs across the focal plane.  A much larger effect ($>0.15$ mag) was seen in R14 due to the fast-varying PSF (change of 1 pixel in FWHM over 0.4 deg) near the center of the focal plane that was not accounted for.  Therefore, in R14, SNe near the center ($r<0.4$ deg) of the focal plane were not used in the analysis.  This problem has been fixed by reducing the area for choosing neighboring stars from which to build a PSF near the center of the focal plane.  Furthermore, there is little dependence ($<2$ mmag) on the airmass of the nightly observations.

In S14, the filters used to measure the SN light-curves are those given at the median radial position across the field of view.  From measuring the expected photometry of synthetic SN spectra integrated through the known PS1 passbands at various focal positions, differences in the photometry of the SN dependent on focal plane position increases scatter by $0.01$ mag.  However, S14 showed that there is only a 2 mmag bias with redshift due to the different passbands.  Further corrections based on the airmass of each observation, as done in \cite{Li16}, may be implemented in the future; however it is shown in Fig.~\ref{fig:ps1star_plot} that the impact is on the 1-mmag scale.  All uncertainties are summarized in Table 1.

\section{PS1 Light-curve fitting and simulation}
\label{sec:lcfit}

\begin{figure}

\epsscale{1.15}  
\plotone{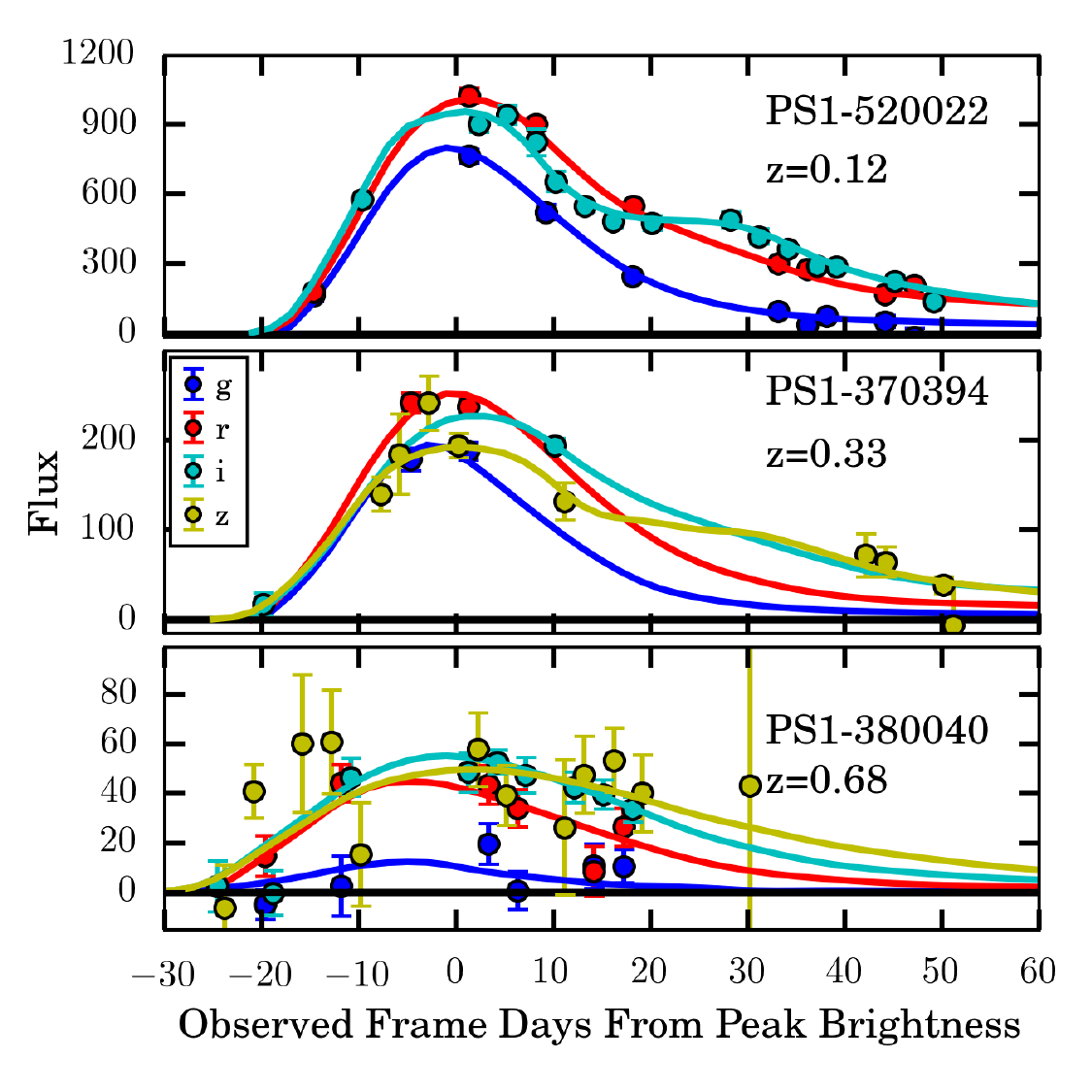}
\caption{Representative light-curves of SNe from the PS1 survey: PS1-520022, PS1-370394, PS1-380040 from top to bottom respectively.  These SNe have redshifts, of $z=0.12$, 0.33, and 0.68. The points shown are data from the PS1 survey and the curves shown are fits using SALT2.  The Flux units are given for a zeropoint of $27.5$ mag in each band.}
\label{fig:lcplot}
\end{figure}

We measured photometry of the total set of \allPS~confirmed SN\,Ia.  In Fig.~\ref{fig:lcplot}, three representative PS1 SN\,Ia light-curves are shown.  All light-curves are available in machine-readable format at \dataset[10.17909/T95Q4X]{http://dx.DOI.org/10.17909/T95Q4X}.

\subsection{Blinding The Analysis}
It is difficult to fully blind an analysis of this sort, since any update in photometry, calibration, etc. of a sample has a direct, and sometimes obvious, impact on recovered cosmological parameters.  As discussed later in this section, we use the BEAMS with Bias Corrections (BBC) method \citep{Kessler16} to recover binned Hubble residuals with redshift, with respect to a reference cosmology.  Therefore, to blind the analysis, the reference cosmology is randomly chosen.  So that a full analysis can be completed without introducing any further SN systematics based on a highly unlikely cosmology, the reference cosmology is randomly chosen from a gaussian distribution of values of the matter density $\Omega_m$ and equation-of-state of dark energy $w$ (discussed in section 5) centered around the recovered values in the B14 analysis of $w=-1.02$ and $\Omega_m=0.307$ with standard deviation of $\sigma_w=0.06$ and $\sigma_{\Omega_m}=0.02$, where the standard deviation is determined from the uncertainties on the cosmological parameters in the B14 analysis.  We use the B14 analysis to choose the blinding parameters rather than R14 because the full sample that this paper will analyze is more consistent with that in B14 than that in R14 and has lower uncertainties.  

\subsection{Light-curve fitting}

While multiple light-curve fitters can be used to determine accurate distances \citep[e.g.,][]{Jha07,Guy10,Mandel11,Burns11}, we use SALT2 \citep{Guy10} for this analysis as it has been trained on the JLA sample (B14) and it is easy to assess various systematic uncertainties with this fitter.  We use the most up-to-date published version of SALT2 presented in B14 and implemented in SNANA\footnote{SNANA\_v10\_50a} \citep{SNANA}.  Differences between the SALT2 spectral model in \cite{Guy10} versus that in B14 are mainly due to calibration errors in the light-curves used for the model training.  The models differ the most at rest frame wavelengths $<4000$\AA~and are described in detail in B14.  Three values are determined in the light-curve fit that are needed to derive a distance: the color $c$, the light-curve shape parameter $x_1$ and the log of the overall flux normalization $m_B$.  The solid lines in Fig.~\ref{fig:lcplot} show the respective light-curve fits with SALT2 for 3 representative PS1 SN\,Ia. 

The SALT2 light-curve fit parameters are transformed into distances using a modified version of the Tripp formula \citep{tripp}
\begin{equation}
  \mu = m_{B} - M + \alpha x_{1} - \beta c + \Delta_M+ \Delta_B, 
  \label{eqn:Tripp}
\end{equation}
where $\mu$ is the distance modulus, $\Delta_M$ is a distance correction based on the host-galaxy mass of the SN and $\Delta_B$ is a distance correction based on predicted biases from simulations.  Furthermore, $\alpha$ is the coefficient of the relation between luminosity
and stretch, $\beta$ is the coefficient of the relation between
luminosity and color and $M$ is the absolute $B$-band magnitude of a
fiducial SN~Ia with $x_1 = 0$ and $c = 0$.  Motivated by
\cite{Schlafly11} and following S14, we modify SALT2 by replacing the
``CCM'' \citep{CCM89} Milky Way (MW) reddening law with that
from \cite{Fitzpatrick99}. 

The total distance error of each SN is 
\begin{equation}
\sigma^2=\sigma^2_{\mathrm{N}}+\sigma^2_{\mathrm{Mass}}+\sigma^2_{\mu-z}+\sigma^2_{\rm lens}+\sigma^2_{\mathrm{int}}+\sigma^2_{\mathrm{Bias}}
\label{eqn:disterror}
\end{equation}
where $\sigma^2_{\mathrm{N}}$ is the photometric error of the SN
distance, $\sigma^2_{\mathrm{Mass}}$ is the distance uncertainty from the mass step correction, $\sigma^2_{\mathrm{Bias}}$ is the uncertainty from the distance bias correction, $\sigma^2_{\mu-z}$ is the uncertainty from the peculiar velocity uncertainty and redshift measurement uncertainty in quadrature, $\sigma^2_{\rm lens}$ is the uncertainty from stochastic gravitational lensing, and $\sigma^2_{\mathrm{int}}$ is the intrinsic scatter. For this analysis, $\sigma_{\rm lens}=0.055z$ as given in \cite{Jonnson10}.

For this analysis, we require every SN\,Ia to have adequate light-curve
coverage to accurately constrain light-curve fit parameters, as well as properties that limit systematic biases
in the recovered distance.  We follow the light-curve requirements in B14 such that the only SNe allowed in the sample have $-3< x_1< 3$,$-0.3< c< 0.3$, $\sigma_{(pkmjd)}< 2$ and $\sigma_{x1} < 1$ (where $\sigma_{(t0)}$ is the uncertainty on the rest-frame peak date and $\sigma_{x1}$ is the uncertainty on $x_1$).  Most of the cuts, as shown in Table~\ref{tab:cuts}, are motivated by B14.  The cuts are somewhat different than those used in R14 that require observations before and after the peak brightness date.  These updated requirements are more stringent than those used in R14, though 3 of the SN\,Ia that don't pass the R14 cuts do pass these new cuts.  These 3 SN\,Ia are all at low-z where it was unclear if there were observations taken before peak due to uncertainty in the peak date, though the $i$ band peak was measured accurately.  A related issue due to uncertainty in the peak date was pointed out in \cite{Dai16}, which finds $\sim 10$ SNe with double-peak probability distribution functions of the light-curve parameters of the SALT2 fits.  We find that many (8/10) of these SNe would be removed from our set if we place an additional cut enforcing observations after post-maximum brightness.  Therefore, we include a cut such that there is an observation at least 5 days after peak-brightness.  B14 also places a requirement for $E(B-V)_{MW} < 0.15$.  This does not apply to the PS1 SN sample but will apply to other samples, as all the Medium Deep fields have low extinction, and as discussed in S14, this constraint is loosened to $E(B-V)_{MW} < 0.20$ due to improved non-linear modeling of high extinction \citep{Schlafly11}.  Furthermore, in B14, a cut on the fit likelihood is placed on the SDSS SN\,Ia but not the SNLS sample in B14.  We follow the strategy for the SDSS sample and place a cut on the  $\chi^2/\textrm{NDOF}<3.0$.  Finally, there is one last additional cut from the BBC method that removes 3 of the SNe because their $x_1$ and $c$ parameters do not fall in the expected distribution - this is discussed in Section 3.5.  Applying these cuts, only \numPS\ SN\,Ia from the initial sample of  \allPS \ spectroscopically-confirmed \PS\ objects remain in our sample for a cosmological analysis.

The SALT2
parameters for the entire set of cosmologically useful SN\,Ia from the
PS1 sample are presented in Appendix A.  A table with the full output of each SNANA fit with SALT2 is included at \dataset[10.17909/T95Q4X]{http://dx.DOI.org/10.17909/T95Q4X}.
\begin{table}[h]
\caption{}
  \begin{center}
  \begin{tabular}{p{3.5cm} l p{1cm}}
\hline
\hline
~ & Discarded & Remaining \\
\hline
Initial & - & \ruta \\
\hline
Quality fit & \rutya & \rutxa \\
$\sigma(x_1) < 1$ & \rutyb & \rutxb \\
$\sigma_{(pkmjd)} < 2 $ & \rutyc & \rutxc \\
$-0.3 < c < 0.3$ & \rutyd & \rutxd \\
$-3 < x_1 < 3$ & \rutye & \rutxe \\
$ E(B-V)_\text{MW}< 0.20$ & 0 & \rutxf \\
$T_{Max}>5$ & $\rutyg$ & $\rutxg$ \\
\hline
BBC cut & 3 & \numPS \\
\hline
  \label{tab:cuts}
    \end{tabular}
  \end{center}
\begin{flushleft} {Notes: Impact of various cuts used for cosmology analysis.  Both the number removed from each cut, and the number remaining after each cut, are shown.  The `Quality fit' includes both the lightcurves that are rejected by the SNANA fitter due to poorly converged fits and those with a fit $\chi^2/\textrm{NDOF}<3.0$.}\end{flushleft}
  \end{table}

\subsection{Survey Simulations}

To correct for biased distance estimates, the PS1 SN survey must be accurately simulated.  Following S14, we simulate the PS1 survey with SNANA using cadences, observing conditions, spectroscopic efficiency, etc. from the data.   Following \cite{Jones16}, we include a complete library of observations of the PS1 survey, noise contributions from the host galaxies of the SN\,Ia as well as a newly modeled SN discovery efficiency.   The noise contributions from the host galaxies were modeled for the PS1 photometric sample, which is a good approximation for the confirmed sample as the average host galaxy magnitudes for the confirmed and photometric samples are within $0.1$ mag.  

\begin{figure}
 \epsscale{1.15}  
\plotone{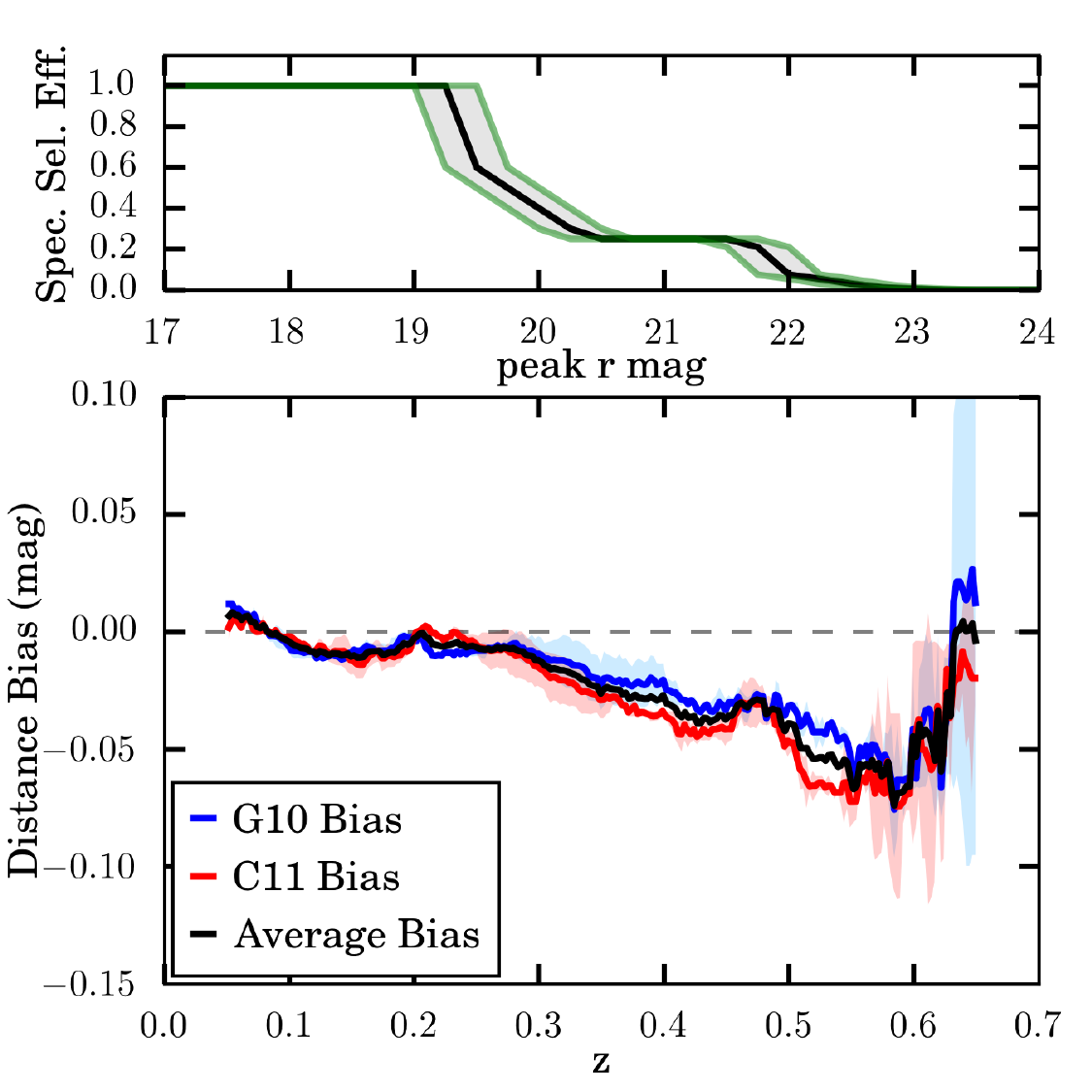}
\caption{(Top) The PS1 spectroscopic selection efficiency as a function of peak $r$ band magnitude.  The shaded band denotes the $1\sigma$ uncertainty on the function.  (Bottom) The predicted distance bias that is caused from the selection effects using the Tripp estimator from simulations with two different intrinsic scatter models.  The average distance bias between the two is also displayed.}
\label{fig:searcheff}
\end{figure}

To model the spectroscopic selection of the PS1 SN survey, an efficiency function must be empirically determined.  Similar to R14, we find it is well-modeled with a dependence on the peak $r$ band magnitude of the SN.  The function is shown in Fig.~\ref{fig:searcheff} (Top).  The method to determine this function is analogous to the approach taken in \cite{Scolnic16} (hereafter SK16) for determining the underlying color populations.  We simulate PS1 without a spectroscopic efficiency function and divide the distribution of $r$ magnitudes at the peak of the light-curves from the data by that from the simulation.  The ratio is the spectroscopic efficiency function and the final curve shown in Fig.~\ref{fig:searcheff} is smoothed from the recovered function.  A coherent shift of the selection function by $0.25$ mag in one direction is found to reduce the match between the predicted and actual redshift distribution by $1\sigma$, and this error is overlaid in Fig.~\ref{fig:searcheff}.

\subsection{Populations and Intrinsic Scatter Models}

\begin{figure}
\epsscale{1.15}  
\plotone{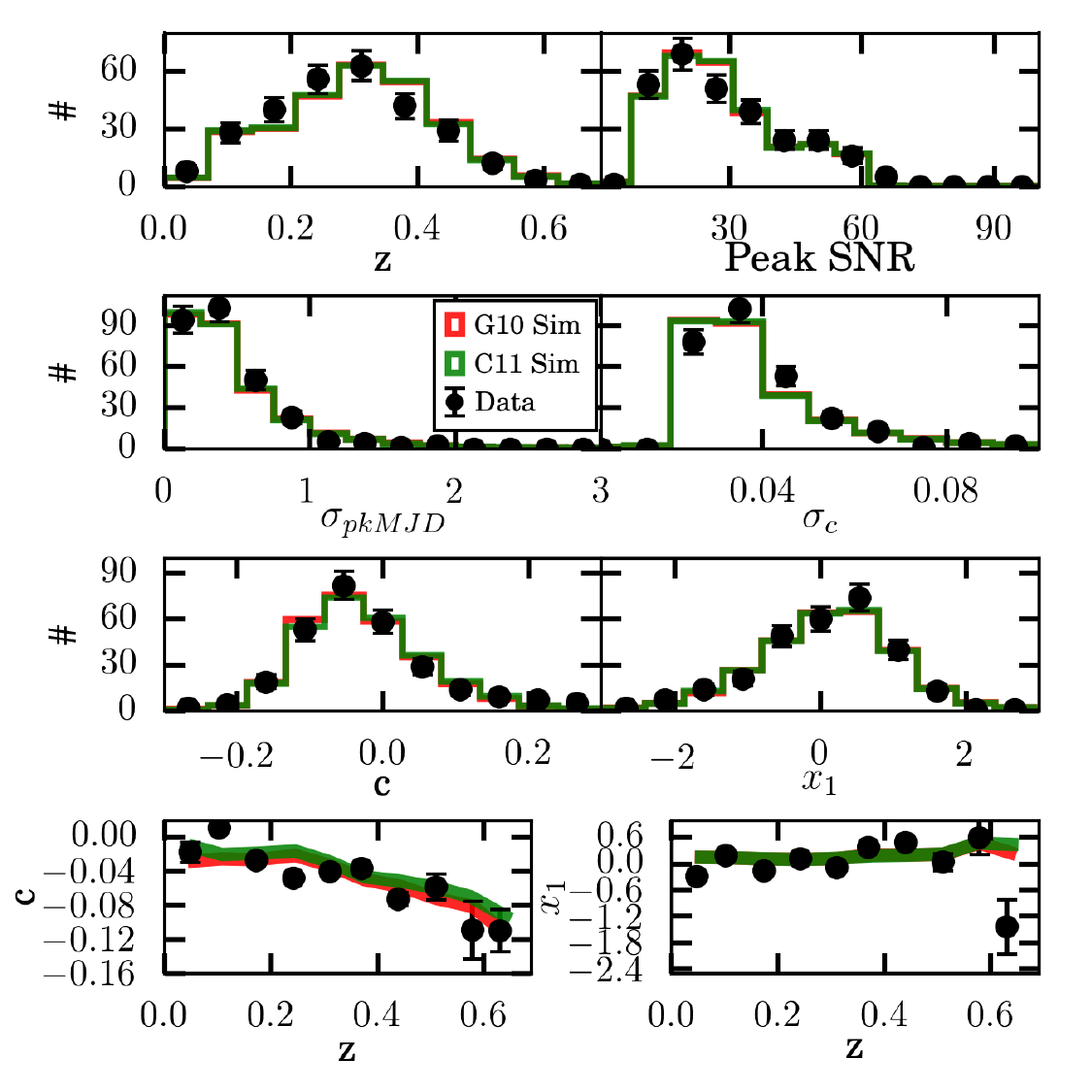}
\caption{Comparison of distributions for PS1 data (points) and 
    simulations (histograms), where each simulation distribution is scaled to 
    have the same sample size as the data.  We show the simulation of the survey assuming a G10 scatter model for the intrinsic dispersion (red) and assuming a C11 scatter model (green).
    The distributions are shown over
    redshift, 
    the error in the Peak MJD, the error in the color $c$, 
    peak SNR of the light-curve, 
    fitted SALT2 color ($c$) and light-curve shape parameter ($x_1$).
    The bottom two panels show the
    SALT2 color ($c$) and shape parameter ($x_1$) versus redshift.
}
\label{fig:bananahistps}
\end{figure}

The underlying population of the stretch and color of the PS1 light-curves is redetermined for the full data sample according to the process described in SK16.  These are given in Table \ref{tab:ps1pop} for two different models of the gaussian intrinsic scatter of SN\,Ia: the `C11' model that is composed of $75\%$ chromatic variation and $25\%$ achromatic variation \citep{Chotard11}, and the `G10' model that is composed of $30\%$ chromatic variation and $70\%$ achromatic variation \citep{Guy10}.  Besides scatter models that have $100\%$ of one type of variation, the C11 and G10 models are the only two published models available for this type of analysis, and either of them may accurately represent the PS1 SN population.  To use these models in simulations, \cite{Kessler13} converts broadband models into spectral variation models.  The color population parameters in Table \ref{tab:ps1pop} show agreement within $1\sigma$ between this analysis and that derived for the PS1 R14 sample (SK16).  The stretch population parameters appear to be slightly discrepant, though this difference is exaggerated because we do not report covariances between  $\bar{x_1}$, $\sigMINUS$ and $\sigPLUS$: the mean of the distribution of recovered $x_1$ values for the full PS1 sample is $\Delta x_1\sim0.03$ from the mean of the $x_1$ distribution from R14.

The population parameters given here are derived from simulations that assume a~\LCDM~model.  SK16 found that changes in the input cosmology within typical statistical uncertainties have a $<0.2\sigma$ effect on the recovered populations. Wolf et al. (in prep.) improves on the analysis of SK16 by attempting to fit for cosmological parameters as well as these population parameters simultaneously.

\begin{table}[ht]
\caption{}
  \begin{center}
    \label{tab:ps1pop}
    \begin{tabular}{ll | c c c }
Analysis &Scat. & $\bar{c}$ & $\sigMINUS$  &$\sigPLUS$   \\
    \hline
This Work & G10 &$ -0.068 \pm 0.023 $&$ 0.034 \pm 0.016 $&$  0.123 \pm 0.022$\\
This Work& C11 &$ -0.100 \pm 0.004 $&$ 0.003 \pm 0.003 $&$  0.134 \pm 0.016$\\

SK16 & G10 &$ -0.077 \pm 0.023 $&$ 0.029 \pm 0.016 $&$  0.121 \pm 0.019$\\
SK16 & C11 &$ -0.103 \pm 0.003 $&$ 0.003 \pm 0.003 $&$  0.129 \pm 0.014$\\

\hline

~ & ~ & ~ & ~ & ~ \\

    \hline
~  &~& $\bar{x_1}$ & $\sigMINUS$  &$\sigPLUS$   \\
    \hline

This Work & G10 &$ 0.365 \pm 0.208 $&$ 0.963 \pm 0.162 $&$  0.514 \pm 0.140$\\
This Work & C11 &$ 0.384 \pm 0.200 $&$ 0.987 \pm 0.155 $&$  0.505 \pm 0.135$\\

SK16 & G10 &$ 0.604 \pm 0.183 $&$ 1.029 \pm 0.138 $&$  0.363 \pm 0.121$\\
SK16 & C11 &$ 0.589 \pm 0.179 $&$ 1.026 \pm 0.137 $&$  0.381 \pm 0.117$\\
\hline

\end{tabular}
  \end{center}
{Notes:  Underlying populations of SN\,Ia $x_1$ and $c$ parameters for the full PS1 sample and those found in SK16.  The first column shows the analysis and second column shows the scatter model used in the simulation.  The first half of the table shows the recovered values of the underlying color ($c$) population and the second part of the table shows the recovered values of the underlying stretch ($x_1$) distribution.  These parameters define the asymmetric Gaussian for the color and light-curve shape distributions : $e^{ [-(x - \bar{x})^2/2\sigMINUS^2] } $ for $x < \bar{x}$ and $e^{ [-(x - \bar{x})^2/2\sigPLUS^2] }$ for $x > \bar{x}$.  }
\end{table}

Fig.~\ref{fig:bananahistps} shows how well simulations model the data by comparing the distribution of redshift, constraint on time of maximum light, color error, and peak SNR distribution compared to the data.  Comparisons of the color and stretch distributions as well as their trends with redshift are also shown.  There is substantial improvement from S14 in how well the simulations and data match due to more statistics in our sample and better modeling methods.

\subsection{BBC Method}

SK16 and \cite{Kessler16} (hereafter KS17) show that the Tripp estimator does not account for distance biases due to intrinsic scatter and selection effects.  KS17 introduces the BBC Method to properly correct these expected biases and simultaneously fit for the $\alpha$ and $\beta$ parameters from Eq. \ref{eqn:Tripp}.  The method relies heavily on \cite{Marriner11} but includes extensive simulations to correct the SALT2 fit parameters $m_B$, $c$ and $x_1$.  In Eq. 1, this correction is expressed as $\Delta_B$ which is actually a function of $\alpha$, $\beta$, $\Delta_{mB}$, $\Delta_{c}$ and $\Delta_{x1}$ that follows the same Tripp format such that $\Delta_B=\Delta_{mB}-\beta \times \Delta_{c} + \alpha \times \Delta_{x1}$.  Furthermore, the measurement uncertainty $\sigma_{\mathrm{N}}$ in Eq. 2 is similarly corrected according to predictions from simulations because KS17 shows that the fit with SALT2 regularly overestimates the uncertainties of the fit parameters. Finally, the BBC method requires that the properties of a SN in the data sample are well represented in a simulation of 500,000 SNe, so any SNe with $z$, $c$ and $x_1$ properties that are not found within $99.999\%$ of the simulated sample will not pass the BBC cut. The impact of the BBC cut is given in Table \ref{tab:cuts}.  The 3 SNe that are cut have $x_1$ and/or $c$ values removed from the distribution as shown in Fig.~\ref{fig:bananahistps}.  They have $(x_1,c)$ values of (-2.915, 0.083),(-1.702, 0.271) and (-0.893, 0.298).  A simpler cut would be to shrink the current $x_1$ range from (-3,3) and the $c$ range of (-0.3,0.3) to narrower ranges, and this will be studied in the future.

Given accurate simulations of the survey, the BBC method retrieves the nuisance parameters $\alpha$ and $\beta$ from Eq.~\ref{eqn:Tripp} and derives the distances of each SN.  As discussed in  KS17, the recovered nuisance parameters depend on assumptions about the intrinsic scatter model. The method returns $\alpha=$\salphanump~, $\beta=$\sbetanump~ and $\sigma_{int}=$\sintnump~ when assuming the G10 scatter model, and  $\alpha=\calphanump$, $\beta=\cbetanump$ and $\sigma_{int}=\cintnump$ when assuming the C11 scatter model.  The difference in $\sigma_{int}$ values is related to the assumed variation in the scatter model; the dispersion of Hubble residuals from both of these fits is about equal at $\sigma_{tot}=0.14$ mag.  

We can calculate the dependence of the bias in recovered distance on redshift by simulating the survey with both scatter models and measuring the difference between the true and recovered distances.   In Fig.~\ref{fig:searcheff} (bottom), we show the distance bias when we simulate the two different scatter models with their associated $\beta$ values (G10: $\sbetanump$; C11: $\cbetanump$), but assume that the true scatter model was the G10 model (effectively determining distances with $\beta=3.0$).  The biases calculated using the two different scatter models are within 5 mmags for almost the entire redshift range until $z\sim0.6$.  The uptick at high-z is due the interplay between color and brightness selection, and is discussed further in Section 5.

\subsection{Comparisons with R14}

\begin{figure}
\centering
\epsscale{1.15}  
\plotone{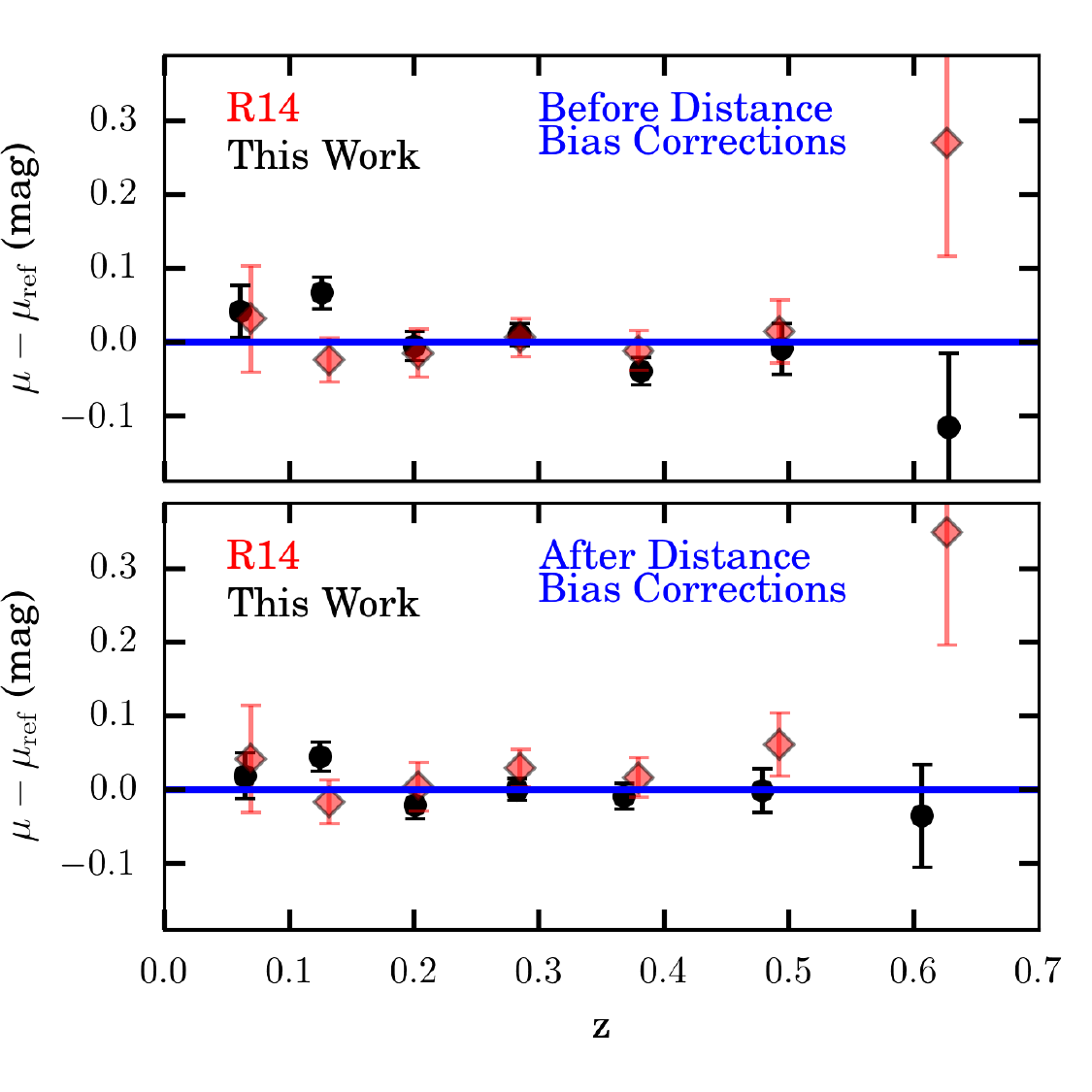}
\caption{Hubble residuals of the data to a reference cosmology for R14 and our analysis.  The residuals are shown before any distance bias corrections (Top) and after distance bias corrections (Bottom).  The top panel shows differences due to improved statistics, calibration and photometry.  The bottom panel shows further differences due to the improved bias corrections.  All bins with $>0$ SNe are shown and differences at high-z are driven by changes in photometry and different selection cuts.  The center of the redshift bins with BBC method are re-weighted using the SN distance uncertainties.}
\label{fig:mu_bias}
\end{figure}

Fig.~\ref{fig:mu_bias} shows a comparison of the mean distances in redshift bins between the R14 sample and our sample.  This is shown for before and after distance bias corrections are applied.  For the BBC method, the average of the corrections from the two scatter models is used.  Our sample has roughly twice as many SNe as the R14 sample, so the comparisons show both statistical differences and systematic differences between the two samples.  Relative to a reference cosmology, one indication of a trend is observed in the R14 sample but not in our sample: an increasing positive distance bias with redshift.  This trend appears significant in the R14 sample due to the highest-z bin, which has a positive residual of $\sim0.3$ mag.  That residual is driven by only two SNe, both with high ($>0.25$ mag) residuals.  In our sample, one of these SNe is cut due to the selection cuts and one of them has significantly changed photometry by $0.3$ mag due to low SNR and poor astrometry.  Smaller differences are driven by changes in the calibration of both the PS1 system and the SALT2 model.

\subsection{Mass determination}
\label{ssec:PS1Mass}
Multiple SN\,Ia analyses (discussed below) show that there is a correlation between luminosity of the SNe and properties of the host galaxies of the SN\,Ia.  This effect is important for measuring cosmological parameters as the demographics of SNe with certain host galaxy properties may change with redshift, and also because correcting for the effect may reduce the scatter of the distances.  Correlations between luminosity and the host galaxy mass (e.g., \citealp{Kelly10,Lampeitl10,Sullivan10}), age, metallicity and SFR (e.g. \citealp{Hayden12,Roman17}) have all been shown.  So far, the relation with mass appears to be the strongest of the correlations, possibly because it is easier to measure the mass of galaxies than the other properties, so in this analysis we use the mass dependence.  In S14, the difference in the mean Hubble residual for SNe in galaxies with high versus low masses (at split of $\textrm{log}(M_{\odot})=10$) was found to be $0.037\pm0.032$ mag, which is consistent with a null result as well as with the B14 statistical result of $0.06\pm0.012$ mag.  In this analysis, our statistics are a factor of $2\times$ larger and allow us to better measure the step.

The masses of PS1 host galaxies are derived similarly to S14 and follow the approach in \cite{Pan14}.  Here we use the seasonal templates, discussed in Section 2, to measure host galaxy photometry, and combine PS1 observations with $u$ band data from SDSS \citep{SDSS15} where available.  SExtractor's  FLUX\_AUTO \citep{Bertin96} was used to determine the flux values in \gps,\rps,\ips,\zps,\yps.  The measured magnitudes are analyzed with the photometric redshift code Z-PEG \citep{LeBorgne02}, which is based on the PEGASE.2 spectral synthesis code \citep{Fioc99}, and follows \cite{daCunha11} to calculate the stellar masses of host galaxies.  This is very similar analysis code to what is used to determine the masses of the host galaxies in the JLA sample (B14).  Further details of the assumptions used to run the code are discussed in \cite{Pan14}.

\begin{figure}

\epsscale{1.15}  
\plotone{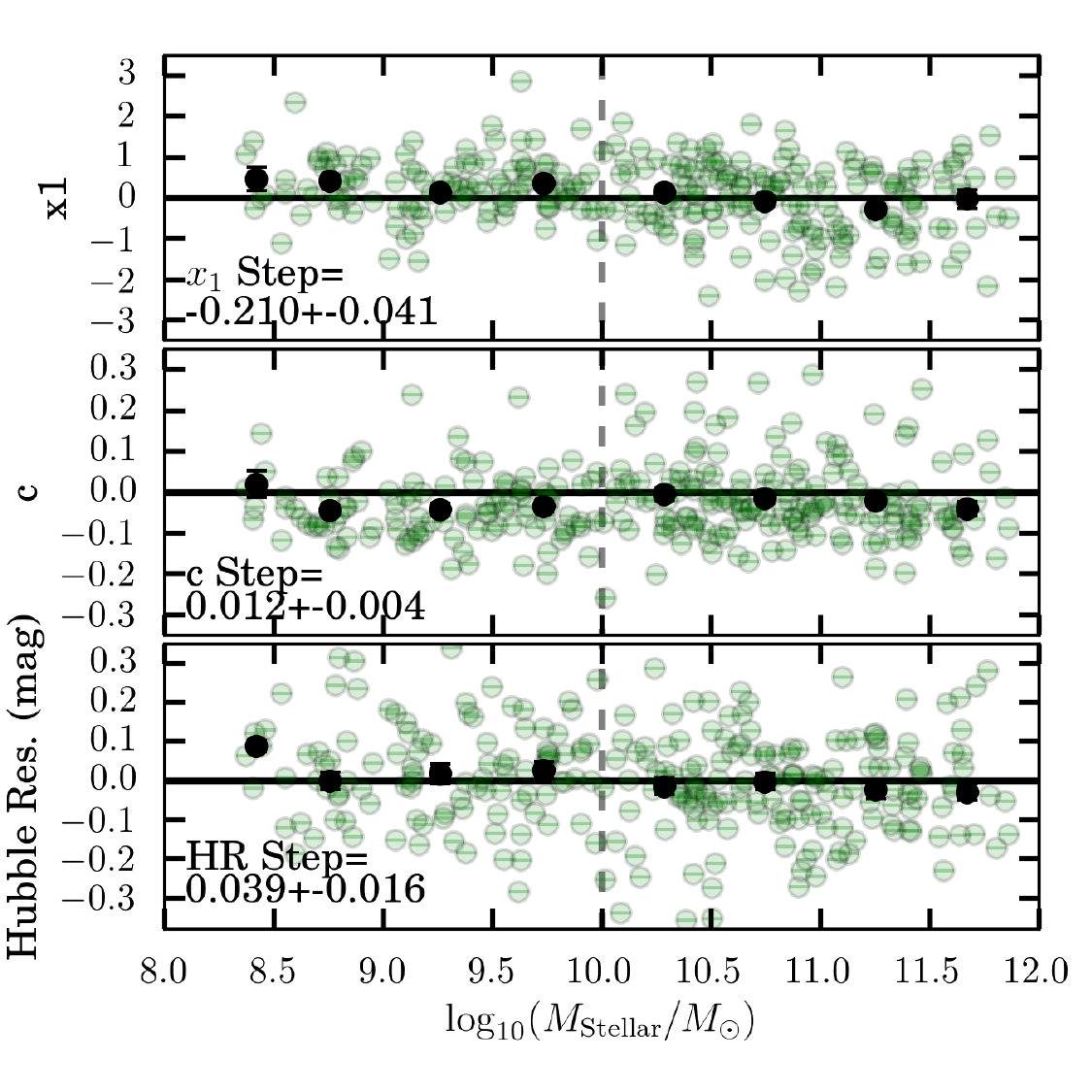}
\caption{Correlations in the data between color, stretch and Hubble Residuals with host galaxy mass.  A vertical line is shown at a host galaxy mass equal to $\log_{10}(M_{Stellar}/M_{\odot})=10$.  Steps are expressed as parameters for the higher mass group minus the lower mass group.}
\label{fig:pshost}
\end{figure}

Masses are determined by the Z-PEG code for all but 4 of the \numPS~PS1 host galaxies.  For the 4 SNe without matched host galaxies, no host galaxy was detected near the SN.  The masses of the hosts of these SNe are placed in the lowest mass bin (as done in B14).  The mass step is typically placed at $10^{10}(M_{\odot})$, and we find that there are~\masspslow~host galaxies with masses higher than the split value and~\masspshigh~with masses lower than the split value.  In Fig.~\ref{fig:pshost}, we show relations between the stretch, color and Hubble residuals of the SN\,Ia with the mass of the host galaxies.  We find no trend with color such that the mean color of SNe in low-mass hosts is $-0.001\pm0.004$ mag smaller than the color for SNe with high-mass hosts.  We also recover the typical trend such that SNe with lower stretch values are more often found in high-mass hosts, with a median difference in stretch values between high and low-mass hosts of $\Delta x_1=0.210\pm0.041$.

The split in luminosity with mass is determined by three parameters: a relative offset in luminosity, a mass step for the split, and an exponential transition term in a Fermi function that describes the relative probability of masses being on one side or the other of the split:
\begin{equation}
\Delta_M=\gamma \times [1+e^{(-(m-m_{\rm step})/ \tau )}]^{-1}~~~.
\label{eqn:mass}
\end{equation}
The Fermi function that is chosen here is used to allow for both uncertainty in the mass step and uncertainty in the host masses themselves.  For the PS1 sample, $\gamma=\sgammanump$~mag, $m_{\rm step}=10.02 \pm 0.06$ and $\tau=0.134 \pm  0.05$.  The step $\gamma=\sgammanump$ is similar to that found in S14, although with a smaller uncertainty.  Interestingly, if we did not apply the BBC method, as done in S14, we find for this sample $\Delta \mu = \ogammanump$~mag. This is roughly $1\sigma$ larger than with the BBC method and is more consistent with the mass step recovered in B14 of $0.06\pm0.012$ mag, which also did not implement the BBC method.  

To test how the BBC method accounts for a relation between mass and luminosity, we created new simulations with host galaxy mass properties assigned to every SN.  We assigned a host-mass to each SN so that the simulated sample replicates the trends of mass with $c$ and $x_1$ seen in Fig.~\ref{fig:pshost}.  Applying Eq.~\ref{eqn:mass} to the simulations using the BBC method, we see a bias of only $0.0035$ mags in the recovered value of $\gamma$ given an input value of $\gamma=0.08$ mag.  Furthermore, we find that including a mass step has only a $0.001$ mag effect on the distance bias corrections shown in Fig.~\ref{fig:searcheff} and therefore has a limited impact on our analysis.

\section{Combining Multiple SN Samples}
\label{sec:Data}

The PS1 survey is the latest in a long line of programs designed to build up a set of cosmologically useful SN Ia. To optimally constrain the cosmological parameters, we supplement the PS1 data with available SN\,Ia samples: CfA1-CfA4,
\cite{Riess99,Jha06,Hicken09a,Hicken09b,Hicken12}; CSP,
\cite{Contreras10,Folatelli10,Stritzinger11}; SNLS \citep{Conley11,Sullivan11}, SDSS \citep{Frieman08, Kessler09} and high-z data ($z>1.0$) from the SCP survey \citep{Suzuki12}, GOODS \citep{Riess07} and CANDELS/CLASH survey \citep{Rodney14,Graur14,Riess17}.  We do not include samples like Calan Tololo which were not in \cite{Supercal} and following \cite{Supercal}, we separate CfA3 into two sub-surveys CfA3K and CfA3S and CfA4 into two periods CfA4p1 and CfA4p2.  These surveys extend the Hubble diagram from $z\sim0.01$ out to $z\sim2$.   We note that because of the difficulty of high-z spectroscopic identification, the confidence in the spectroscopic identification of the $z>1$ HST SNe is not quite as high as for the $z<1$ SNe, but this is addressed in each of the papers above and we only include SNe that are at a `Gold'-like level.  In total, there are \numTOT~SNe that are used in our cosmology analysis and we refer to this sample as the `Pantheon sample'.  The numbers of SNe from each subsample that are used in our cosmology analysis are shown in Table \ref{tab:vari_surv}.  The differences in the number of low-z SNe that pass the cuts compared to R14 (given in number of SNe from R14 minus number of SNe from our Pantheon sample) are:  CSP  (\rumCSP), CfA1 (\rumCFAo), CfA2 (\rumCFAw), CfA3 (\rumCFAt), CfA4 (\rumCFAf).  The largest difference here is from CSP, which may have underestimated their photometric error uncertainties so that the $\chi^2/NDOF$ values returned are typically too high to pass the quality cut.  This is briefly discussed in Appendix C.  

One of the main differences between the cuts used in R14 versus the current analysis is that we now require that the uncertainty of the date of peak ($\sigma_{pkmjd}$) is $<2$ days.  In R14, we required that there were observations of the SN taken before the date of peak SN brightness.  Here we require that there are observations taken at least 5 days after peak.  To understand the impact of this change, we simulated whether there is any bias in the recovered distances of SNe for which there are no observations before the date of peak brightness.  In simulations of 20,000 SNe, we found that any bias is $<1$ mmag.  Following B14, we do not place a further goodness-of-fit cut on the light-curve fit for the SNLS sample as those with a poor goodness-of-fit pass visual inspection except for single-observation outliers that are not removed in the SNLS light-curves.  For PS1, SDSS and the low-z samples, we include the goodness-of-fit cut; however, this is after removing photometric data points that are outliers ($>4\sigma$) from the light-curves.  Similar to the analysis of the PS1 sample, the BBC method cuts on SNe with $c$ and/or $x_1$ values outside the expected color and stretch distributions.  While the median absolute values of the $x_1$ and $c$ values for the entire Pantheon sample are $x_1=0.70$ and $c=0.06$, the median absolute values of the SNe that are cut when applying the BBC method are $x_1=1.7$ and $c=0.21$.  A total of 19 SNe are cut from the BBC method, which is discussed in more detail later in this section.

\begin{table}
   \caption{}
 \centering 
  \begin{tabular}{lll}
\hline
Sample & Number  & Mean $z$\\
CSP & 26 & 0.024 \\ 
CFA3 & 78 & 0.031 \\ 
CFA4 & 41 & 0.030 \\ 
CFA1 & 9 & 0.024 \\ 
CFA2 & 18 & 0.021 \\ 
SDSS & 335 & 0.202 \\ 
PS1 & 279 & 0.292 \\ 
SNLS & 236 & 0.640 \\ 
SCP & 3 & 1.092 \\ 
GOODS & 15 & 1.120 \\ 
CANDELS & 6& 1.732 \\ 
CLASH & 2& 1.555 \\ 
\hline  
Tot & 1048 &  

  \end{tabular}
\begin{flushleft}{Notes: Total numbers of SN Ia from surveys included in the Pantheon sample after all sample selection cuts for cosmological analysis are applied, as well as the mean redshift of each subsample.}\end{flushleft}
    \label{tab:vari_surv}
\end{table}

\begin{figure}

\epsscale{1.15}  
\plotone{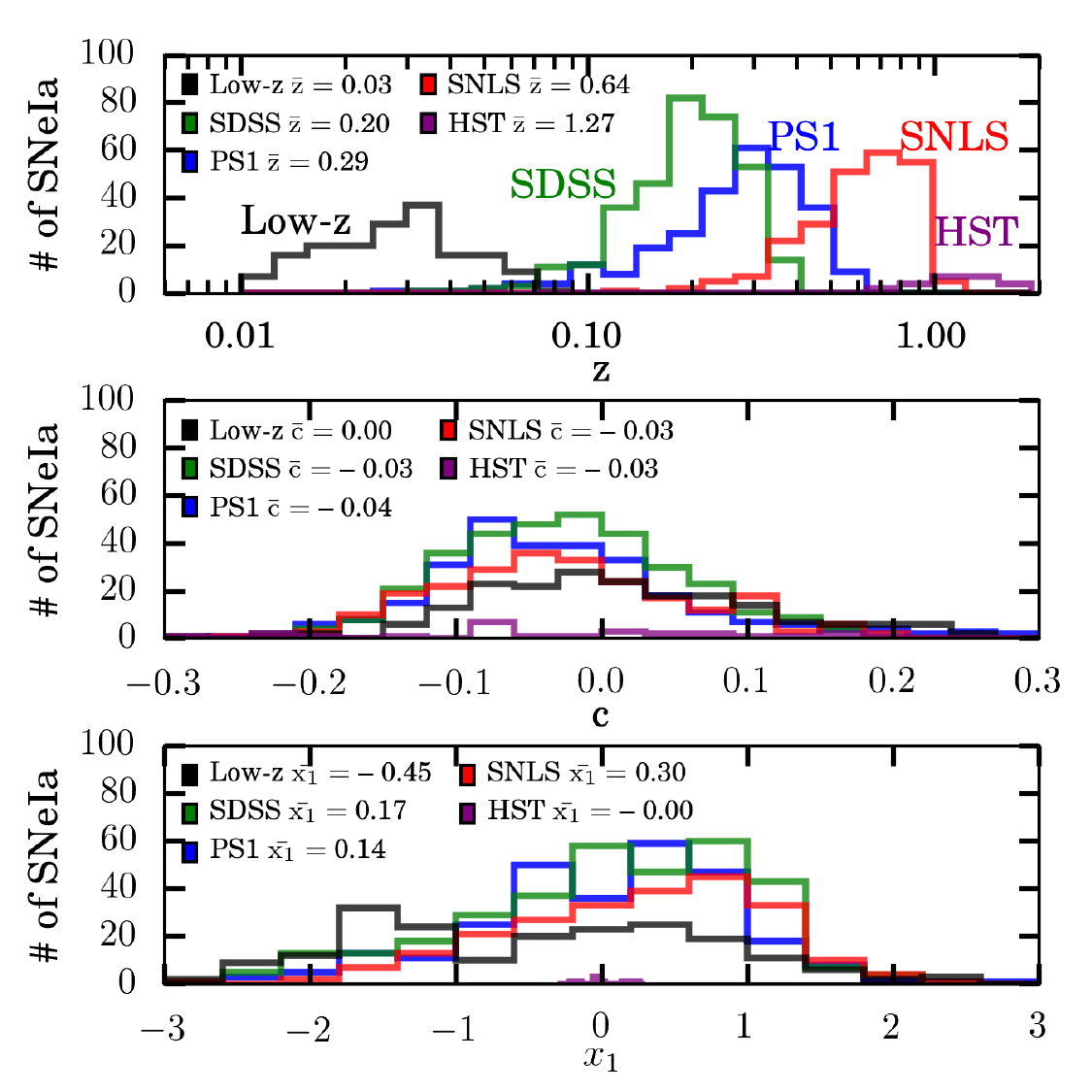}
\caption{Histograms of the redshift, color and stretch for each of the subsamples of the data.  The mean of each distribution is given in the legend.}
\label{fig:csz_hist}
\end{figure}

We fit all of the SNe in the same manner as for the PS1 sample described in Section 3.1; various aspects of this treatment for the non-PS1 samples are discussed in the following section.  We separate the full Pantheon sample into five subsamples: PS1, SDSS, SNLS, Low-z and HST where Low-z is the compilation of all the smaller Low-z surveys and HST is the compilation of all the HST surveys.  Histograms of the redshift, color and stretch for each of these five subsamples are shown in Fig.~\ref{fig:csz_hist}.  The subsamples cover a redshift range from $0.01<z<2.3$.  

\section{Analysis Framework}
\label{sec:Analysis}

\begin{figure*}
\epsscale{1.15}  
\plotone{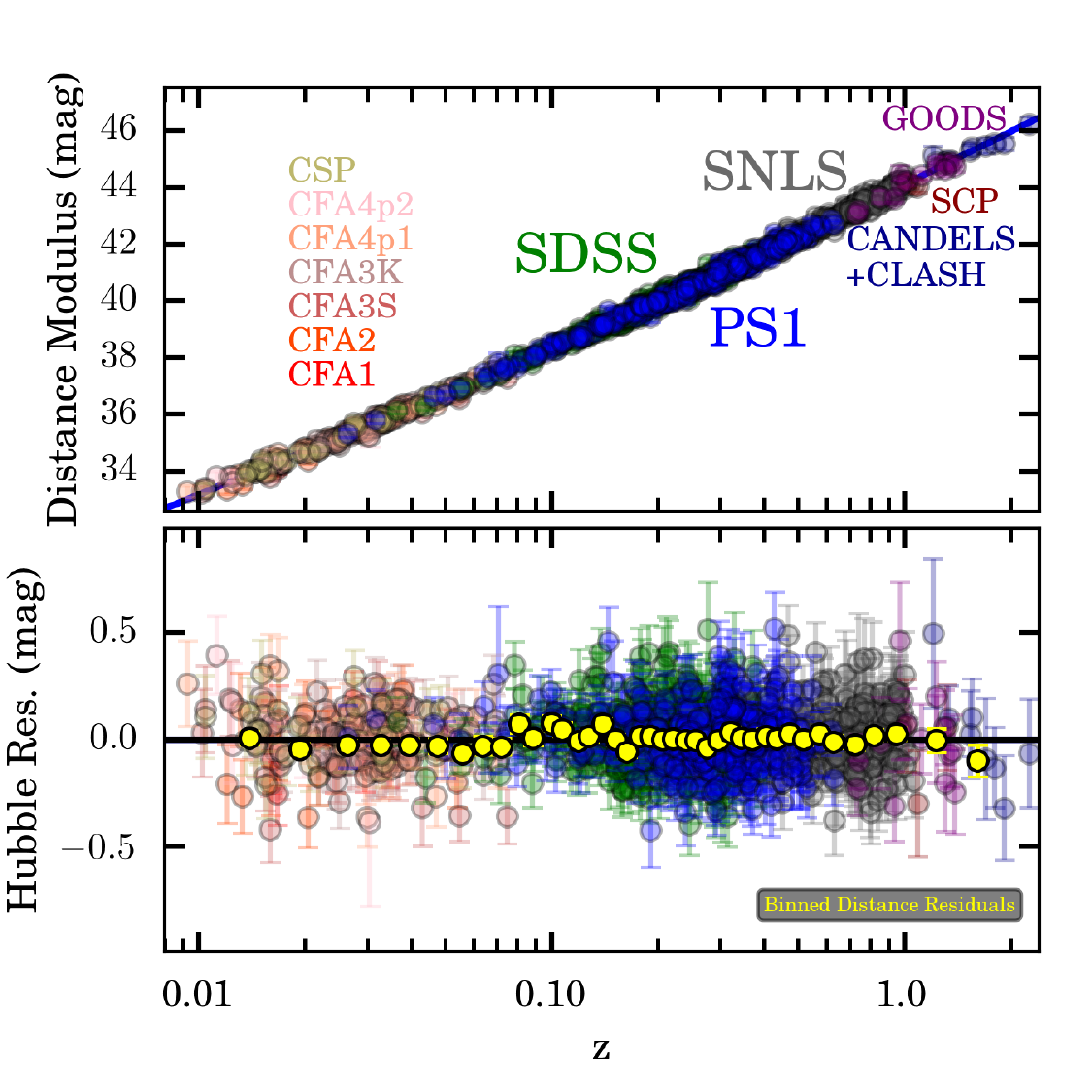}
\caption{The Hubble diagram for the Pantheon sample.  The top panel shows the distance modulus for each SN; the bottom panel shows residuals to the best fit cosmology.  Distance modulus values are shown using G10 scatter model.}
\label{fig:hubble}
\end{figure*}

The main steps of the analysis are calibration, distance bias corrections, MW extinction correction and coherent flow correction.  Furthermore, as described in Sec. 3.1, this analysis is blinded.  The Hubble diagram for the combined sample is shown in Fig.~\ref{fig:hubble}.  The distances for each of these SNe are determined after fitting the SN light-curves with SALT2, then applying the BBC method to determine the nuisance parameters and adding the distance bias corrections.  

Following \cite{Conley11}, the systematic uncertainties are propagated through a systematic uncertainty matrix. An uncertainty matrix
$\mathbf{C}$ is defined such that 
\begin{equation}
\mathbf{C} =\mathbf{D}_{\mathrm{stat}} + \mathbf{C}_{\mathrm{sys}}.
 \label{eqn:cdef}
\end{equation}
The statistical matrix
$\mathbf{D}_{\mathrm{stat}}$ has only a diagonal component that
includes errors defined in Eq. \ref{eqn:disterror}.  Since the BBC method produces distances from the fit parameters directly, there is only a single systematic covariance matrix for $\mu$ instead of the 6 parameter systematic covariance matrices ($m_b, x_1, c, m_bc,x_1m_b,x_1c$) for each of the SALT2 fit parameters \citep{Conley11}. We apply a series of systematics to the analysis and run BBC,  which produces binned distances over discrete redshift bins.
Therefore, the systematic covariance $\mathbf{C}_{\mathrm{sys}}$, for a vector of binned distances $\vec{\mu}$, between the ith and jth redshift bin is calculated as:
\begin{equation}
 \mathbf{C}_{ij,\mathrm{sys}} = \sum_{k=1}^K
   \left( \frac{ \partial \mu_{i} }{ \partial S_k } \right)
   \left( \frac{ \partial \mu_{j} }{ \partial S_k } \right)
   \left( \sigma_{S_k} \right)^2,
\label{eqn:covar}
\end{equation}
where the sum is over the $K$ systematics - each denoted by $S_k$, $\sigma_{ S_k}$ is the
magnitude of each systematic error, and
$ \partial \mu$ is defined as the difference in binned distance values after changing one of the systematic parameters.  

Given a vector of binned distance residuals of the SN sample that may be expressed as $\Delta \vec{\mathbf{\mu}} = \vec{\mu} - \vec{\mu}_{\rm model}$ (as shown in Fig.~\ref{fig:hubble} (bottom)) where $\vec{\mu}_{\rm model}$ is a vector of distances from a cosmological model, then the $\chi^2$ of the model fit is
expressed as
\begin{equation}
 \chi^2 = \Delta \vec{\mu}^T \cdot \mathbf{C}^{-1} \cdot  \Delta \vec{\mu} .
  \label{eqn:chieqn}
\end{equation}

Here we review each step of the analysis of the Pantheon sample and their associated systematic uncertainties.

\subsection{Calibration}

\begin{table*}[ht]
\caption{}
\centering
\begin{tabular}{l|ccccc}
\hline \hline
Survey & Filters & S15 zpt. offsets& ZP Err & Eff. Wave. Err.  & Ref. \\
~&~& [mmag] &  [mmag]  & [nm] & ~\\
\hline
PS1 & griz  & [-4,-7,-4,8] & [2,2,2,2] & [0.7,0.7,0.7,0.7] & \cite{Tonry12,Supercal}\\
SNLS & griz &  [7,-1,-6,2]&[2,2,2,2] & [0.3,1.0,3.1,0.6]  & \cite{Betoule12}\\
SDSS & griz&  [-3,4,1,-8] & [2,2,2,2] & [0.6,0.6,0.6,0.6]  & \cite{DOI10,Betoule12}\\
CfA1 & BVRI&  [33,4,0,-7] & [10,10,10,10]  & [1.2,1.2,2.5,2.5] &\cite{Landolt92}\\
CfA2 & BVRI&  [-2,0,0,-7] & [10,10,10,10]  & [1.2,1.2,2.5,2.5]  &\cite{Landolt92}\\
CSP & griBV & [9,1,-16,-8,2] &[4,3,5,5,5]&[0.8,0.4,0.2,0.7,0.3] & \cite{Contreras10}\\
CfA3Kep & riBV & [6,-3,-31,-6] & [3,5,6,4]  & [0.7,0.7,0.7,0.7] & \cite{Hicken09a}\\
CfA3S & BVRI& [-34,-9,-20,-14] & [6,4,3,5]  & [0.7,0.7,0.7,0.7]  & \cite{Hicken09a}\\
CfA4 & riBV& [6,-3,-31,-6] & [3,5,6,4] & [0.7,0.7,0.7,0.7] &\cite{Hicken09a}\\
\hline
\end{tabular}
\begin{flushleft}{Notes: Summary of various surveys used in this analysis.  The columns are: \textbf{Filters} used for observations, \textbf{S15 calibration offsets} to correct defined calibration zeropoints so that each system is tied to the homogeneous Supercal calibration, \textbf{zeropoint error} from S15, \textbf{uncertainty in the mean effective wavelength} of the filter bandpasses, and \textbf{reference} for calibration.  $U$ and $u$ passbands are not used in the fitting of light-curves.  The \textit{HST} system is defined in \cite{Bohlin14} with uncertainties therein.}\end{flushleft}
 \label{tab:calib}
\end{table*}

The `Supercal' calibration of all the samples in this analysis is presented in S15.  S15 takes advantage of the sub-$1\%$ relative calibration of PS1 \citep{Schlafly12} across $3\pi$ steradians of sky to compare photometry of tertiary standards from each survey.   S15 measures percent-level discrepancies between the defined calibration of each survey by determining the measured brightness differences of stars observed by a single survey and PS1 and comparing this with predicted brightness differences of main sequence stars using a spectral library.  The largest calibration discrepancies found were in the $B$ band of the Low-z photometric systems: CfA3 and CfA4 showed calibration offsets relative to PS1 of $2-4\%$.  

In S15, calibration offsets from each system relative to PS1 were given.  However, since there is uncertainty in the AB zeropoints of PS1 and both SDSS and SNLS attempt to tie their calibration to HST Calspec standards, we adjust the PS1 zeropoints to reduce discrepancies in the cross-calibration with SDSS and SNLS; we average the absolute calibration offsets that are given in S15 from SNLS, SDSS and PS1 (PS1 has by definition offsets of 0.0).  DOIng so, we subtract the following calibration offsets from PS1 catalog magnitudes: $\Delta g=-0.004, \Delta_r=-0.007, \Delta_i=-0.004$, and $\Delta_z=0.008$.  Calibration offsets of every other survey are corrected accordingly.   These calibration offsets are shown in Table \ref{tab:calib}, in addition to the uncertainties in the S15 zeropoints.  The uncertainties on the mean effective wavelength of the transmission functions of each system are unchanged, except for SNLS $r$ band, which in B14 has a stated uncertainty of $3.7$ nm.  The recovered calibration discrepancy found in S15 for the SNLS $r$ band is $<1$ nm so we conservatively prescribe a $1$ nm uncertainty to this band.  

This work does not achieve the maximum possible reduction of systematic biases from the Supercal approach, because the SALT2 model was trained and calibrated using a SN sample that was not recalibrated using the Supercal method.  Therefore, the SALT2 model itself propagates calibration uncertainties with values that are assigned by B14.  To account for the possibility of calibration biases in the SALT2 model, we fitted our SN sample with multiple iterations of the SALT2 model that were made in B14 by propagating systematic uncertainties in the calibration of each sample used for training. Furthermore, we estimate an additional systematic uncertainty of the whole Supercal process to be $1/3$ of the Supercal correction as the correction is dominated by discrepancies of $B-V$ to $3\%$, and we are confident to roughly $1\%$.

The calibration uncertainty from the HST Calspec standards is described in \cite{Bohlin14}.  A relative flux uncertainty as a function of wavelength is determined by a comparison of pure hydrogen models of different white dwarfs to observed spectra, and is set such that the relative flux uncertainty is 0 at $5556~$\AA.  Roughly, the uncertainty is $5$ mmag for every $7000~$\AA.  There is an additional absolute uncertainty from \cite{Bohlin14} of $5$ mmag coherent across all wavelengths, however this uncertainty has no impact since all subsamples are tied to the same system.  In follow-up analyses, we will include a new network of WD standards from \cite{Narayan16b}.

\subsection{Distance Bias Corrections}

Following the method described in Sec. 3.5, to model the dependence of distances on assumptions about SN color and selection effects, the BBC method is applied with two different intrinsic scatter models to determine distances.  The population parameters for each non-PS1 sample are given in SK16.  For this baseline analysis, we don't allow for any evolution in $\alpha$, $\beta$ or $\gamma$.

The simulations for SDSS and SNLS are described in B14 and S15, and the Low-z simulations and selection effects are described in Appendix C.  The HST simulations are made in the same way as the PS1 and Low-z simulations so that they directly represent the data in the SCP Cluster survey, GOODS and CANDELS/CLASH surveys, but the spectroscopic selection efficiency was set equal to unity for these surveys.

The recovered nuisance parameters $\alpha$ and $\beta$ from the BBC method are given in Table~\ref{tab:nuisance} for both scatter models.  For the G10 and C11 models, values from each survey of $\alpha$ and $\beta$ are within $1\sigma$ of the combined Pantheon sample.  The recovered $\beta$ values are slightly less consistent using the C11 model, with a range from $\beta=\cbetanums$ from SNLS and $\beta=\cbetanumd$ from SDSS, but are all still near $1\sigma$ of the mean. These higher values of $\beta$, when using the C11 model, are consistent with recent analyses \citep{Scolnic14a,Mosher14,Mandel16} that find larger $\beta$ dependent on various assumptions about the intrinsic scatter of SN\,Ia.  

\begin{figure*}
\epsscale{1.15}  
\plotone{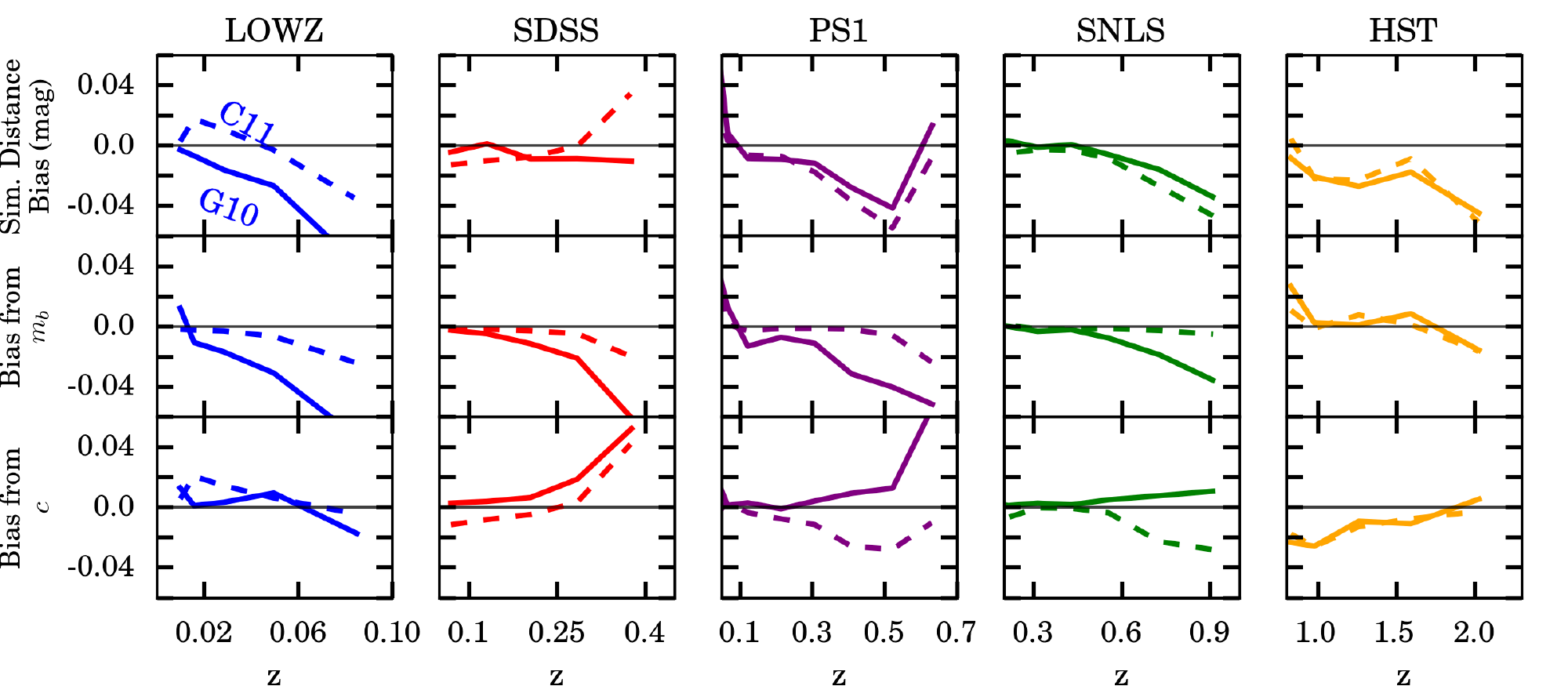}
\caption{(Top) Predicted bias in distance $\mu$ for each survey (Low-z, SDSS, PS1, SNLS, HST) and for the two scatter models (G10, C11).  (Middle) This row shows the predicted bias in $m_B$.   Combined, the middle and bottom panels roughly add up to the top panel.  The bias in $x_1$ is not shown because the bias propagated to distance is $<10\%$ of the total distance bias. (Bottom)  This row shows the predicted bias in $c$, but converted to distance units via Eq.~\ref{eqn:Tripp} and assuming $\beta=3.1$. 
}
\label{fig:oned_bias}
\end{figure*}

The predicted distance bias for each survey, using simulations of $>100,000$ SNe for each, is shown in Fig.~\ref{fig:oned_bias}. For display purposes, these biases are shown after simulating both assumptions about the scatter model but then assuming the `G10' scatter model is correct in the analysis and assuming a $\beta$ value of 3.1.  It is instructive to compare the biases in $m_B$ and $c$ for the different scatter models (the distance bias from $x_1$ is typically $<10\%$ of the total distance bias).  A key difference due to the scatter models is whether the selection effects at relatively high-z favor SNe with bluer color values or brighter peak brightness values (e.g., the SNLS $m_b$ panel and $c$ panel in Fig.~\ref{fig:oned_bias}).  For the G10 scatter model, the relative color bias with redshift is small compared to the relative $m_B$ bias with redshift.  For the C11 scatter model, the opposite is true.  The distance biases, when applying the two scatter models, agree to within 1\% for PS1, SNLS and HST, but as shown in Fig.~\ref{fig:oned_bias}, diverge more significantly for SDSS at high-z.  This difference for SDSS is likely due to its preferential selection of SNe based on both the magnitude and color of the SNe, rather than just the magnitude, as is the case for the other surveys.

The most significant difference in the distance biases between the two scatter models is for the Low-z sample, which as seen in Fig.~\ref{fig:oned_bias}, has a $\sim 0.03$ mag difference in the predicted distance bias for the two scatter models.  This large offset, relative to the other surveys, is due to the different distribution of color for this sample, as shown in Fig.~\ref{fig:csz_hist}: the mean color in the Low-z sample is $\Delta c\sim0.025$ mag redder than the other three samples.  The difference in bias between C11 and G10 for the Low-z sample comes from the difference in $\beta$ of $\Delta \beta=0.7$ from the two models multiplied the difference in the mean color of the sample relative to the higher-z samples of $\Delta c\sim 0.035$.  

Table~\ref{tab:nuisance} also shows the nuisance parameters and Hubble residual dispersion ($\sigma_{tot}$, different from the intrinsic scatter $\sigma_{int}$) for each subsample using both the BBC and the conventional method from B14 and S14.  It is clear from the dispersion values given in Table~\ref{tab:nuisance} that the BBC method reduces the dispersion of each subsample significantly.  A comparison with the conventional model is discussed more in Section 7.  The impact of the BBC method is higher when the measurement noise is greater.  For example, the RMS of SNLS distance residuals decreases from $0.18$ mag to $0.14$ mag, and this reduction can be seen higher on the higher-z SNe. Furthermore, the RMS reduction from bias corrections is least significant for the Low-z sample because the widths of the underlying $c$ and $x_1$ distributions are larger than the noise and intrinsic scatter of these parameters.  

\begin{figure}
\epsscale{1.15}  
\plotone{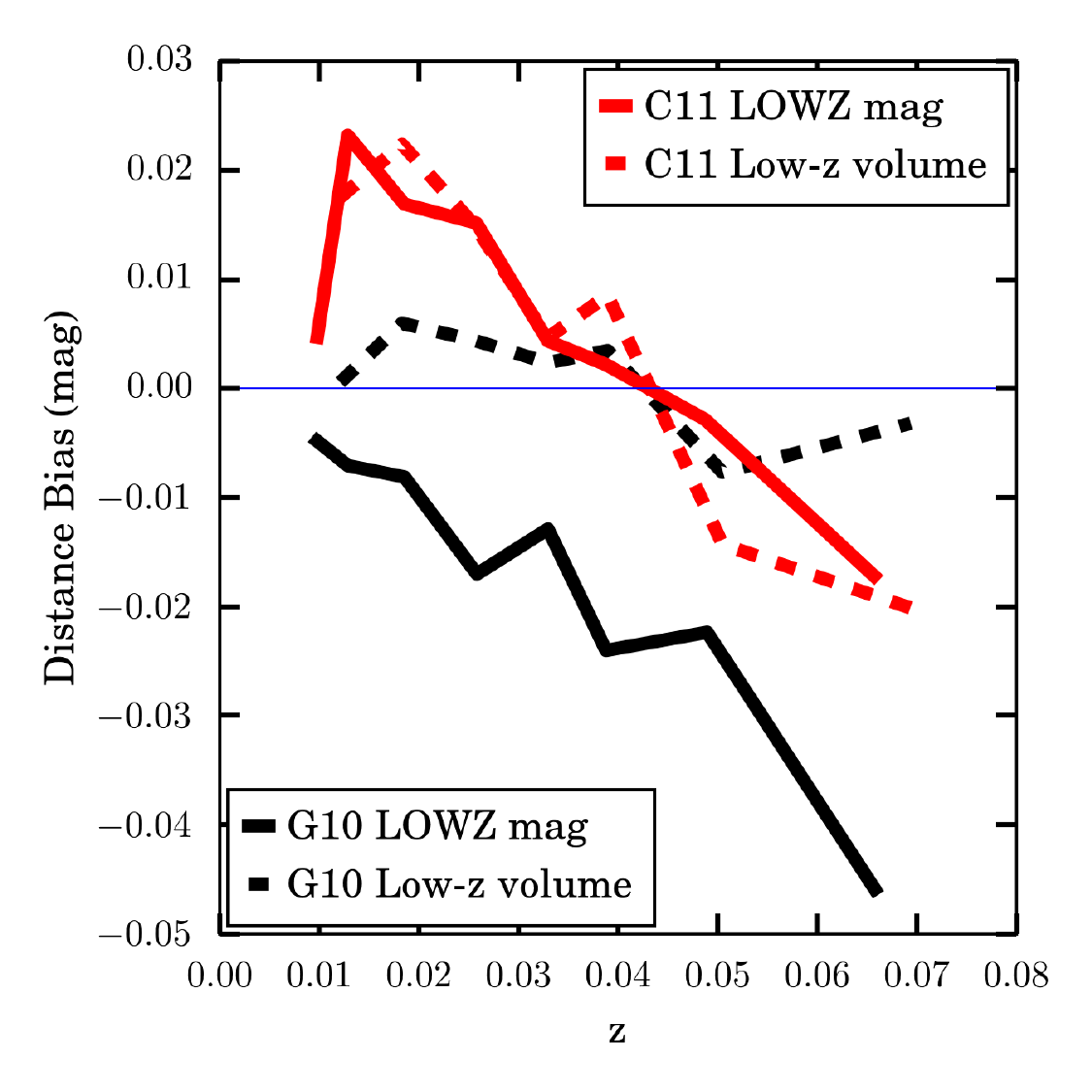}
\caption{Predicted bias in distance $\mu$ for different assumptions of the Low-z sample.  The plot is similar in form to the top left of Fig.~\ref{fig:oned_bias}.}
\label{fig:lowz_bias}
\end{figure}

To correct the distances for this analysis, we take the average of the G10 and C11 bias corrections.  The systematic uncertainty is half the difference between the two models.  For each survey, there is an additional systematic uncertainty in the distance bias corrections due to the uncertainty of the selection function of each survey.  As shown for the PS1 simulations in Fig.~\ref{fig:searcheff}, this uncertainty is determined by varying the selection function so that the $\chi^2$ agreement of the simulated redshift distribution and observed redshift distribution is reduced by $1\sigma$.  Understanding the uncertainty of the Low-z selection is most difficult because it is unclear to what extent the discoveries were magnitude-limited or volume-limited (see S14 appendix).  The differences in distance biases with redshift for the volume-limited and magnitude-limited assumptions are shown for the different cases in Fig.~\ref{fig:lowz_bias}.  For the volume-limited case, we prescribe a mean $c$ and $x_1$ dependence on $z$ in our simulations to mimic the trends seen in the data.  The differences in bias corrections at the high end of the redshift range can be as much as $0.03$ mag.  We use the magnitude-limited case in our baseline analysis and the volume-limited as our systematic, and this is discussed further in the Appendix C.

\begin{table}
\caption
 \centering 
  \begin{tabular}{l|cccc}
\hline

Survey & $\alpha$ & $\beta$ & $\gamma$ & $\sigma_{int}$[$\sigma_{tot}$] \\

\hline
\multicolumn{5}{c}{G10 with BBC}\\
Pantheon & $\salphanumg$ & $\sbetanumg$ & $\sgammanumg$ & $\sintnumg$[$\sdispnumg$] \\
Low-z & $\salphanuml$ & $\sbetanuml$ & $\sgammanuml$ & $\sintnuml$[$\sdispnuml$] \\
SDSS & $\salphanumd$ & $\sbetanumd$ & $\sgammanumd$ & $\sintnumd$[$\sdispnumd$] \\
PS1 & $\salphanump$ & $\sbetanump$ & $\sgammanump$ & $\sintnump$[$\sdispnump$] \\
SNLS & $\salphanums$ & $\sbetanums$ & $\sgammanums$ & $\sintnums$[$\sdispnums$] \\
\multicolumn{5}{c}{C11 with BBC}\\
Pantheon & $\calphanumg$ & $\cbetanumg$ & $\cgammanumg$ & $\cintnumg$[$\cdispnumg$] \\
Low-z & $\calphanuml$ & $\cbetanuml$ & $\cgammanuml$ & $\cintnuml$[$\cdispnuml$] \\
SDSS & $\calphanumd$ & $\cbetanumd$ & $\cgammanumd$ & $\cintnumd$[$\cdispnumd$] \\
PS1 & $\calphanump$ & $\cbetanump$ & $\cgammanump$ & $\cintnump$[$\cdispnump$] \\
SNLS & $\calphanums$ & $\cbetanums$ & $\cgammanums$ & $\cintnums$[$\cdispnums$] \\
\multicolumn{5}{c}{G10 with conventional fitting}\\
Pantheon & $\oalphanumg$ & $\obetanumg$ & $\ogammanumg$ & $\ointnumg$[$\odispnumg$] \\
Low-z & $\oalphanuml$ & $\obetanuml$ & $\ogammanuml$ & $\ointnuml$[$\odispnuml$] \\
SDSS & $\oalphanumd$ & $\obetanumd$ & $\ogammanumd$ & $\ointnumd$[$\odispnumd$] \\
PS1 & $\oalphanump$ & $\obetanump$ & $\ogammanump$ & $\ointnump$[$\odispnump$] \\
SNLS & $\oalphanums$ & $\obetanums$ & $\ogammanums$ & $\ointnums$[$\odispnums$] \\
\hline
  \end{tabular}
  \begin{flushleft}{Notes: Nuisance parameters $\alpha$ and $\beta$ of each sample, as well as the derived mass step, the intrinsic scatter $\sigma_{int}$ and the total Hubble residual RMS $\sigma_{tot}$.  Values are given for each sample, as well as the combined Pantheon sample.  Furthermore, values are given for the two different scatter model assumptions, G10 and C11, using the BBC method, as well as for the conventional method assuming all residual scatter after the SALT2 fits are due to luminosity variation.}\end{flushleft}
\label{tab:nuisance}
\end{table}

\subsection{Host Galaxy Mass -- Pantheon Sample}

\begin{figure}

\epsscale{1.15}  
\plotone{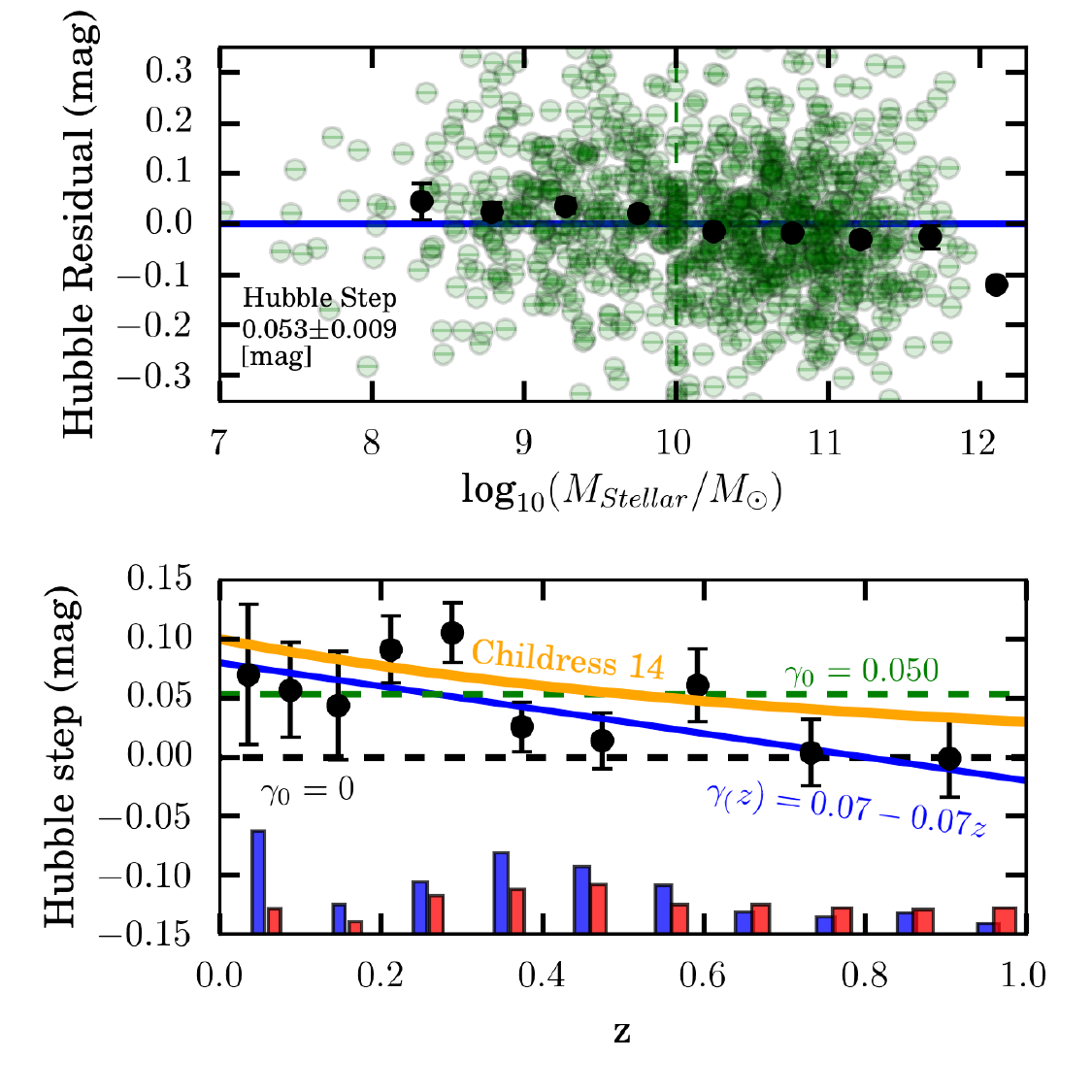}
\caption{(Top) Correlations of the data between mass and luminosity for the full Pantheon sample.  (Bottom) The Hubble step as a function of redshifts from $0<z<1$.  The bar graph shows the relative numbers of high ($M_{stellar}>10^{10}M_{\odot}$, solid-blue) and low ($M_{stellar}<10^{10}M_{\odot}$, solid-red) masses.  The dashed-green line shows the best fit constant offset of the Hubble step and the solid-blue line shows the best fit evolution of the Hubble step for high and low-mass galaxies (non-zero slope significant at $\sim1.75\sigma$ as given in Table~\ref{tab:Evolution}).  The orange line shows the prediction from \cite{Childress14}, see also \cite{Rigault13} for a similar prediction.}
\label{fig:mass_global}
\end{figure}

For the global analysis, we include mass estimates for all of the host galaxies of the SNe in each of the samples.  For the PS1 sample, these estimates are discussed in Sec.~\ref{ssec:PS1Mass}, and the estimates for the SDSS and SNLS sample are provided in B14.  We select 70 galaxies in the SDSS sample to compare our mass procedure to that done in B14, and we find a difference of {$\Delta \mathrm{log}_{10}(M_{stellar}/M_{\odot})=0.08\pm0.04$} with no dependance on galaxy mass.  For the Low-z sample, we redetermine masses for all the host galaxies using the same procedure as done for the PS1 sample and with photometry in $ugrizBVRIJHK$ that is available from the 2MASS data release \citep{2mass} and the SDSS Photometric Catalog, Data Release 12 \citep{SDSS15}.  For hosts that are too faint for survey depth, we assign them a mass in the lowest mass bin. 

The correlation between host mass and Hubble residual is shown in Fig.~\ref{fig:mass_global}.  Fitting Eq.~\ref{eqn:mass} to the Pantheon sample, we find that $m_{\rm step}=10.13\pm0.02$ and $\tau=0.001\pm0.071$.  We assume these values for this analysis and the difference with the fiducial $m_{\rm step}=10.0$ as a systematic uncertainty.  The inferred value of $\gamma$ and its uncertainty given the change in location of mass step is $<1\%$.  We find a mass step of  $\gamma=\sgammanumg$~mag and  $\gamma=\cgammanumg$~mag for the Pantheon sample using the G10 and C11 scatter models respectively.  This is smaller than the offset of $\gamma=\ogammanumg$~mag using no bias corrections.  
 
 As shown in Table~\ref{tab:nuisance}, the HR mass steps found for each subsample are all consistent to $\sim1\sigma$.  For the Pantheon sample, there are \massalllow~host galaxies with $\mathrm{log}_{10}(M_{stellar}/M_{\odot})<10$ and~\massallhigh~host galaxies with $\mathrm{log}_{10}(M_{stellar}/M_{\odot})>10$.  The relative splits of low/high-mass galaxies for each subsample are: PS1 (\masspslow / \masspshigh), SNLS (\masssnlslow / \masssnlshigh), SDSS (\masssdsslow / \masssdsshigh), and Low-z (\masslowzlow / \masslowzhigh).  While the Low-z sample has the highest offset of $\sgammanuml$~mag, the significance is $\sim2\sigma$ because there is a large imbalance in the number of high-mass and low-mass galaxies.  As shown in Fig.~\ref{fig:mass_global}, the relative numbers of high-mass and low-mass galaxies are within a factor of $2$ for all redshift bins $z>0.1$.  A change in host galaxy demographics with redshift is expected due to galaxy evolution, though the galaxy-targeted nature of the Low-z sample exaggerates the effect seen in Fig.~\ref{fig:mass_global}.  This is discussed further in Appendix C.

One possible systematic uncertainty from our baseline analysis is if the mass-luminosity relation itself changes with redshift, as predicted by \cite{Rigault13} and \cite{Childress14}.  These predictions are due to the inference that $\Delta_M$ from Eq.~\ref{eqn:mass} is due to a dependence of SN luminosity on the age of the SN progenitor. Since the correlation between host mass and progenitor age evolves with cosmic time, $\Delta_M$ should also change with redshift.  The modeled change in Hubble residuals due to this transition for \cite{Childress14} is shown in Fig.~\ref{fig:mass_global} - the magnitude of the Hubble step decreases with redshift.  We find the best fit line, solved simultaneously with the other nuisance parameters in the BBC fit, is $\gammanumc + \gammanumzc \times z$ for the G10 scatter model, and it is roughly the same for the C11 scatter model.  Since our measured slope appears to be roughly consistent ($2\sigma$) with the prediction of \cite{Childress14} and \cite{Rigault13}, we include it as a systematic  uncertainty. 

As done for PS1, we simulate host galaxy masses for each sample to determine if there are any biases in the recovered host mass-luminosity relation.  We do not see any discrepancies between input and output values of $\gamma$ from simulations beyond $3$ mmags, much smaller than the uncertainty on $\gamma$ values reported in Table~\ref{tab:nuisance}. 

\subsection{Demographic Changes}

Any change in the standardization parameters can produce systematic uncertainties in the measurements of cosmological parameters \citep{Conley11}.  We define $\beta(z)=\beta_0+\beta_1 \times z$, and similarly for $\alpha(z)$.  Before discussing the recovered $\alpha$ and $\beta$ values with redshift, we note that in our simulations of the Pantheon sample, using 30 simulations of $\sim1000$ SNe, when we input $\beta_1=0$, we recover a biased value of $\beta_1$ ($\beta_1=-0.35\pm0.06$ for the G10 model and $\beta_1=-0.7\pm0.10$ for the C11 model).  This bias is not present when we just fit for $\beta(z)=\beta_0$ and the problem is predicted in \cite{Kessler13} due to problems with using the correct intrinsic-scatter matrix in the fit.  We find similar issues with or without the BBC method.  Therefore, we subtract out the evolution bias predicted from the simulations in our fits for evolution.  

\begin{figure}
\epsscale{1.15}  
\plotone{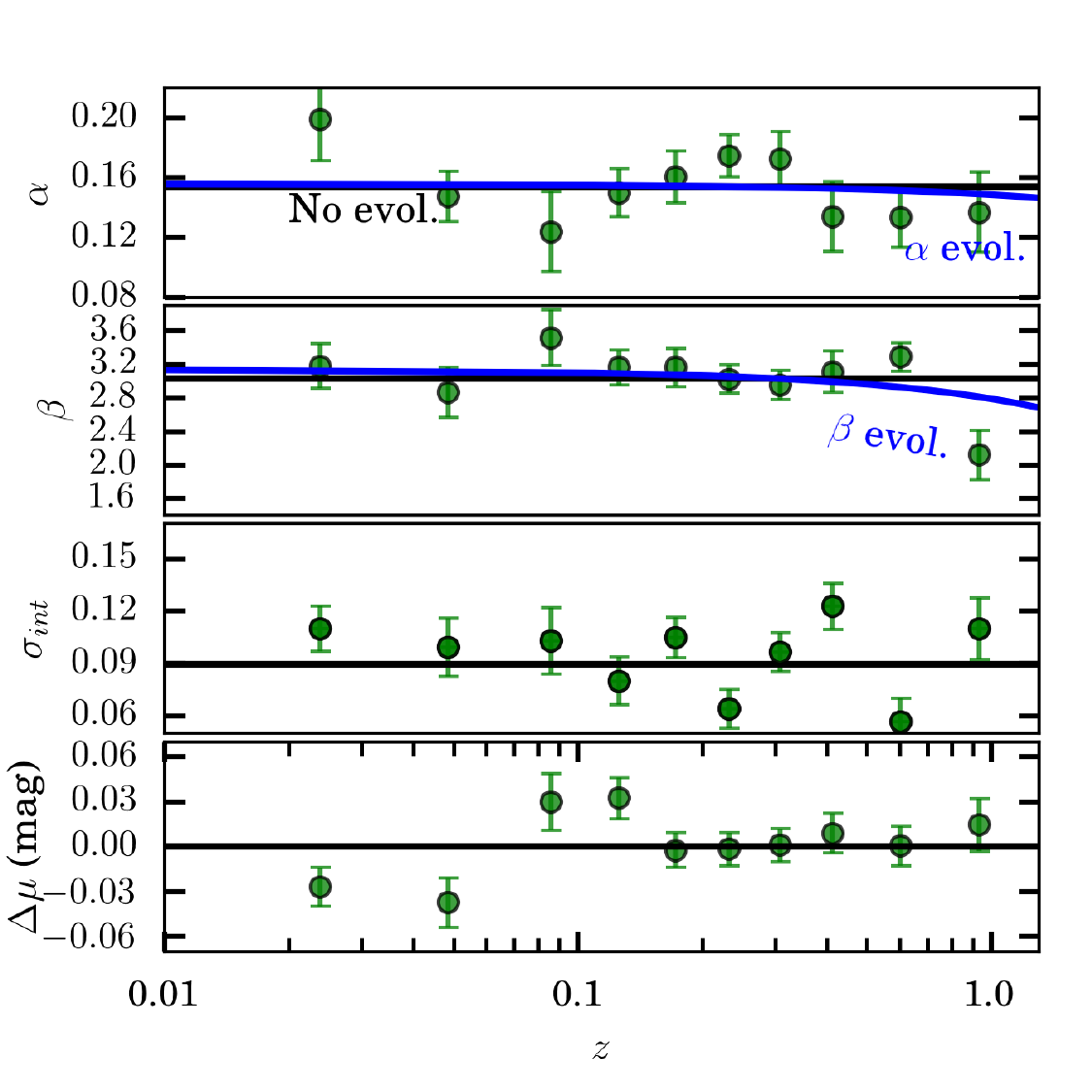}
\caption{Values of $\alpha$, $\beta$ and $\sigma_{int}$ for discrete redshift bins using the Pantheon sample.  A fit using an additional parameter to describe the evolution is shown in blue for the upper two panels and the baseline fit excluding evolution is shown in black for each panel.  The uncertainties of the slopes are given in Table~\ref{tab:Evolution}.}
\label{fig:param_evol}
\end{figure}

In Fig.~\ref{fig:param_evol}, the values of $\alpha$, $\beta$ and $\sigma_{int}$ are all shown for discrete redshift bins (only for G10 model for simplicity).  Table~\ref{tab:Evolution} reports the parameters of the best-fit lines to the evolution shown in Fig.~\ref{fig:param_evol}.   We do not see convincing evidence of $\alpha$ or $\beta$ evolution except in the highest redshift bin in the $\beta$ evolution plot where the SNe have the largest uncertainties.   We therefore choose for the baseline analysis to have $\beta_1=0$ and $\alpha_1=0$, though we allow for there to be a $\beta_1$ systematic equal to the size of the uncertainty in the $\beta_1$ measurement, treating it like a statistical uncertainty.  In past analyses (e.g., \citealp{Conley11}), different values of the $\sigma_{int}$ are used for different samples; however as shown in Fig.~\ref{fig:param_evol} we find consistency across samples and fix one value.

One related issue to the parameter evolution is possible population drift of the underlying $c$ and $x_1$ populations with redshift.  As shown in \cite{Rubin16}, not accounting for this drift yields very large differences in the inferences of cosmological parameters.  This drift is accounted for using the BBC method, since it accounts for selection effects and allows for different underlying light-curve parameter distributions for each subsample as presented in SK16.

\begin{table*}
\caption{}
 \centering 
 
  \begin{tabular}{l|cccccc}

\hline
\hline
~ & $\alpha_0$ & $\alpha_1$ &  $\beta_0$ &  $\beta_1$ & $\gamma_0$ & $\gamma_1$ \\
\hline
Baseline & \alphanumn  & 0 & \betanumn  & 0 & \gammanumn & 0 \\
$\alpha$ Evol. & \alphanuma & \alphanumza & \betanuma & 0 & \gammanuma &0\\
$\beta$ Evol. & \alphanumb & 0 & \betanumb & \betanumzb & \gammanumb & 0 \\
$\gamma$ Evol. & \alphanumc & 0 & \betanumc & 0 & \gammanumc &  \gammanumzc  \\
$\alpha,\beta,\gamma$ Evol. & \alphanumd & \alphanumzd  & \betanumd & \betanumzd  &  \gammanumd & \gammanumzd  \\
\hline
  \end{tabular}
    \begin{flushleft}{Notes: Recovered evolution values from the data assuming the G10 scatter model.  $\alpha$, $\beta$ and $\gamma$ evolution are each fit separately, as well as together.  The zeroth and first order component of each term is given, such that e.g., $\alpha(z)=\alpha_0+\alpha_1 \times z$.} \label{tab:Evolution}\end{flushleft}
\end{table*}

\subsection{Host Galaxy Extinction}
For each SN, we use an estimate of the extinction from dust along the line of sight determined from \cite{Schlafly11}.  Following S14, for the systematic uncertainty we adopt a global $5\%$ scaling of $E(B-V)$ as the systematic uncertainty.  A systematic bias in the SN distances due to uncertainty in the MW extinction is partially mitigated from fitting of the light-curve, as the light-curve color parameter may absorb some of the effects of the uncorrected MW extinction.  However, the impact on recovered cosmology may still be significant because the average MW $E(B-V)$ (mag) value per survey is different: 0.033 for Low-z, 0.037 for SDSS, 0.029 for PS1, 0.018 for SNLS, 0.010 for HST.  Additionally, extinction is treated differently in rest-frame versus redshift-frame.  Finally, we note that the SALT2 model was trained with the \cite{1998ApJ...500..525S} extinction model, which has been improved.  It is unclear how large of a systematic another retraining would yield, though we estimate that this is subdominant to the calibration and intrinsic scatter systematics.

\subsection{Coherent Flow Corrections}

The motion of SN host galaxies from coherent flows, like dipole or bulk flows, are corrected to reduce biases in cosmological parameters.  Past analyses, like S14 and B14, use the velocity field in \cite{Hudson04} which is derived using the galaxy density field from the IRAS PSCz redshift survey \citep{Branchini99}.  The same method is applied here using a map of the matter density field calibrated by the 2M++ catalogue\footnote{http://cosmicflows.iap.fr/} out to $z\sim0.05$, with a light-to-matter bias parameter\footnote{This has no relation to $\beta$ from Eqn. \ref{eqn:Tripp}.} of $\beta = 0.43$ and a dipole as described in \cite{Carrick15}.  

The impact on the dispersion of distance residuals due to the coherent flow corrections is relatively small.  The dispersion of distance residuals for SNe with $z<0.1$ is \pdispnump  ~mag after the corrections but \pdispnumn~mag before the corrections.  \cite{Carrick15} states that the uncertainty in galaxy velocities after the corrections is $150$ km/s, though it is unclear if that magnitude fully describes the uncertainty in redshift estimates of the SN host galaxies.  To test this issue, one may compare the intrinsic scatter of SN distances for low-z SNe to high-z SNe.  However, this test is complicated because there are many key differences between these subsamples, as shown in Fig.~\ref{fig:csz_hist} and discussed throughout this analysis.  Instead, we can compare the intrinsic scatter of distances of SNe from $0.01<z<0.03$ to distances of SNe from $0.03<z<0.06$ so that the subsamples are much more consistent.  DOIng so, we find the difference in intrinsic scatter between these two subsamples indicates an individual redshift uncertainty of $250$ km/s.  This is higher than the estimate from \cite{Carrick15}, but we use it for our analysis.  Furthermore, if we did not apply the coherent flow corrections, we find the individual redshift uncertainty must be $260$ km/s so that the sample has the same intrinsic scatter as the sample with coherent flow corrections.

The systematic uncertainty of the coherent flow corrections should account for covariance between the velocities of the SN host galaxies \citep{Hui06}.  This covariance matrix is modeled explicitly in \cite{Huterer16} using the Low-z sample from \cite{Supercal}, which is nearly identical to the Low-z sample used here.   Analyzing the SNe and galaxy data from 6dFGS separately, \cite{Huterer16} show both samples are consistent with the peculiar velocity signal of a fiducial~\LCDM~model.  Instead of implementing the full covariance matrix from \cite{Huterer16} which does not account for corrections of bulk flows, following \cite{Zhang17} we account for a systematic uncertainty in the coherent flow corrections by shifting the light-to-matter bias parameter $\beta$ by $10\%$ and redetermining the velocity corrections.  

\subsection{Summary of Systematic Uncertainties}

\begin{figure}
\epsscale{1.15}  
\plotone{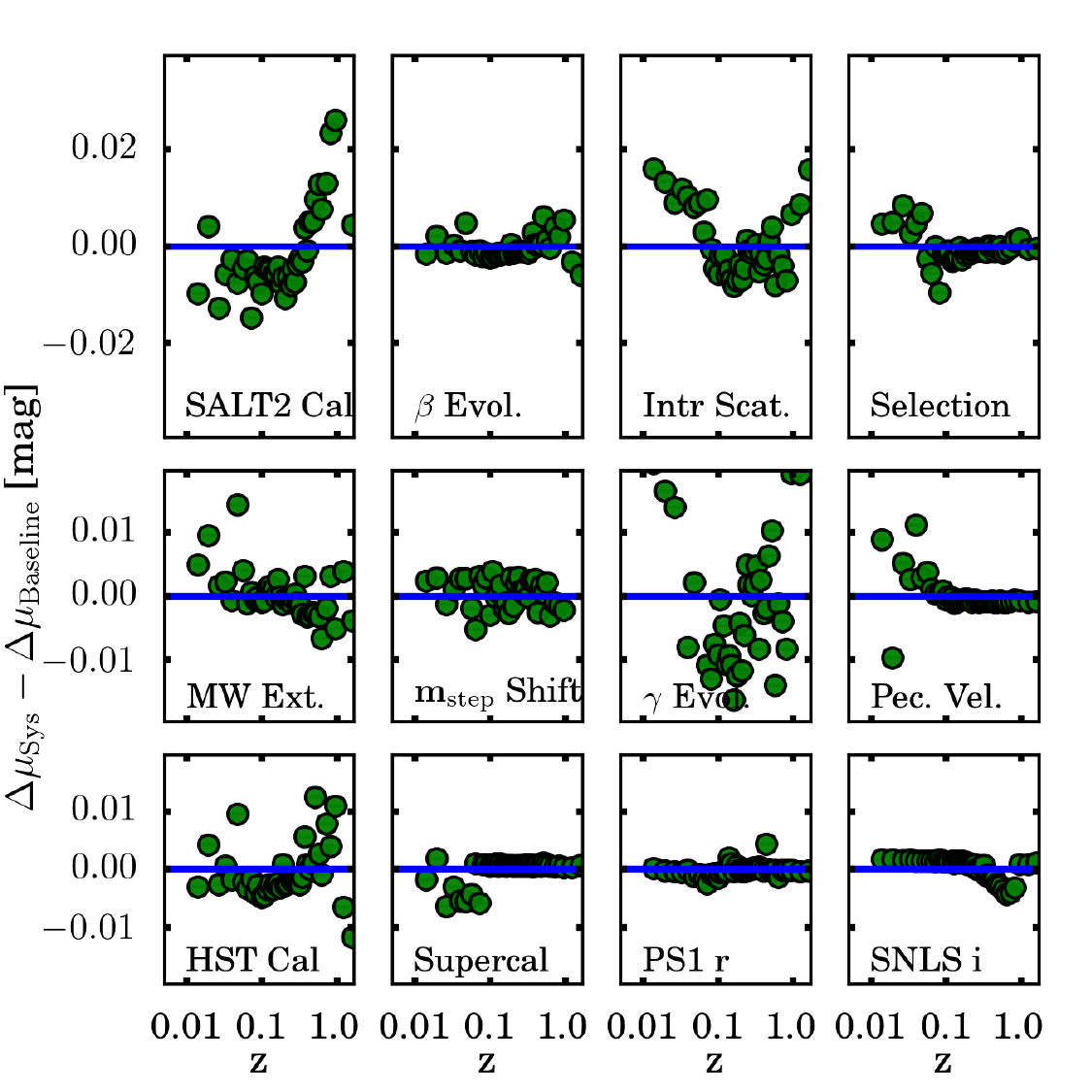}
\caption{Mean Hubble residuals differences relative to the binned distances of the Pantheon sample after individual systematic uncertainties are propagated.  The calibration uncertainties for various bands of various surveys are a representative selection of the survey uncertainties.}
\label{fig:sysgrid}
\end{figure}

\begin{figure}

\epsscale{1.15}  
\plotone{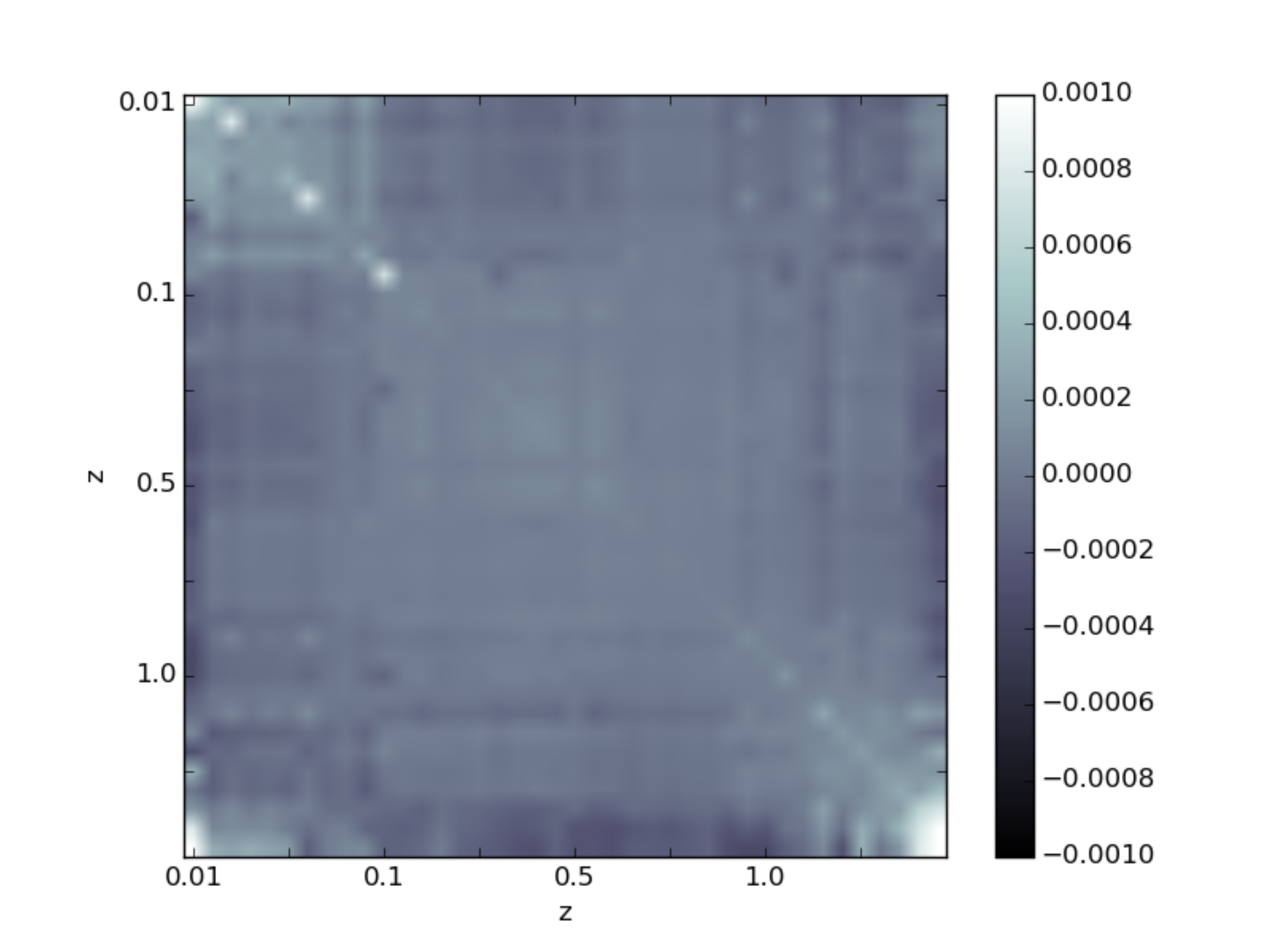}
\caption{Visualization of the covariance matrix for the full Pantheon SN Sample. }
\label{fig:covplot}
\end{figure}

There is a large list of systematic uncertainties associated with the various analysis steps in this section.  In total, there are 85 separate systematic uncertainties, though 74 are related to calibration.  All of the main systematic uncertainties on the binned distances is shown in Fig.~\ref{fig:sysgrid}; here we show the change in binned distances if we vary a given systematic by $1\sigma$.  For the survey-calibration uncertainties, we only show two of them, a systematic from SNLS and PS1, and there are roughly 8 of these uncertainties per survey as described in Table~\ref{tab:calib}.  The full systematic covariance matrix is shown in Fig.~\ref{fig:covplot}.

\section{Results}
\label{sec:Results}

\subsection{Fitting For Cosmological Parameters and the Impact of Systematic Uncertainties}

To determine cosmological parameters, each
measured distance modulus ($\mu$) from Eq.~\ref{eqn:Tripp} is compared to a model distance that depends on redshift
and cosmological parameters, $\mumodel = {+5}\log(d_L/10{\rm pc})$,
such that,
%
 \begin{eqnarray}
d_L(z) = \frac{c}{H_0} \ \lim_{\Omega_k' \to \Omega_k} \frac{1 + z}{\sqrt{\Omega_k'}} \, \sinh \left[\sqrt{\Omega_k'} \int_0^z \frac{dz'}{E(z')} \right]~~~~~\\
E(z) = \sqrt{\Omega_m (1 + z)^3 + \Omega_\Lambda (1 + z)^{3(1 + w)} + \Omega_k (1 + z)^2}~~~~~~~
\end{eqnarray}
    \label{eqn:cosmodel}
A grid of cosmological models is fitted over to minimize the $\chi^2$ in Eq.~\ref{eqn:chieqn}.
We explore four cosmological models: a flat~\LCDM~model ($w=-1$, $\Omega_K=0$), a non-flat~\OCDM~model ($w=-1$, $\Omega_K$ varies), a flat~\WCDM~model ($w_0$ varies, $w_a=0$), and a flat~\WACDM~model ($w_0$, $w_a$ both vary, $\Omega_K=0$).  All of the calculations are done with CosmoMC \citep{Lewis02}.  While the BBC method produces binned distances over redshift, for cosmological fitting we use the unbinned, full SN dataset.  We use the un-binned dataset mainly to be in-line with general community reproducibility.  We still use the binned distances to generate the systematic covariance matrix, which is used as a 2d 40-bin interpolation grid to create a covariance matrix for the full SN dataset.  Diagonal uncertainties from the individual distances can be added together with the full systematic matrix following Eq.~\ref{eqn:cdef}.  Differences in $w$ between the binned and un-binned datasets are at a $<1/16\sigma$ level for the statistical measurements, and $<1/8\sigma$ when including the systematic covariance matrix.

\begin{figure}

\epsscale{1.15}  
\plotone{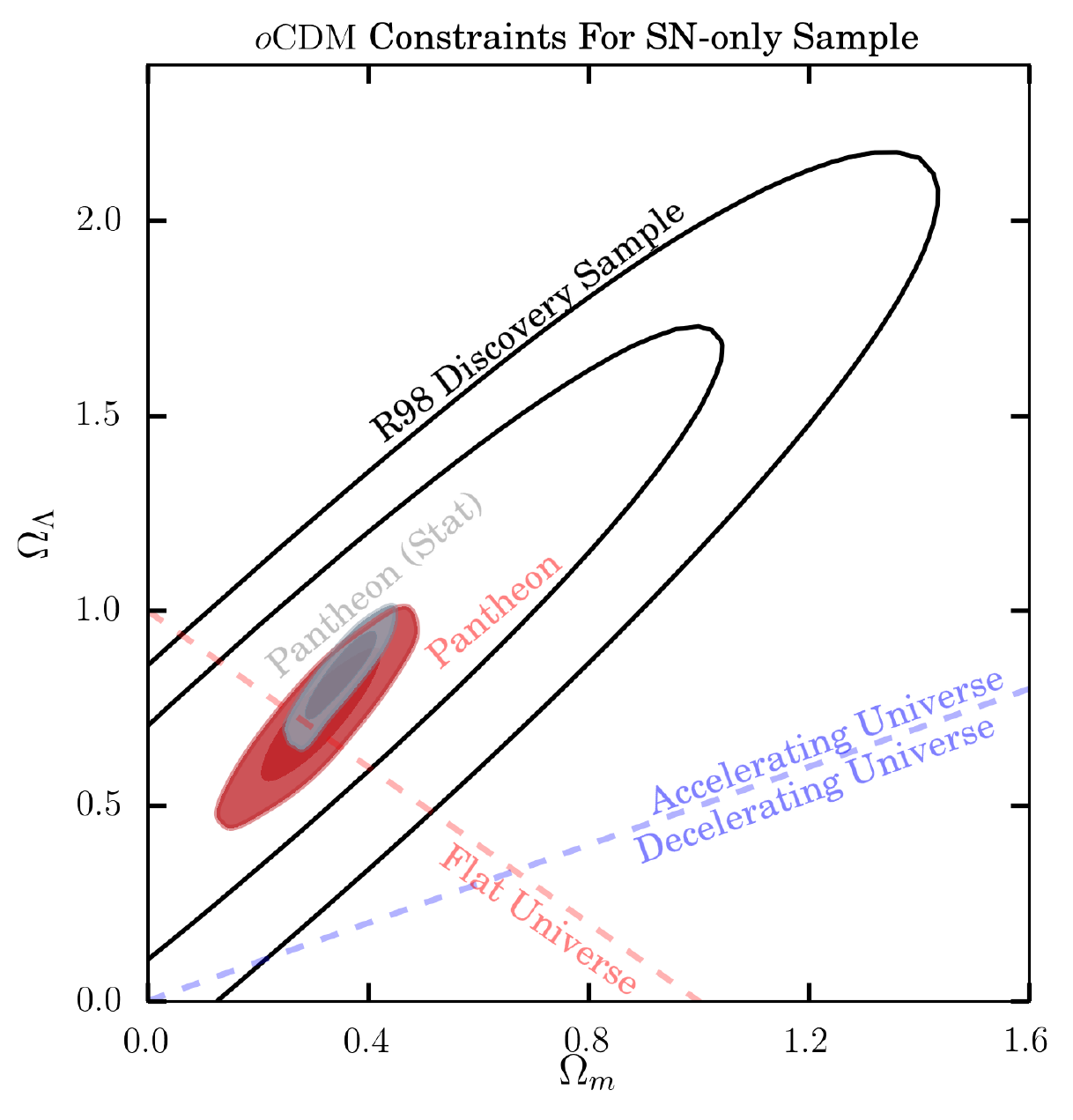}
\caption{Evidence for dark energy from SN-only constraints.  Here we show confidence contours at 68\% and 95\% for the $\Omega_m$ and $\Omega_\Lambda$ cosmological parameters
for the~\OCDM~model for both the \cite{Riess98} discovery sample and the Pantheon sample.  The Pantheon constrains with systematic uncertainties  are shown in red and with only statistical uncertainties are shown in gray (line).  }
\label{fig:plot_omol}
\end{figure}

The cosmological fits to the SN-only sample are shown in Table~\ref{tab:subs4} with and without systematic uncertainties.  Using our full SN sample with systematic uncertainties, with no external priors, we find $\Omega_m=\SNlOM$.  Without systematic uncertainties, the uncertainty on $\Omega_m$ is roughly $2\times$ smaller.  When not assuming a flat universe, we combine various probes together to constrain the~\OCDM~model.    When using SN alone, we find that $\Omega_m=\SNoOM$ and $\Omega_L=
\SNoOL$.  We find the evidence for non-zero $\Omega_\Lambda$ from the SN-only sample is $>6\sigma$ when including all systematic uncertainties.  As shown in Fig.~\ref{fig:plot_omol}, this is a factor of $\sim20$ improvement over the \cite{Riess98} constraints in this plane.  Furthermore, the significance for non-zero $\Omega_\Lambda$ is much higher than the $<3\sigma$ effect quoted by \cite{Nielsen16} which re-analyzed the B14 sample though their analysis technique is disputed by \citep{Rubin16}.  A study using the Pantheon sample and null tests done in this analysis to examine non-standard cosmological results like those from \cite{Nielsen16} and \cite{Dam2017} is currently in prep. (Shafer et al. in prep.).

\begin{table}
\caption{}
  \centering
  \begin{tabular}{l|lccc}

\hline
\hline
Analysis & Model & $w$ & $\Omega_m$  &  $\Omega_\Lambda$ \\

\hline
SN-stat & \LCDM && $ 0.284 \pm 0.012$ & $ 0.716 \pm 0.012 $\\
SN-stat & \OCDM && $ 0.348 \pm 0.040$ & $ 0.827 \pm 0.068 $\\
SN-stat & \WCDM & $ -1.251 \pm 0.144$ & $ 0.350 \pm 0.035$& \\
SN & \LCDM && $ 0.298 \pm 0.022$ & $ 0.702 \pm 0.022 $\\
SN & \OCDM && $ 0.319 \pm 0.070$ & $ 0.733 \pm 0.113 $\\
SN & \WCDM & $ -1.090 \pm 0.220$ & $ 0.316 \pm 0.072$& \\

\hline
   
  \end{tabular}
\begin{flushleft}  {Notes: Cosmological constraints for the SN-only sample with and without systematic uncertainties.  Values are given for three separate cosmological models:~\LCDM,~\OCDM~and~\WCDM.} \label{tab:subs4}\end{flushleft}
\end{table}

To evaluate the impact of the systematic uncertainties, we combine constraints from the Pantheon SN sample with those from the compressed likelihood of the CMB from \cite{Planck16} and measure $\Omega_m$ and $w$ in the \WCDM~model. Constraints from BAO and $H_0$ measurements are included later in this section.  The impact of systematic uncertainties is shown in terms of the relative size of the uncertainty of $w$ in Table \ref{tab:system}.  We find that the systematic uncertainty ($\sigma_w=\syssys$) is smaller than the statistical uncertainty ($\sigma_w=\sysstat$).  Unlike previous analyses (e.g., B14 and S14) that found that calibration uncertainties made up $>80\%$ of the systematic error budget, we find a more even split between the various systematics. The calibration uncertainties are due to uncertainties of the individual photometric systems of each sample as well as the calibration uncertainties propagated through the SALT2 model.  We find that the SALT2 calibration uncertainty is larger in magnitude than the combined impact from all the various systems, which are reduced by S15 and are independent of each other.    Still, all of the systematic uncertainties related to calibration have a net effect of roughly $66\%$ of the total systematic error.

\begin{table}[!htb]
\caption{}
  \centering

  \begin{tabular}{l|ccc}
\hline
~ & $w$ shift & $\sigma_w^{\rm syst}$ & Fraction of $\sigma_w^{(\rm stat)}$ \\
\hline
\hline
Stat. Uncertainty & +0.000 & 0.031 & 1.000 \\ 
Total Sys Uncertainty & +0.031& 0.025& 0.814\\ 
~~Calibration&~&~&\\ 
~~SALT2 Cal & -0.002& 0.014 & 0.457\\ 
~~Survey Cal & +0.006& 0.009 & 0.285\\ 
~~HST Cal & -0.006& 0.006 & 0.177\\ 
~~Supercal & +0.002& 0.003 & 0.098\\ 
~SN Modeling&~&~&\\ 
~~Selection & +0.010& 0.007 & 0.233\\ 
~~Intrinsic Scatter & +0.019& 0.005 & 0.170\\ 
~~$\beta$ Evol. & -0.001& 0.007 & 0.238\\ 
~~$\gamma$ Evol. & -0.002& 0.000 & 0.000\\ 
~~$m_{\rm step}$Shift & -0.002& 0.002 & 0.064\\ 
~External&~&~&\\ 
~~MW Extinction & +0.010& 0.008 & 0.262\\ 
~~Pec. Vel. & +0.000& 0.003 & 0.103\\ 

\hline
  \end{tabular}
  \begin{flushleft}{Notes: The dominant systematic uncertainties in the Pantheon SN sample with respect to $w$ while solving for a~\WCDM~model.  The $w$ shift is defined relative to the statistical value and $\sigma_w^{\rm syst}$ is defined to be $\sqrt{\sigma_w^2-\sigma_{w-stat}^2}$ when a specific systematic uncertainty is applied.}\label{tab:system}\end{flushleft}
\end{table}

The systematic uncertainties increase the uncertainties of the best fit parameters, and also shift the best fit parameters by reweighting the pulls of each SN in the fit.  These two impacts are shown in Table \ref{tab:system} as both the best-fit value of $w$ is shifted and the uncertainty on $w$ is increased.  The shifts are mainly due to systematic uncertainties that most strongly affect the low-z sample: calibration, MW extinction, intrinsic-scatter and selection. 

Of the non-calibration uncertainties listed in Table ~\ref{tab:system}, the uncertainty due to selection, MW extinction, intrinsic scatter and $\beta$ evolution are all similarly large at $\sim0.25$ of the statistical error. While the impact of the systematic uncertainty due to intrinsic scatter is not the largest, including it shifts the best-fit value of $w$ by $\Delta w =0.019$.  The systematic uncertainty effectively deweights the low-z sample and would have been even larger if we hadn't reduced the impact by a factor of $2\times$ by averaging the distance biases from the G10 and C11 model.   We find little impact on $w$ from the location of the mass step or the possibility that the magnitude of the host mass-luminosity relation is changing with redshift.  We see negligible impact from the peculiar velocity corrections.

\begin{table}
\caption{}
  \centering
  \begin{tabular}{l|cc}

\hline
\hline
Variant & $\Delta w$ & $\Delta \Omega_m$  \\

\hline
No Bias Corr. & $ +0.068$ & $ +0.015$\\ 
No Mass Corr. & $ -0.023$ & $ -0.007$\\ 
No Supercal Corr. & $ +0.024$ & $ +0.004$\\ 
No PV Corr. & $ +0.009$ & $ +0.001$\\ 

\hline
   
  \end{tabular}
\begin{flushleft}  {Notes: Differences in recovered values of $w$ and $\Omega_m$ with the~\WCDM~model of the main corrections in the analysis are omitted.  } \label{tab:subs2}\end{flushleft}
\end{table}

A useful guide to understand the possible scale of systematic uncertainties is to redetermine the cosmological parameters without the main sample-corrections.  This is shown in Table~\ref{tab:subs2} for an analysis only considering statistical uncertainties for cases where no distance bias correction is applied, no mass correction is applied, no Supercal correction is applied and no peculiar velocity correction is applied.  While the size of these shifts is larger than any of the systematics, they give a sense of where possible systematic uncertainties could reside.  We find that the distance bias correction changes the value of $w$ by $\sim7\%$, larger than any of the other corrections.  The mass and Supercal corrections are both large ($\Delta w \sim 0.024$), due to their impact on the Low-z sample.  As shown in Fig.~\ref{fig:mass_global}, the change in demographics of the host galaxies with redshift makes the recovered cosmological values from the sample sensitive to the mass correction.  The Supercal correction is $\Delta w =0.024$ because there is a recalibration of $B-V$ zeropoints in the low-z sample which shift the average distances by $\sim 0.02$ mag.  The peculiar velocity correction is small, on the order of $1\%$ in $w$.

\begin{figure}
\epsscale{1.15}  
\plotone{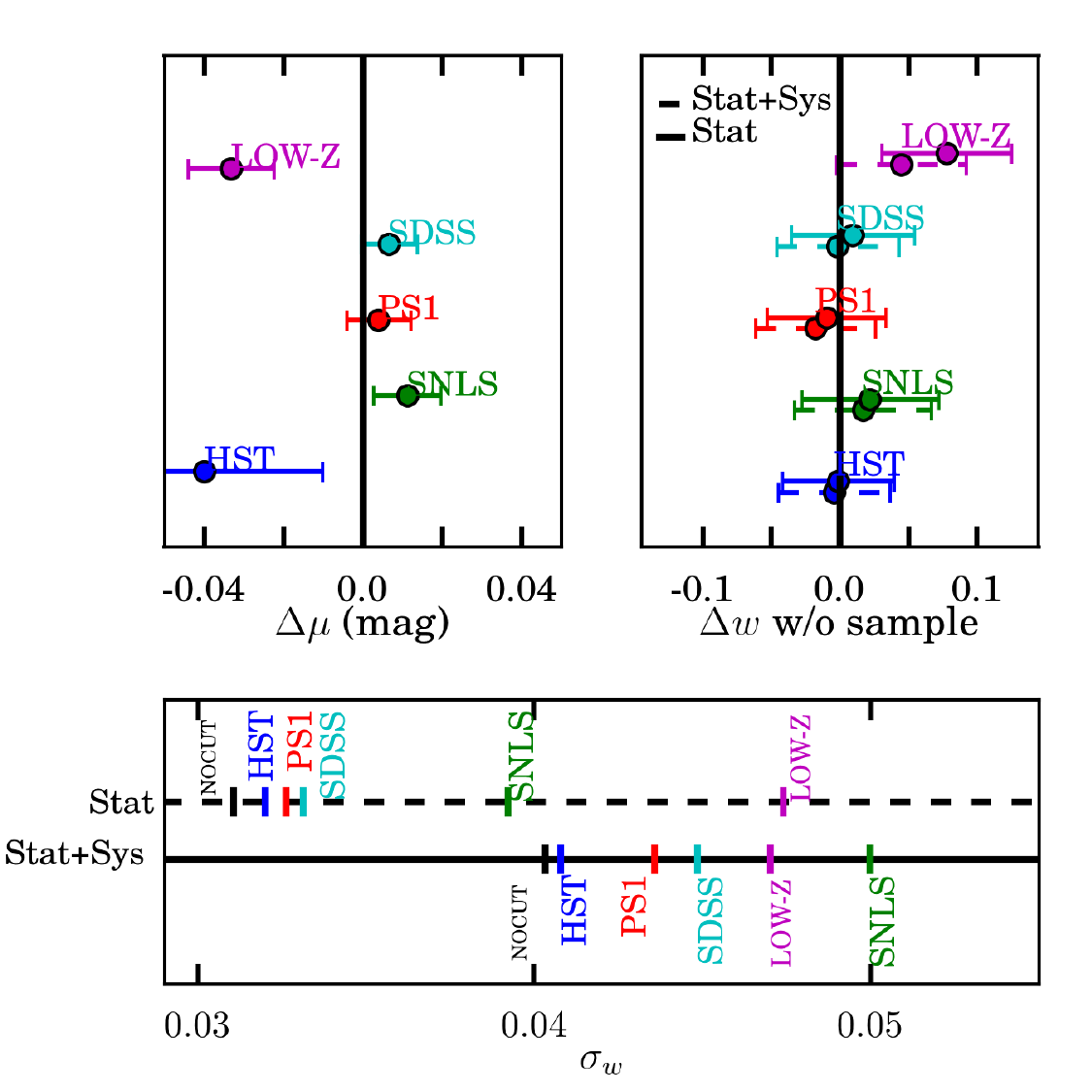}
\caption{(Top Left) Mean Hubble residuals for each sample relative to the cosmological fit for the baseline analysis. (Top Right) Impact on $w$ of removing a set from a sample; the change is expressed as the recovered value of $w$ after a sample is removed \textit{minus} the value of $w$ derived using the full Pantheon sample.  (Bottom) The uncertainty in $w$ after a sample is removed.  For both the middle and bottom panels, the impact of removing a sample is shown for both the statistical uncertainty only case and the statistical and systematic uncertainty case.}\label{fig:lcdm_resid}
\end{figure}

Furthermore, it is instructive to compare the relative pulls on the distances and recovered $w$ values from each subsample.  
As shown in Fig.~\ref{fig:lcdm_resid}, we find that the mean distance residual relative to the best-fit cosmology is $0.02$ mag or lower.  Upon removing any single subsample from the analysis, the Low-z has the largest impact on $w$ and causes a change of $\Delta w \sim0.07$ if only the statistical uncertainties are included.  When including systematic uncertainties, the Low-z sample causes a change of $\Delta w\sim0.04$.  The pulls of the other samples are all within $| \Delta w|=0.02$.  Finally, we also compare the impact on the uncertainty on $w$ when each subsample is removed for the both the statistical and statistical+systematic analyses.  The Low-z sample has the strongest impact on the statistical uncertainty, but not on the total uncertainty, because of its large systematics.  The SNLS has the strongest impact on the total uncertainty, likely because it is at high-redshift and has small systematic uncertainties.  We find that the small high-z sample from HST has a minimal impact on our measurement of $w$.  This is likely due to the sample size and this parameterization of dark energy which assumes $w(z)$ is constant; this is discussed in detail in Riess et al. (2017), which uses the Pantheon sample and varies the parameterization of dark energy.

The Low-z sample has an outsized impact on a number of the variants shown in Table~\ref{tab:subs2}.  Interestingly, we find that when applying the C11 scatter model with the BBC method, we recover a difference in $w$ of $0.065$ depending on whether we include SNe with $z<0.1$, but when we apply the G10 scatter model with the BBC method, we recover a difference in $w$ of $0.010$ depending on whether we include SNe with $z<0.1$.  This effect can be traced back to the $3\%$ offset in distance biases for the Low-z sample shown in Fig.~\ref{fig:oned_bias}.  This issue must be resolved in future analyses to continue using the Low-z sample.

\pagebreak
\clearpage

\subsection{Combining probes and understanding cosmological models}

\begin{table*}
\caption{}
  \centering
  \label{tab:cosmo_omol}
  \begin{tabular}{l|cccc}
\hline
Sample & $\Omega_m$ & $\Omega_\Lambda$ & $\Omega_K$ & $H_0$ \\
\hline
CMB+BAO & $ 0.310 \pm 0.008$ & $ 0.689 \pm 0.008 $& $ 0.001 \pm 0.003 $&$ 67.900\pm0.747$\\
CMB+H0 & $ 0.266 \pm 0.014$ & $ 0.723 \pm 0.012 $& $ 0.010 \pm 0.003 $&$ 73.205\pm1.788$\\
CMB+BAO+H0 & $ 0.303 \pm 0.007$ & $ 0.694 \pm 0.007 $& $ 0.003 \pm 0.002 $&$ 68.723\pm0.675$\\
SN+CMB & $ 0.299 \pm 0.024$ & $ 0.698 \pm 0.019 $& $ 0.003 \pm 0.006 $&$ 69.192\pm2.815$\\
SN+CMB+BAO & $ 0.309 \pm 0.007$ & $ 0.690 \pm 0.007 $& $ 0.001 \pm 0.002 $&$ 67.985\pm0.699$\\
SN+CMB+H0 & $ 0.274 \pm 0.012$ & $ 0.717 \pm 0.011 $& $ 0.009 \pm 0.003 $&$ 72.236\pm1.572$\\
SN+CMB+BAO+H0 & $ 0.303 \pm 0.007$ & $ 0.695 \pm 0.007 $& $ 0.003 \pm 0.002 $&$ 68.745\pm0.684$\\

  \end{tabular}\\
  \begin{flushleft}{Notes:  Cosmological constraints from different combinations of probes when assuming the~\OCDM~model.}\end{flushleft}
\end{table*}

To better determine cosmological parameters, we include constraints from measurements of the CMB from \cite{Planck15}, measurements of local value of $H_0$ from \citep{Riess16}, and measurements of baryon acoustic oscillations from the SDSS Main Galaxy Sample \citep{Ross15}, the Baryon Oscillation Spectroscopic Survey and CMASS survey \citep{Anderson13}.  These BAO measurements set the BAO scale at $z = 0.106, 0.35$, and 0.57.  For all CMB constraints, we include data from the \textit{Planck} temperature power spectrum and low-$\ell$ polarization (\textit{Planck} TT + lowP).

Before combining constraints from different probes, we can compare constraints on $\Omega_m$ when we assume the universe is flat, $w_0=-1$, and $w_a=0$.  Using our full SN sample with systematic uncertainties, with no external priors except flatness, we find $\Omega_m=\SNlOM$.  This is similar to the value determined from \cite{Planck15} of $0.315 \pm 0.013$ and the value from BAO of $0.310\pm0.005$ \citep{Alam17}.  Using only SNe, there is no constraint on $H_0$ since $H_0$ and $\mathcal{M}$ from Eq.~\ref{eqn:Tripp} are degenerate.  Constraints on $H_0$ from data that includes SN measurements only come indirectly from the SN component in that the SN measurements constrain parameters like $\Omega_m$ and $w$ which have covariance with $H_0$.  Since the low-z SNe in this sample and the one used in \cite{Riess16} are very similar, there may be some common systematics that affect both probes, though this is likely to be small as \cite{Riess16} compare SNe in the Hubble flow to SNe with $z<0.01$ whereas our analysis compares SNe in the Hubble flow to SNe with $z>0.1$.

\begin{figure}

\epsscale{1.15}  
\plotone{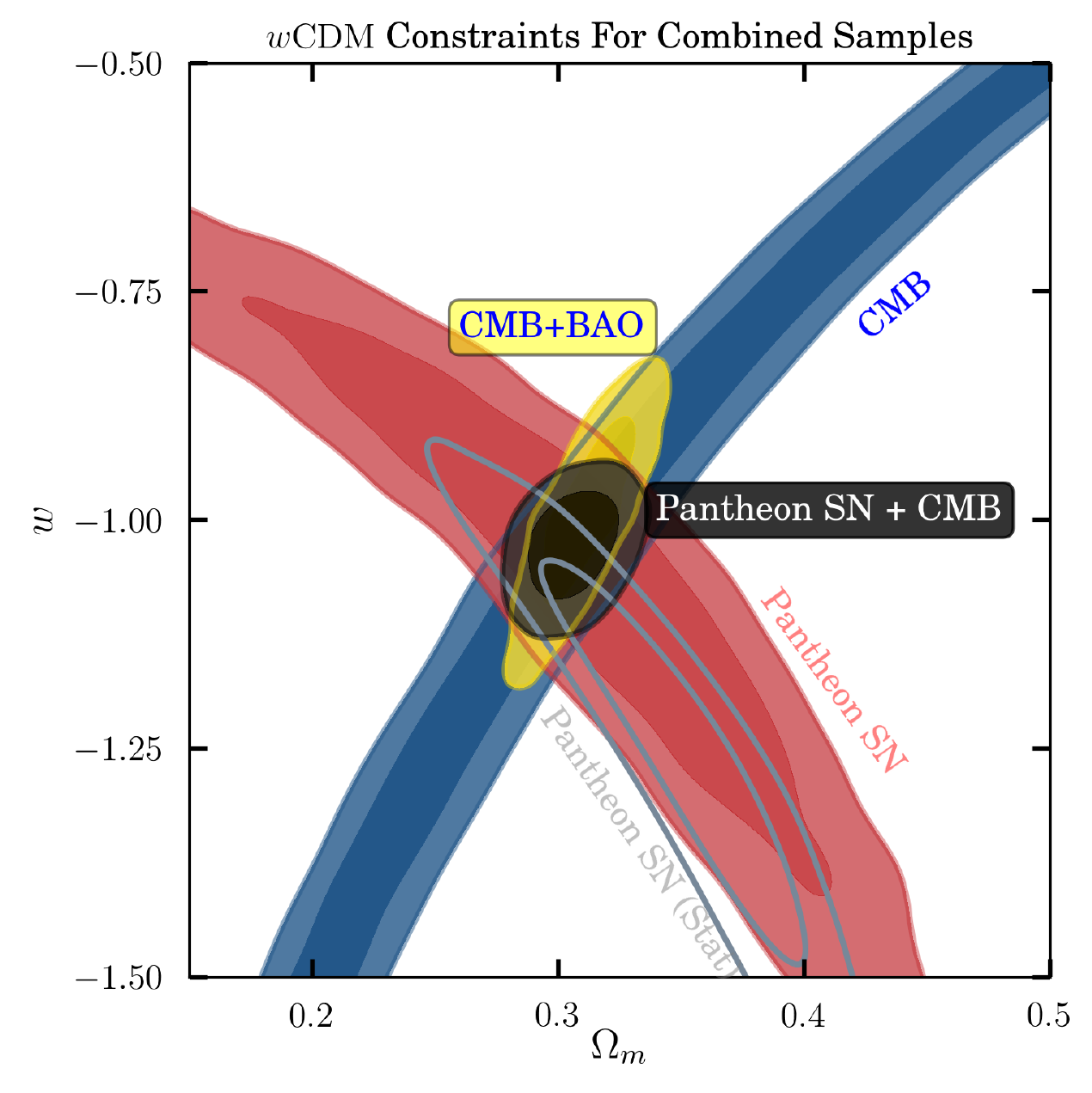}
\caption{Confidence contours at 68\% and 95\% for the $\Omega_m$ and $w$ cosmological parameters
for the~\WCDM~model.  Constraints from CMB (blue), SN - with systematic uncertainties (red), SN - with only statistical uncertainties (gray-line), and SN\plus CMB (black) are shown.}
\label{fig:plot_omw}
\end{figure}

\begin{table*}
\caption{}
  \centering
  \label{tab:cosmo_omw}
  \begin{tabular}{l|ccc}
\hline
Sample & $w$ & $\Omega_m$ & $H_0$ \\
\hline
CMB+BAO & $ -0.991 \pm 0.074$ & $ 0.312 \pm 0.013 $&$ 67.508\pm1.633$\\
CMB+H0 & $ -1.188 \pm 0.062$ & $ 0.265 \pm 0.013 $&$ 73.332\pm1.729$\\
CMB+BAO+H0 & $ -1.119 \pm 0.068$ & $ 0.289 \pm 0.011 $&$ 70.539\pm1.425$\\
SN+CMB & $ -1.026 \pm 0.041$ & $ 0.307 \pm 0.012 $&$ 68.183\pm1.114$\\
SN+CMB+BAO & $ -1.014 \pm 0.040$ & $ 0.307 \pm 0.008 $&$ 68.027\pm0.859$\\
SN+CMB+H0 & $ -1.056 \pm 0.038$ & $ 0.293 \pm 0.010 $&$ 69.618\pm0.969$\\
SN+CMB+BAO+H0 & $ -1.047 \pm 0.038$ & $ 0.299 \pm 0.007 $&$ 69.013\pm0.791$\\

  \end{tabular}\\
\begin{flushleft}{Notes: Cosmological constraints from different combinations of probes when assuming the~\WCDM~model.  The value of $w=-1$ corresponds to the cosmological constant hypothesis.}\end{flushleft}
\end{table*}

Relaxing the assumption of a cosmological constant, we measure $w$, the dark energy equation-of-state parameter.  For these~\WCDM~models, we assume a flat universe ($\Omega_{k} = 0$).  In Table~\ref{tab:cosmo_omw}, we
compare how the different cosmological probes impact the constraints
on $\Omega_m$ and $w$.  As shown in Figure~\ref{fig:plot_omw}, combining Planck and SN measurements, we find $\Omega_m=\SNCMBxOM$ and $w=\SNCMBxW$.  This is to date the tightest constraint on dark energy, and we find that it is consistent with the cosmological constant model.  These values are more precise than, though consistent with, the values from combining Planck and BAO measurements which are $\Omega_m=\CMBBAOxOM$ and $w=\CMBBAOxW$.  Combining SN, BAO, Planck and H0 measurements yield $\Omega_m=\SNCMBBAOHSTxOM$ and $w=\SNCMBBAOHSTxW$, similar to the results of just SN\plus Planck.  If we replace constraints from Planck with those from WMAP9 \citep{Hinshaw13}, we see a shift of $\Delta w\sim+0.04$ seen in past studies (e.g., B14 or R14) which does not change any of our conclusions.

\begin{figure}

\epsscale{1.15}  
\plotone{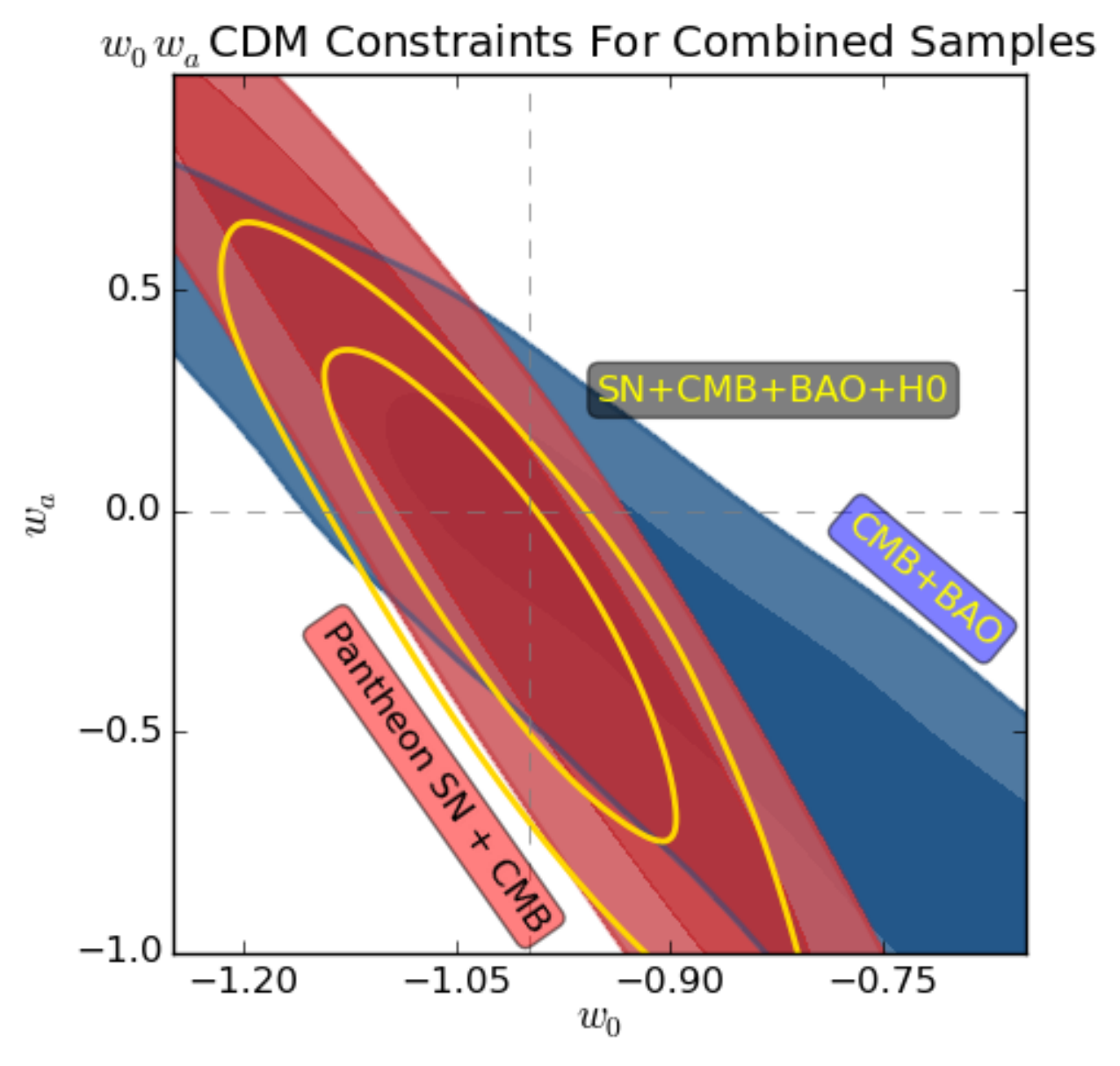}
\caption{Confidence contours at 68\% and 95\% (including systematic
uncertainty for SNe) for the $w$ and $w_a$ cosmological parameters
for the~\WACDM~model.  Constraints from BAO\plus CMB (blue), SN\plus CMB (red), SN\plus CMB\plus BAO (yellow) and SN\plus CMB\plus BAO\plus HST (yellow) are shown.  }
\label{fig:plot_wwa}
\end{figure}

In Table~\ref{tab:cosmo_wwa}, we
compare how the different cosmological probes impact the constraints
on $w_0$ and $w_a$.  We show in
Figure~\ref{fig:plot_wwa}, the constraints of various combinations of the different probes given the~\WACDM~model.   We find that combining SN, BAO, Planck and H0 measurements, $w_0=\SNCMBBAOHSTwWO$ and $w_a=\SNCMBBAOHSTwWA$.  These values are consistent with the cosmological constant model of dark energy such that $w_0$ is consistent with $-1$ and $w_a$ is consistent with 0, or no evolution of the equation-of-state of dark energy.  

\subsection{Comparison of Cosmological Results to R14 and B14}

Comparisons between the results from R14 and B14 with the results from this analysis are shown in Table \ref{tab:cosmo_comp}. R14 used a sample of 112 PS1 SNe and 180 Low-z SNe to measure cosmological parameters, and found for the~\WCDM~model a $\sim2\sigma$ deviation from $w=-1$ when combining SN and Planck measurements.  With a larger sample of PS1 SNe and an improved analysis, we find no hints of tension with a cosmological constant from the parameters derived for the PS1+Low-z sample.  

As can be seen in Table \ref{tab:cosmo_comp}, the statistical-only constraints from the improved PS1+Low-z sample are consistent with those from R14 and the constraints on $\Omega_m$ and $w$ are tighter.  However, accounting for systematic uncertainties cause the best-fit parameters of this analysis to diverge from R14.  One of the main reasons for this is that compared to the analysis of S14, the systematics of the PS1 sample are smaller but the systematics of the Low-z sample are larger, thereby effectively down-weighting the Low-z sample with respect to the PS1 sample.  

There are no large differences between the constraints from our full Pantheon sample and that from the B14 analysis.  The reason for this is shown in Fig.~\ref{fig:lcdm_resid} - even though our Low-z sample is much larger, our systematic uncertainties on the Low-z bias correction are also much larger.  Furthermore, the addition of the PS1 sample does not have much pull as it is consistent with SNLS and SDSS. This subsample also occupies a redshift range in between those the SNLS and SDSS subsamples.  Still, we note the 30\% decrease in total uncertainties from B14 and our analysis.

\begin{table*}
  \centering
  \caption{$w_0$ versus $w_a$}
  \label{tab:cosmo_wwa}
  \begin{tabular}{l|ccccc}
\hline
Sample & $w_0$ & $w_a$ & $\Omega_m$  & $H_0$ & FoM\\
\hline
CMB+BAO & $ -0.616 \pm 0.262$ & $ -1.108 \pm 0.771$ & $ 0.343 \pm 0.025 $&$ 64.614\pm2.447$& 14.5\\
CMB+H0 & $ -1.024 \pm 0.347$ & $ -0.789 \pm 1.338$ & $ 0.265 \pm 0.015 $&$ 73.397\pm1.961$& 9.1\\
CMB+BAO+H0 & $ -0.619 \pm 0.270$ & $ -1.098 \pm 0.781$ & $ 0.343 \pm 0.026 $&$ 64.666\pm2.526$& 14.5\\
SN+CMB & $ -1.009 \pm 0.159$ & $ -0.129 \pm 0.755$ & $ 0.308 \pm 0.018 $&$ 68.188\pm1.768$& 31.4\\
SN+CMB+BAO & $ -0.993 \pm 0.087$ & $ -0.126 \pm 0.384$ & $ 0.308 \pm 0.008 $&$ 68.076\pm0.858$& 65.0\\
SN+CMB+H0 & $ -0.905 \pm 0.101$ & $ -0.742 \pm 0.465$ & $ 0.287 \pm 0.011 $&$ 70.393\pm1.079$& 54.2\\
SN+CMB+BAO+H0 & $ -1.007 \pm 0.089$ & $ -0.222 \pm 0.407$ & $ 0.300 \pm 0.008 $&$ 69.057\pm0.796$& 63.2\\

  \end{tabular}
 \begin{flushleft}{Notes: Cosmological constraints from different combinations of probes when assuming the~\WACDM~model.  The FoM (Figure-of-Merit) is defined in \cite{Wang08}. }\end{flushleft}
\end{table*}

 \begin{table*}
\caption{}
  \centering
  \label{tab:cosmo_comp}
  \begin{tabular}{l|ccc}
\hline
Analysis & $w$ & $\Omega_m$ & Sample \\
\hline
Here &$-1.041\pm0.046$  & $0.304\pm0.014$  & PS1 [279] +Low-z [172] -- Stat Only \\
Here & $-0.990\pm0.063$  & $0.317\pm0.019$  & PS1 [279] +Low-z [172] -- Stat+Sys \\

R14/S15 & $-1.102\pm0.058$ &  $0.289\pm0.017$ & PS1 [112] + Low-z [180] -- Stat Only \\
R14/S15 & $-1.136\pm0.078$ &  $0.281\pm0.020$ & PS1 [112] + Low-z [180] -- Stat+Sys \\
\hline
Here & $-1.061\pm0.031$  & $0.301\pm0.009$ & Full Pantheon -- Stat Only \\
Here & $-1.026\pm0.041$  & $0.307\pm0.012$ & Full Pantheon -- Stat+Sys \\

B14 & $-1.018\pm0.057$ &  $0.307\pm0.017$ & SDSS [374] + SNLS [239]+Low-z [118]+ HST[9] -- Stat+Sys \\

  \end{tabular}\\
\begin{flushleft}{Notes: Cosmological constraints from combined Planck and SN data for the~\WCDM~model from different analyses and different samples.  For JLA, the statistical-only constraints were not given.}\end{flushleft}
\end{table*}

\section{Discussion}
\label{sec:Discussion}

Here we discuss specific areas of this analysis that require further analysis or future study.

\subsection{Low-z Samples}
Each aspect of the analysis from R14 and S14 has been improved for the present analysis, though we find here that the Low-z sample must be better modeled in order to realize the significant gains from the larger statistics and smaller systematics in the high-z SN samples. 
Since there are $\sim180$ Low-z SNe each with a distance modulus precision of $0.15$ mag, the standard error on the sample is $0.011$ mag.  Therefore, systematics that affect the Low-z sample relative to the high-z sample at the $1\%$~will significantly diminish the impact of the Low-z sample.  There are a series of systematics on this level that affect the Low-z sample more than other samples: intrinsic scatter, selection, MW extinction and calibration.  The impact is higher for the Low-z sample because the Low-z sample has redder SNe on average (by 0.03) than each of the higher-z samples, the MW extinction at the location of the SNe is higher (by 0.05) on average than in the higher-z samples, the selection effects are more difficult to model because there is uncertainty in whether the selection was volume or magnitude-limited, and the calibration uncertainties are $2\times$ as large as the high-z samples.  

While there are some Low-z data samples not included here (e.g., \citealp{Ganeshalingam13}), other low-z samples face similar issues and will likely not improve the cosmological constraint without improving the systematic uncertainties. The most helpful Low-z sample would be one which was based off a rolling survey so that selection effects are well understood and the color distribution is similar to that of the high-z samples, and one in which the calibration of the sample is on the level of the high-z samples.  This can be expected from the Foundation SN sample \citep{Foley17}, which uses the PS1 telescope to follow-up SNe discovered by rolling surveys.  Other possible low-z samples based on rolling surveys, like ATLAS \citep{TonryAtlas}, may further help this issue.

\subsection{Comparison of BBC method with older methods}

There's a fundamental difference in the approach of applying bias corrections between this analysis and that of B14 and S14.  Both B14 and S14 use a redshift-dependent distance bias correction as shown in Fig.~\ref{fig:oned_bias}, though both these analyses use inaccurate underlying $c$ and $x_1$ populations for their simulations (see SK16 for a review).  KS17 showed that with very large statistics and the same underlying populations, only very slight mmag-level differences are expected between a redshift-dependent distance bias correction and the more complex BBC method if $\alpha$ and $\beta$ are known a priori.  However, using incorrect $\alpha$ and $\beta$ introduces small biases in the method of B14 and S14 because $\alpha$ and $\beta$ are not solved simultaneously when measuring the distance biases.  

Both B14 and S14 consider the G10 and C11 scatter models, though while S14 and our analysis average the two, B14 chooses the G10 scatter model for its baseline analysis.  We do not choose one model or the other as there is insufficient empirical evidence to favor either model.  Somewhat implicit in the choice of scatter model is the assumption of a single $\sigma_{int}$ value.  Both B14 and S14 determine separate $\sigma_{int}$ values for the high-z and low-z sample, but we find a single value characterizes the full sample.  This is likely due to our higher value of the peculiar velocity uncertainty used in our analysis (S14 and B14 used 150 km/s, we use 250 km/s) and that the BBC method corrects for the overestimation of the fit parameter errors which has led to an incorrect assessment of $\sigma_{int}$ in past analyses \citep{Kessler16}.  
 
All of these analyses are still limited in measurements of the evolution of standardization parameters.  In this analysis, we find a $1\sigma$ signal for evolution of the $\beta$ parameter.  Interestingly, the analysis of \cite{Jones17} finds $\beta$ evolution of $-1.28\pm0.49$ with a sample $3\times$ larger (though with its own systematic uncertainties from contamination).  PS1 is not ideal for determining this evolution because its maximum redshift is $\sim0.6$.   Additionally, we find $\sim2\sigma$ evolution of the $\gamma$ parameter.  If the BBC method is not applied, we recover a measured slope of $\ngammanumzd$, which is a $\sim1.5\sigma$ effect.  B14 saw no evidence of evolution. 

New releases from SNLS and DES should help settle this question of parameter evolution.  High-z SNe observed by HST should provide excellent leverage to determine evolution of the nuisance parameters.  However, there is currently not enough data at high-z to provide tight constraints.  There is a similar issue in trying to use HST SNe for constraining $w_0-w_a$ \citep{Riess17}.  However, WFIRST  \citep{Hounsell17,Spergel15} should provide sub-percent level distance constraints of SNe at $z\sim1.5$ and significantly improve constraints on both evolution systematics and dark energy models.

\subsection{Further Examples of Population Drift}

There could be further evolution in the mean of the color variation of the intrinsic scatter model; for both C11 and G10 scatter models, the color scatter is centered around $c=0$ and this is assumed to not change with redshift.  Various analyses \citep{Foley11,Mandel14} have hypothesized evolution of the mean of the color scatter may be possible, however it is unclear with what significance and how well current data already constrain it.  It is not included as a systematic here, but studies like the one by \cite{Mandel16} may be able to isolate the effect so that we can put it into our simulations.  

A related evolution uncertainty is due to a possible bimodal population of SNe when considering their UV flux \citep{Milne15}.  If the relative fractions of the different UV subclasses changes with redshift, it would propagate to systematic biases in the recovery of SN color with redshift which would itself propagate to errors in the recovered cosmology.  \cite{Cinabro16} simulated simplistic models of different UV subclasses of SNe inferred from \cite{Milne15} and compared the output light-curves real SDSS and SNLS samples and did not see consistency.  We did not include it as a systematic uncertainty, but more UV data would serve the double purpose of clarifying this issue and helping with SALT2 training.

\section{Conclusion}
\label{sec:Conclusion}

We have presented a cosmological analysis of \numPS~spectroscopically confirmed Type Ia Supernovae ($0.03 < z <0.65$) discovered by the \PS\ Medium Deep Survey.  Combined with the set of cosmologically useful SN\,Ia from SDSS, SNLS, Low-z and HST samples, this is the largest combined sample of SN\,Ia.   This analysis uses the PS1 Supercal process, which determines a global calibration solution to combine 13 different SN samples.  Furthermore, it corrects for expected biases in light-curve fit parameters and their errors using the BBC Method.  We find that these improvements have substantially reduced the systematic uncertainties related to photometric calibration, which have long dominated the systematic error budget.  Those calibration uncertainties are now similar in magnitude to uncertainties related to the underlying physics of the SN population, such as the intrinsic scatter of SN\,Ia distances and the possible evolution of the correlation between SN color and luminosity.  The systematic uncertainties on our measurements of dark energy parameters are now smaller than the statistical uncertainties.  The cosmological fit to \numTOT \ SN\,Ia using SNe combined with constraints from Planck CMB measurements gives $\Omega_m=\SNCMBxOM$
and $w=\SNCMBxW$.  When the SN and Planck CMB constraints are combined with constraints from BAO and local $H_0$ measurements, the analysis yields $w_0=\SNCMBBAOHSTwWO$ and
$w_a=\SNCMBBAOHSTwWA$ including all identified systematics. Tension with a cosmological constant model, previously seen in an analysis of PS1 and low-z SNe, is not seen here.   This analysis presents the most precise measurements of dark energy to date and we find no hint of tension with the current \LCDM~model.  As there is still no plausible theoretical explanation of this model, observations should continue to probe this outstanding mystery.
\clearpage

\facility{PS1 (GPC1), Gemini:South (GMOS), Gemini:North (GMOS), MMT (Blue Channel spectrograph), MMT (Hectospec), Magellan:Baade (IMACS), Magellan:Clay (LDSS3), APO (DIS)}.

\acknowledgments

The Pan-STARRS1 Surveys have been made possible through
contributions of the Institute for Astronomy, the University of
Hawaii, the Pan-STARRS Project Office, the Max-Planck Society and its
participating institutes, the Max Planck Institute for Astronomy,
Heidelberg and the Max Planck Institute for Extraterrestrial Physics,
Garching, The Johns Hopkins University, Durham University, the
University of Edinburgh, Queen's University Belfast, the
Harvard-Smithsonian Center for Astrophysics, the Las Cumbres
Observatory Global Telescope Network Incorporated, the National
Central University of Taiwan, the Space Telescope Science Institute,
the National Aeronautics and Space Administration under Grant
No. NNX08AR22G issued through the Planetary Science Division of the
NASA Science Mission Directorate, the National Science Foundation
under Grant No. AST-1238877, the University of Maryland, and Eotvos
Lorand University (ELTE).
Some observations reported here were obtained at the MMT Observatory, a joint facility of the Smithsonian Institution and the University of Arizona. 
Based on observations obtained at the Gemini Observatory, which is operated by the 
Association of Universities for Research in Astronomy, Inc., under a cooperative agreement 
with the NSF on behalf of the Gemini partnership: the National Science Foundation 
(United States), the National Research Council (Canada), CONICYT (Chile), the Australian 
Research Council (Australia), Minist\'{e}rio da Ci\^{e}ncia, Tecnologia e Inova\c{c}\~{a}o 
(Brazil) and Ministerio de Ciencia, Tecnolog\'{i}a e Innovaci\'{o}n Productiva (Argentina).
This paper includes data gathered with the 6.5-m Magellan Telescopes located at Las Campanas Observatory, Chile. 
Based on observations obtained with the Apache Point Observatory 3.5-meter telescope, which is owned and operated by the Astrophysical Research Consortium.
CWS and GN thank the DOE Office of Science for their support under grant ER41843.
Partial support for this work was provided by National Science
Foundation grant AST-1009749.
The ESSENCE/SuperMACHO data reduction pipeline {\it photpipe}
was developed with support from National Science
Foundation grant AST-0507574, and {\it HST} programs GO-10583 and GO-10903.
RPKs supernova research is supported in part by NSF Grant AST-1211196 and {\it HST} program GO-13046.
Some of the computations in this paper were run on the Odyssey cluster
supported by the FAS Science Division Research Computing Group at
Harvard University.
Much of the analysis was done using the Midway-RCC computing cluster at University of Chicago. 
This research has made use of the CfA Supernova Archive, which is
funded in part by the National Science Foundation through grant AST
0907903.
This research has made use of NASA's Astrophysics Data System.
This work was generated as part of NASA WFIRST Preparatory Science program 14-WPS14-0048 and is supported in part by the U.S. Department of Energy under Contract DE-AC02-76CH03000.
This work was supported in part by the Kavli Institute
for Cosmological Physics at the University of Chicago
through grant NSF PHY-1125897 and an endowment
from the Kavli Foundation and its founder Fred Kavli.
We gratefully acknowledge support from NASA grant
14-WPS14-0048. 
D.S. is supported by NASA through
Hubble Fellowship grant HST-HF2-51383.001 awarded
by the Space Telescope Science Institute, which is operated
by the Association of Universities for Research
in Astronomy, Inc., for NASA, under contract NAS 5-
26555.  D.O.J. is supported by the Gordon and Betty Moore Foundation
postdoctoral fellowship at the University of California, Santa Cruz.  R.C. thanks the Kavli Institute for Theoretical Physics for its hospitality while this work was in the final stages of preparation. This research was supported in part by the National Science Foundation under grant no. NSF PHY11-25915.  R.K. is supported by DOE grant DEAC02-76CH03000.
The computations in this paper used a combination 
of three computing clusters.  The bulk of the final analysis was performed 
using the University of Chicago Research Computing Center 
and the earlier analysis was done at the Odyssey cluster at Harvard University.   The 
Odyssey cluster is supported by the FAS Division of Science, Research 
Computing Group at Harvard University.  Supernova light curve 
reprocessing would not have been possible without the Data-Scope project 
at the Institute for Data Intensive Engineering and Science at 
Johns Hopkins University.
\bibliographystyle{apj}

\clearpage

\appendix
\section{Data Tables and Code Repository}
Upon publication, we will release \dataset[10.17909/T95Q4X]{http://dx.DOI.org/10.17909/T95Q4X} a suite of data files, coding routines and supplementary tables to replicate this analysis.  This includes:
\begin{itemize}
\item A table of the spectroscopic observations of each SN in the PS1 sample that includes their ID, date of observation, telescope observed and measured redshift.  A shortened version is included below.
\item A table of key recovered parameters from the light-curve fits for the full Pantheon sample.  A shortened version of this is shown below in Table \ref{tab:Ancillary}.  We also include a full output table from the SNANA fitter of a thorough listing of fitted parameters and other properties of the light-curves.  Final redshifts and distances are also given - a shortened version is shown in Table \ref{tab:final}.
\item A table of binned distance estimates over redshift for a compressed version of the dataset
\item A full systematic covariance matrix for the binned and unbinned versions.
\item Stellar catalogs of the MD fields.
\item Necessary files to use with the CosmoMC or CosmoSIS software with instructions.
\item A folder of all the SNANA set-up scripts to fit each sample.  A folder of all the SNANA set-up scripts to simulate each sample.
\item Output tables for 30 simulated samples used to test external methods on perform null tests on this dataset.
\item Code for remaking all figures in this paper.

\end{itemize}

\begin{table}[ht]
\caption{}
\centering
\begin{tabular}{cccc}
\hline \hline
PS1-ID & Spec. Date & Telescope  & z-helio  \\
\hline
PS1 110716 & 55570 & MMT & 0.315(0.001) \\
PS1 110721 & 55570 & MMT & 0.56(0.01) \\
PS1 110734 & 55570 & MMT & 0.401(0.001) \\
PS1 120085 & 55570 & MMT & 0.32(0.01) \\
PS1 120143 & 55571 & MMT & 0.173(0.001) \\
PS1 120225 & 55571 & MMT & 0.106(0.001) \\
PS1 120243 & 55570 & MMT & 0.34(0.01) \\
PS1 130150 & 55614 & MMT & 0.21(0.01) \\
PS1 130283 & 55614 & MMT & 0.076(0.001) \\
PS1 130308 & 55614 & MMT & 0.081(0.001) \\
PS1 130755 & 55615 & MMT & 0.292(0.001) \\
PS1 130862 & 55615 & MMT & 0.332(0.001) \\
PS1 130943 & 55615 & MMT & 0.301(0.001) \\
PS1 130945 & 55615 & MMT & 0.266(0.001) \\
PS1 140152 & 55687 & MMT & 0.208(0.001) \\
\hline
\end{tabular}
\begin{flushleft}{Notes:  Spectroscopic information for all spectroscopically classified
Pan-STARRS1 SN Ia from 2011 June to 2014 September.  Redshifts are given in the heliocentric frame.  A redshift uncertainty of $0.001$ means that the redshift is acquired from the host.  A redshift uncertainty of $0.01$ means the redshift is acquired from the SN itself.  A full version of this table can be found \dataset[10.17909/T95Q4X]{http://dx.DOI.org/10.17909/T95Q4X}.}\end{flushleft}
 \label{tab:redshifts}
\end{table}

\begin{table}[ht]
\caption{}
\label{tab:Ancillary}
\centering
\begin{tabular}{lcccccccc}
\hline \hline
SN & Subsample & $z$  & $m_b$ & $x_1$ & $c$ & $\mu_{Corr} $ & Mass \\
\hline
170428  & PS1 & 0.3001 & $21.81\pm0.04$ & $-0.99\pm0.23$ & $-0.01\pm0.03$ & $-0.07\pm0.00$ & $9.11\pm0.10$\\
180166  & PS1 & 0.1476 & $19.90\pm0.03$ & $0.60\pm0.11$ & $0.04\pm0.03$ & $-0.02\pm0.00$ & $10.88\pm0.08$\\
180561  & PS1 & 0.2288 & $21.43\pm0.04$ & $0.01\pm0.21$ & $0.14\pm0.03$ & $-0.04\pm0.01$ & $8.44\pm0.72$\\
190230  & PS1 & 0.1388 & $19.81\pm0.04$ & $-1.41\pm0.14$ & $-0.07\pm0.03$ & $-0.01\pm0.01$ & $10.87\pm0.03$\\
190260  & PS1 & 0.1436 & $19.58\pm0.06$ & $0.98\pm0.12$ & $-0.04\pm0.03$ & $-0.01\pm0.00$ & $10.83\pm0.04$\\
300105  & PS1 & 0.0919 & $19.17\pm0.04$ & $0.33\pm0.09$ & $0.08\pm0.03$ & $-0.02\pm0.00$ & $9.87\pm0.06$\\
310025  & PS1 & 0.1568 & $19.72\pm0.05$ & $1.85\pm0.25$ & $-0.07\pm0.03$ & $0.05\pm0.01$ & $10.09\pm0.16$\\
310042  & PS1 & 0.2388 & $20.88\pm0.21$ & $1.28\pm0.61$ & $-0.05\pm0.14$ & $0.03\pm0.01$ & $11.64\pm6.78$\\
310073  & PS1 & 0.1496 & $19.69\pm0.05$ & $0.38\pm0.23$ & $-0.19\pm0.03$ & $0.09\pm0.01$ & $9.31\pm0.10$\\
310091  & PS1 & 0.5078 & $22.73\pm0.05$ & $-0.25\pm0.47$ & $-0.09\pm0.04$ & $-0.08\pm0.01$ & $9.07\pm1.30$\\
310161  & PS1 & 0.2528 & $21.13\pm0.04$ & $0.13\pm0.18$ & $-0.03\pm0.03$ & $-0.01\pm0.00$ & $9.58\pm0.35$\\
310238  & PS1 & 0.2842 & $21.17\pm0.03$ & $-0.29\pm0.17$ & $-0.07\pm0.03$ & $-0.01\pm0.00$ & $9.91\pm0.17$\\
310574  & PS1 & 0.2368 & $21.00\pm0.04$ & $-0.38\pm0.16$ & $-0.02\pm0.03$ & $-0.02\pm0.00$ & $10.23\pm0.14$\\
320258  & PS1 & 0.3412 & $21.88\pm0.05$ & $-1.86\pm0.39$ & $-0.06\pm0.04$ & $-0.11\pm0.01$ & $10.96\pm0.04$\\
330022  & PS1 & 0.2641 & $21.24\pm0.06$ & $-1.25\pm0.28$ & $0.02\pm0.04$ & $-0.06\pm0.01$ & $10.98\pm0.09$\\

\hline

\end{tabular}
\begin{flushleft}{Notes:  SN ID, subsample, fit parameters, mass and distance corrections of SNe after cuts.   Full versions of this table can be found \dataset[10.17909/T95Q4X]{http://dx.DOI.org/10.17909/T95Q4X} when using either the G10 and C11 scatter model as well ancillary information including covariance between fit parameters, bias correction information, RA, DEC and further material.}\end{flushleft}
 \label{tab:fitparam}
\end{table}

\begin{table}[ht]
\caption{}
\label{tab:final}
\centering
\begin{tabular}{lcc}
\hline \hline
SN & $z$  & $\mu+$M$ $ \\
\hline

170428  & 0.30012 & $21.71\pm0.12$\\
180166  & 0.14761 & $19.89\pm0.11$\\
180561  & 0.22853 & $20.97\pm0.13$\\
190230  & 0.1388 & $19.82\pm0.11$\\
190260  & 0.14343 & $19.88\pm0.11$\\
300105  & 0.09201 & $18.95\pm0.11$\\
310025  & 0.1568 & $20.12\pm0.13$\\
310042  & 0.23851 & $21.20\pm0.25$\\
310073  & 0.14949 & $20.15\pm0.17$\\
310091  & 0.50718 & $22.99\pm0.14$\\
310161  & 0.25249 & $21.21\pm0.11$\\
310238  & 0.28397 & $21.32\pm0.11$\\
310574  & 0.2368 & $21.02\pm0.12$\\
320258  & 0.34092 & $21.89\pm0.15$\\
330022  & 0.26388 & $21.06\pm0.14$\\

\hline

\end{tabular}
\begin{flushleft}{Notes:  Final redshifts and corrected magnitudes used to measure cosmological parameters. Since the absolute magnitude of an SNIa is degenerate with $H_0$, only the corrected magnitudes are given here.  A full version of this table can be found \dataset[10.17909/T95Q4X]{http://dx.DOI.org/10.17909/T95Q4X}.}\end{flushleft}
 \label{tab:fitparam}
\end{table}

\clearpage

\section{Template construction}
\label{subsubsec:Template}

In order to separate the SN flux from the SN host galaxy, R14 creates a template from stacking multiple images where there is no SN light, convolving template to match image with SN light, and then subtracting the convolved template from the nightly image where there is SN light.  In Photpipe, the seasonal template is constructed by combining nightly stacks weighted by the product of the inverse variance and the inverse area of the PSF.  To better understand the systematic uncertainties in the photometry, we implement an independent photometry routine that constructs light-curves using a `scene modeling' algorithm based on the method presented in \cite{Holtzman08} (hereafter H08).  The purpose of the scene modeling for this analysis is to determine photometry of the SNe on the nightly images without stacking multiple images for the template and without the need to spatially resample a template image.  The process also has many similarities to the method first presented in \cite{Astier06}, with recent updates in \cite{Astier14}.  H08 explains that the largest benefits to the scene modeling approach over the conventional template construction (e.g. as in Photpipe) are when the depth or PSF size of the template images is less than or equal to the depth or seeing of the SN images.  This is not a major issue for the PS1 analysis because the deep seasonal templates can be degraded or resampled to the depth and resolution of the SN images without introducing much correlated noise. However, a secondary photometry pipeline provides an independent crosscheck of the accuracy of the photometry.

\begin{figure*}

\epsscale{1.15}  
\plotone{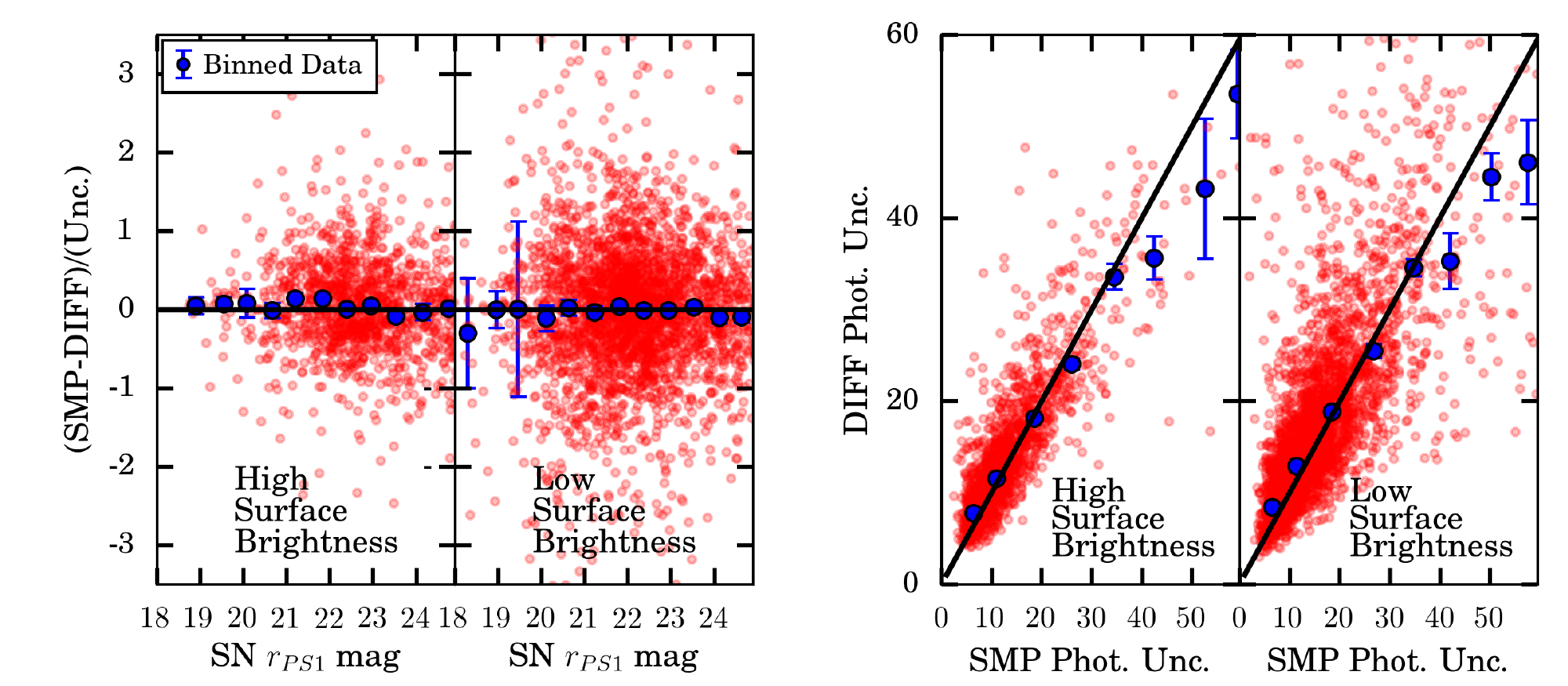}
\caption{(Left) The difference in recovered SN magnitudes using the scene modeling pipeline (SMP) versus the nominal difference imaging approach from Photpipe.  The difference is divided by the uncertainty from the difference imaging approach, rather than the combined uncertainties from both approaches.  Red points show individual photometry observations and blue points are binned estimates. (Right) The correlation of the uncertainties of recovered photometry between the two approaches.  A line set such that the uncertainties are \textit{equal} is drawn to show agreement between the uncertainties. For each grouping of panels, the left and right panels show the correlation for SNe located near high, local surface brightness of the host galaxy ($m_r<21.0$ mag/arcsec$^2$) and low, local, surface brightness  ($m_r>21.0$ magarcsec$^2$) respectively.} 
\label{fig:smp_comp_flux}
\end{figure*}

The process of the scene modeling algorithm is to create a pixel-based map of a temporally-constant galaxy and a temporally-varying SN.  While the galaxy is modeled as a grid of pixels that each have an independent brightness value, each SN is modeled as a point source with variable brightness. Being able to empirically determine a galaxy model and use the nightly PSF to project the model onto the nightly image allows for a robust assessment of the photometric uncertainties (H08).

The general formalization is to fit a completely unparameterized galaxy model and a temporally varying SN brightness to the observed data. In a given filter, the flux is modeled at each pixel with coordinates $(x,y)$ such that,
\begin{eqnarray}
M(x,y)  =  \textrm{sky} + S [ I_{SN}PSF(x-x_{SN},y-y_{SN}) +   \nonumber \\ 
 \sum_{x_g,y_g} {G}(x_g,y_g) PSF(x-x_g,y-y_g) ]~~~.
\end{eqnarray}
Here, $M(x,y)$ is the total model intensity (DN) at each pixel, 
$I_{SN}$ is the unknown total calibrated SN intensity, 
$PSF(\Delta X,\Delta Y)$ is the measured fraction of
light from a star as a function of the distance of each pixel from the
central position, ${G}(x_g,y_g)$ represents the unknown grid of galaxy 
intensities,
and $\textrm sky$ is the measured background value at each pixel.  
The position ($x_{SN},y_{SN}$) is the pixel coordinates of the SN that have been already astrometrically aligned as part of the main pipeline.  $S$ is the scaling factor so that a galaxy with non-varying flux will have the same total magnitude for each image.  This is set for each image by the nightly zeropoint.

The fits are weighted by the expected errors from photon statistics
and readout noise.  Since the gain for the PS1 images is unity and read-out noise is negligible, a minimization is done for:
\begin{equation}
  \chi^2 = \sum_{xy} {(O(x,y) - M(x,y))^2 \over (M(x,y) )}
\end{equation}
where $O(x,y)$ is the observed value at each pixel.  

While the scene modeling technique used here primarily follows that of H08, we alter their procedure to better isolate certain systematic uncertainties and to better incorporate the procedure in the PS1 photometry pipeline.  Specifically, a WCS solution is determined for each image from the main pipeline, and the PSF and sky value near the SN that were determined from the main pipeline are used in the scene modeling process.  For consistency, the zeropoint of the nightly image is redetermined using the same minimization technique (Levenberg-Marquardt) as what is used in the scene-modeling minimization.  

As only flux from or nearby the SN is important for this analysis, a small $45 \times 45$ pixel image subsection is extracted around the position of the SN in every frame.  The sub-image size is chosen to be larger than the largest host galaxy (10 arcsec across) in our sample.  Furthermore, unlike in H08 which groups pixels into $2\times2$ bins, each pixel is independent.  All observations more than 90 days before peak or 270 days after peak are constrained to to have zero SN flux in the fit.  The SN peak is estimated from the search photometry.  A single SN position is fit to the entire stack and the galaxy model is initialized as a point source.

A comparison between the scene modeling photometry and difference imaging photometry is shown in Fig.~\ref{fig:smp_comp_flux}.  The comparison is separated for SNe located near  high surface brightness regions of the host galaxy ($m_r<21.0$ mag/arcsec$^2$) and low surface brightness regions of the host galaxy ($m_r>21.0$ mag/arcsec$^2$).  The surface brightness of the host galaxy is measured in a fixed 1 square-arcsecond aperture. The agreement in the final photometry is better than $0.2\sigma$ for SNe on top of areas with high underlying surface brightness  and better than $0.1\sigma$ for SNe on top of areas with low underlying surface brightness.   On average, absolute difference in the photometry between the two approaches for SNe in high local surface brightness is $2.0\pm0.5$ mmag and the difference for low local surface brightness is $0.4\pm0.2$ mmag.  These differences are subdominant in the uncertainty budget summarized in Table 1.

The treatment of the photometric uncertainties is checked by comparing the recovered uncertainties from the scene modeling photometry and the difference imaging.  As shown in Fig.~\ref{fig:smp_comp_flux}, there is excellent agreement (a Pearson correlation coefficient of $0.997$) between the uncertainties from the two methods.  Recent analyses of both PS1 photometry (R14, \citealp{Jones16}) and DES photometry \citep{Kessler15} show that photometric uncertainties are underestimated when SNe are located on top of areas of high surface brightness.  For PS1, \cite{Jones16} shows that photometric uncertainties near bright local galaxy flux of $20$ mag may be underestimated by a factor of $2.5$.  While it was possible that the scene modeling approach could remove the dependence of the underestimation of the SN uncertainties based on the underlying galaxy brightness, Fig.~\ref{fig:smp_comp_flux} shows this is not the case. Instead, both the scene modeling and template algorithms recover similar errors for areas with both high and low surface brightness.  To account for the dependence of the photometric uncertainties on the local surface brightness, we increase, by addition in quadrature, the photometric uncertainties of observations of an individual SN observed in one passband such that the reduced $\chi^2$ of the photometry of the SN pre- and post- explosion epochs is unity.  This same process is done in R14.  For the simulations discussed in section 3, this dependence of photometric uncertainties on host galaxy properties is included.

\section{Low-z Simulations}

\begin{figure}
\epsscale{1.15}  
\plotone{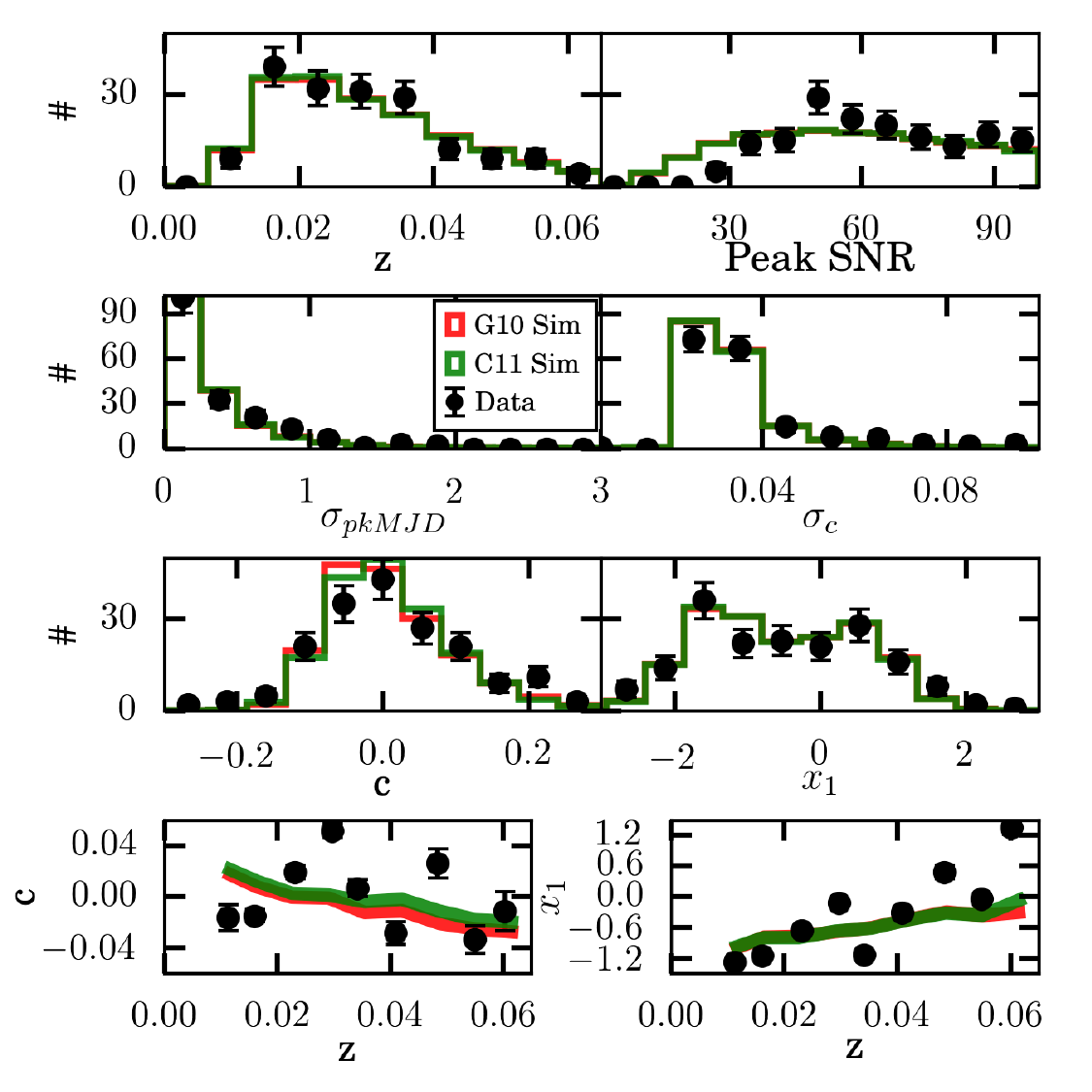}
\caption{Same as Fig.~\ref{fig:bananahistps} but for the combined Low-z sample.}
\label{fig:snana_hist_lowz}
\end{figure}

The Low-z sample is a compilation of subsamples from different samples.  Here we attempt to model the selection effects that went into the following samples: CfA1$\&$CfA2, CfA3, CfA4 and CSP\footnote{CSP SN photometry had unrealistically small (down to 0) photometric uncertainties, so we added an error floor of $0.01$ mag consistent with other surveys.}.  We also must determine the combined discovery and follow-up efficiency for each survey.  For these different samples, the survey that discovered the SNe was almost always not the survey that acquired a light-curve for the samples above.  The best way forward would therefore attempt to separate the Low-z sample according to the actual discovery surveys so that you could apply different selection functions to the different surveys.  This approach would reveal various selection biases beyond just volume-limited versus magnitude-limited, like selection based on host-galaxy type \citep{Leaman11}.  However, modeling Low-z sample from discovery to follow-up is too challenging for this analysis.

Instead, similar to the modeling of the PS1 survey described in Section 3, we determine selection effects from the data itself.   In S14 and B14, a single selection function was determined for the full low-z sample.  However, we found that DOIng so did not produce high-quality matches between the data and monte-carlo (MC) when analyzing the SNR and distance uncertainties.  We determine the combined efficiency for each subsample by comparing simulations without any efficiency cut to the data.  We fit a one-sided Gaussian to describe selection efficiency as a function of peak-$B$ photometric magnitudes such that\\
\begin{equation}
\begin{split}
{\rm Eff}=h_2 &~\textrm{for}~B < h_0 \\
{\rm Eff}=h_2 & ~e^{ [-({B} - h_0)^2/2h_1^2] }~\textrm{for}~B > h_0 . \\
\end{split}
\end{equation}
The $h_0,h_1,h_2$ values for each survey are given in Table \ref{tab:lowz_search}.  The comparison between data and MC for the combined Low-z sample is shown in Fig.~\ref{fig:snana_hist_lowz}.  We could choose to model the surveys based on the brightness in a different passband, but we find $B$ is adequate for these purposes.

Furthermore, we find that to optimize the match of Low-z MC to the data, we must change the functional form that expresses the underlying stretch population.  As discussed for Table~\ref{tab:ps1pop}, we define the population by an asymmetric gaussian.  However, in Fig.~\ref{fig:snana_hist_lowz}, the $x_1$ distribution is bimodal.  Therefore, we express the population as the combination of two asymmetric gaussians.  Using the same process as in SK16, the $x_1$ distributions for the Low-z sample are found to be $\bar{x}=0.703,\sigMINUS=1.0,\sigPLUS=0.47$ for the first mode and $\bar{x}=-1.5,\sigMINUS=1.0,\sigPLUS=0.47$ for the second mode.

\begin{table}[h]
  \centering
  \caption{Efficiency of Low-z Samples}
  \label{tab:lowz_search}
  \begin{tabular}{l|cc}
\hline
Surv &  Efficiency ($h_0,h_1,h_2$) & Sys-Efficiency ($h_0,h_1,h_2$)\\
\hline
CfA3 & 15.1,1.1,0.1 & 14.7, 1.3, 0.1 \\
CfA4 & 13.6, 1.45, 0.3 & 11.2, 2.05, 0.7 \\
CSP & 10.9,2.1,0.9 & 12.5,2.0,0.2 \\
CfA1\&CfA2 & 12.5,1.55,0.9 & 11.9, 1.65, 0.8 \\ 
  \end{tabular}\\
\begin{flushleft}{Notes: The efficiencies of the low-z surveys.  These efficiencies combine the detection and spectroscopic selection efficiencies.  The parameters $h_0,h_1,h_2$ describe the parameters of Appendix C Eq. 1.  The Sys-Efficiency is the $1\sigma$ 3-dimensional shift in best fit parameters.}\end{flushleft}
\end{table}

Finally, the alternative to a magnitude-limited survey is a volume-limited survey.   This is discussed in detail in S14, but to reproduce the trends of $c$ and $x_1$ with redshift shown in Fig.~\ref{fig:snana_hist_lowz} (bottom), we need to put in a redshift dependence of the mean $c$ and $x_1$ that is caused by evolution of demographics.  We find the dependence, which is roughly the slope of the trends in Fig.~\ref{fig:snana_hist_lowz}, to be $\Delta c=-1.0 \times \Delta z$ and  $\Delta {x_1}=25 \times \Delta z$.

One other approach to model the Low-z sample is to study the host galaxy demographics.  As discussed in Section 5.3, there is a large imbalance between high and low-mass host galaxies at low-z.  \cite{Pan14} analyzed the PTF low-z SN sample, which is a rolling and not galaxy-targeted sample, and found that there was a strong imbalance of the number of high-mass and low-mass host galaxies (56/26 for $m_{\rm step}=10.0$ and 48/34 for $m_{\rm step}=10.13$).  It is not as large as that shown for the Low-z sample in Fig.~\ref{fig:mass_global}, however gives a better sense of how much the targeting of galaxies skewed the distribution.  This information can be used with relations between host galaxy mass and SN properties to try to infer characteristics of the Low-z SN sample.

\end{document}